\documentclass{SciPost}

\hypersetup{
    colorlinks,
    linkcolor={red!50!black},
    citecolor={blue!50!black},
    urlcolor={blue!80!black}
}

\usepackage[bitstream-charter]{mathdesign}
\usepackage{bbm}
\usepackage[frozencache]{minted}

\numberwithin{equation}{section}    

\newcommand{\ket}[1]{\left|#1\right\rangle}    
\newcommand{\lmax}{\ell_{\textrm{max}}}

\DeclareSymbolFont{usualmathcal}{OMS}{cmsy}{m}{n}
\DeclareSymbolFontAlphabet{\mathcal}{usualmathcal}

\fancypagestyle{SPstyle}{
\fancyhf{}
\lhead{\colorbox{scipostblue}{\bf \color{white} ~SciPost Physics }}
\rhead{{\bf \color{scipostdeepblue} ~Submission }}

\fancyfoot[C]{\textbf{\thepage}}
}

\DeclareMathOperator{\sinc}{sinc}

\begin{document}

\pagestyle{SPstyle}

\begin{center}{\Large \textbf{\color{scipostdeepblue}{
Vector Spaces for Dark Matter (VSDM): \\[2pt]  Fast Direct Detection Calculations with Python and Julia
}}}\end{center}

\begin{center}\textbf{
Benjamin Lillard\textsuperscript{1$\star$} and
Aria Radick\textsuperscript{1,2$\dagger$} 
}\end{center}

\begin{center}
{\bf 1} Institute for Fundamental Science, University of Oregon, Eugene, OR 97401, USA
\\
{\bf 2} Department of Physics, University of Oregon, Eugene, OR 97401, USA
\\[\baselineskip]
$\star$ \href{mailto:email1}{\small blillard@uoregon.edu}\,,\quad
$\dagger$ \href{mailto:email2}{\small aradick@uoregon.edu}
\end{center}

\section*{\color{scipostdeepblue}{Abstract}}
\textbf{\boldmath{%
Anisotropic target materials are promising candidates for dark matter direct detection experiments, providing a directional sensitivity that can be used to distinguish a dark matter (DM) signal from the various Standard Model backgrounds. In this paper we introduce the Julia and Python implementations of \emph{Vector Spaces for Dark Matter} (VSDM), which handle the difficult scattering rate computation for these rotating, three-dimensional response functions by calculating a partial rate matrix for every combination of DM velocity distribution, material response function, and particle DM properties. 
}}

\vspace{\baselineskip}

\noindent\textcolor{white!90!black}{%
\fbox{\parbox{0.975\linewidth}{%
\textcolor{white!40!black}{\begin{tabular}{lr}%
  \begin{minipage}{0.6\textwidth}%
    {\small Copyright attribution to authors. \newline
    This work is a submission to SciPost Physics. \newline
    License information to appear upon publication. \newline
    Publication information to appear upon publication.}
  \end{minipage} & \begin{minipage}{0.4\textwidth}
    {\small Received Date \newline Accepted Date \newline Published Date}%
  \end{minipage}
\end{tabular}}
}}
}


\vspace{10pt}
\noindent\rule{\textwidth}{1pt}
\tableofcontents
\noindent\rule{\textwidth}{1pt}
\vspace{10pt}

\nocite{Lillard:2023nyj,Lillard:2023cyy} 

\section{Introduction}
\label{sec:intro}

In the field of dark matter direct detection, a need has arisen for a much faster scattering rate calculation.
Refs.~\cite{Lillard:2023nyj,Lillard:2023cyy} solve this problem by projecting the dark matter (DM) velocity distribution and the Standard Model (SM) detector response function onto vector spaces spanned by orthogonal wavelet-harmonic basis functions.
Taking advantage of the symmetries of the DM--SM scattering operator, this vector space integration method calculates the scattering rate using a relatively small number of vector products, reducing the evaluation times for the hardest analyses by up to a factor of $10^8$~\cite{Lillard:2023cyy}. 
In this paper, we present two numerical implementations of this method for public use: VectorSpaceDarkMatter, a Julia package, and \texttt{vsdm}, in Python. 
The two packages are designed with complementary priorities: the Python version emphasizes flexibility in the intermediate stages of the calculation, while the Julia package is designed for speed and simplicity.

If you, the reader, have found yourself spending too much of your time calculating nonrelativistic scattering rates, 
then there is a good chance that VSDM will be useful to you. 
By representing the input functions on an orthogonal basis of functions, and performing the difficult integrals using the basis functions instead, the difficult pieces of an analysis can be factorized from each other. This is particularly helpful when there are multiple versions of each input function. 
Due to some properties of the spherical harmonics outlined in~\cite{Lillard:2023nyj}, the event rate as a function of detector orientation can be entirely represented by an object called the \emph{partial rate matrix}. The main purpose of VSDM is to evaluate this object, and to use it to efficiently calculate the scattering rate. 

The Julia version can be installed by opening a Julia REPL and typing
\begin{align*}
&\texttt{using Pkg}\\
&\texttt{Pkg.add("VectorSpaceDarkMatter")}
\end{align*}
The Python version can be installed from PyPI via the command:
\begin{align*}
\texttt{pip install vsdm}
\end{align*}

After Section~\ref{sec:whyvsdm}, which provides a review of the scaling problem that VSDM is designed to solve,
the remainder of this section is devoted to reviewing the standard literature on direct detection scattering rates.
Sections~\ref{sec:dmsmreview}--\ref{sec:F2DM} present the expression for the DM--SM scattering rate, and its dependence on the details of the DM particle model.   
Section~\ref{sec:fS2} discusses the detector response function in the contexts of DM--nucleus, DM--electron, and DM--phonon scattering. Because several different notations are currently in use, we show explicitly how to convert each response function into our notation.  
Section~\ref{sec:gX} introduces the DM velocity distribution,  the Standard Halo Model, and the integrated velocity distribution or ``inverse average speed'' $\eta$. 
Finally, Section~\ref{sec:etc} describes a few effects (like energy-dependent efficiency factors in the material or detector) that VSDM does not include, and that the user will have to implement themselves.

Section~\ref{sec:method} introduces the vector space version of the rate calculation, largely following the notation of~\cite{Lillard:2023cyy}, but taking care to match the conventions used in the VSDM packages for all of the intermediate steps. Sections~\ref{sec:julia} and~\ref{sec:python} document the Julia and Python packages, respectively, providing examples of how to complete the scattering calculation in each case. 
Section~\ref{sec:performance} compares the evaluation times for the two packages, at each stage of the calculation. The most time-consuming parts of the VSDM calculation are faster by a factor of $10^2$ in the Julia version, making it the best choice for analyses with multiple momentum or velocity distributions, or detailed scans over the particle DM parameter space.
Once the partial rate matrix of~\cite{Lillard:2023nyj} has been evaluated, both versions of the code are extremely efficient at calculating the scattering rate as a function of detector orientation.

\paragraph{Units:} 
Both implementations of the code use units of $\hbar = c = 1$. In this paper, we include the factors of $c$ in relevant equations. 
The Python implementation \texttt{vsdm} uses $c = 1$ as the default option, but it is possible to change the fundamental velocity unit to something else (e.g.~to km/s).

\begin{figure}
    \centering
    \includegraphics[width=0.98\linewidth]{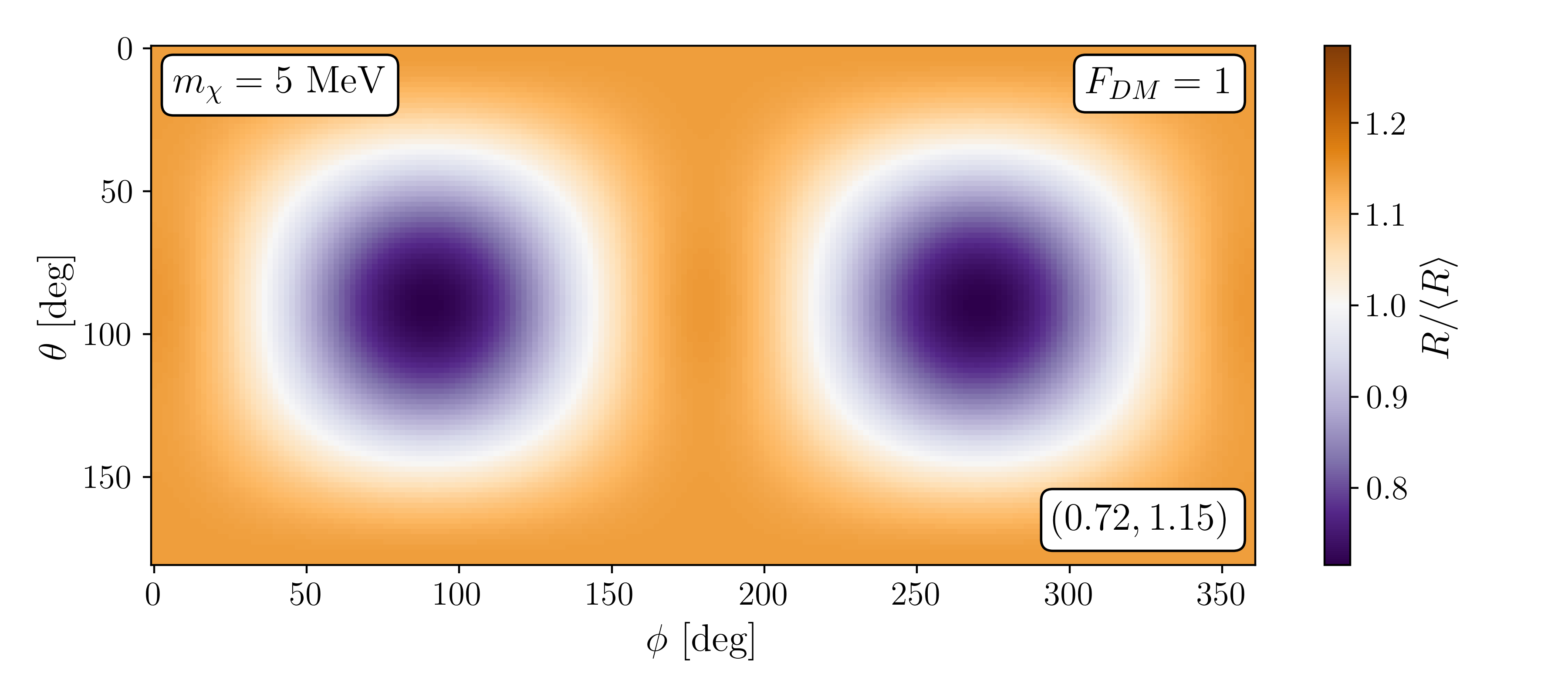}
    
    \vspace{-2ex}
    
    \includegraphics[width=0.98\linewidth]{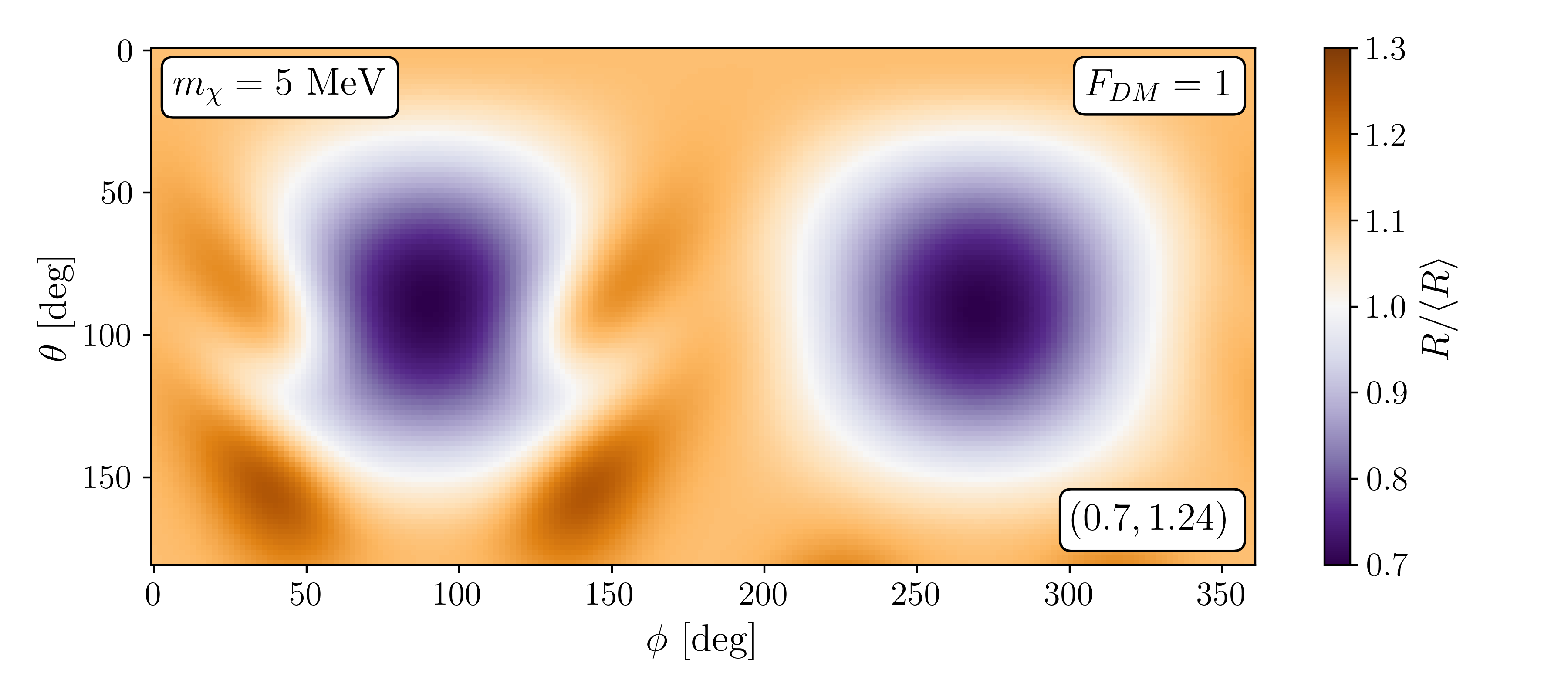}
    \caption{These two plots show a type of analysis that is easy with VSDM, and challenging with traditional methods. Each plot required a scan over $10^4$ detector orientations: with VSDM this is effectively instantaneous. We show the scattering rate $R$ for a set of detector orientations parameterized by $\theta$ and $\phi$, the direction of the Earth velocity vector relative to the detector in spherical coordinates: $\vec v_E = (v_E, \theta, \phi)$.
    Both plots use the same material response function (from from~\cite{Lillard:2023cyy}) and DM model (a 5\,MeV DM candidate with a heavy mediator form factor, $F_\text{DM} = 1$), but the velocity distributions are different. The upper plot assumes the Standard Halo Model, while the lower plot takes 5\% of the DM to be in an off-axis stream with relative speed of 238\,km/s, with the SHM comprising the remaining 95\% of the distribution.
    The bottom right hand corner of each figure denotes the minimum and maximum values of $R$ divided by the angular average rate $\langle R \rangle$.  }
    \label{fig:shm_box_rot}
\end{figure}

\subsection{The Problem That VSDM Solves} \label{sec:whyvsdm}

Except for the very simplest detector systems, predicting the DM scattering rate in a detector requires repeated numerical integration. This is especially challenging for directionally sensitive target materials~\cite{Hochberg:2016ntt,Budnik:2017sbu,Hochberg:2017wce,Coskuner:2019odd,Geilhufe:2019ndy,Griffin:2020lgd,Blanco:2021hlm,Coskuner:2021qxo,Blanco:2022pkt,Boyd:2022tcn,Catena:2023qkj,Catena:2023awl}, but also isotropic semiconductor~\cite{SuperCDMS:2018mne,DAMIC:2019dcn,EDELWEISS:2020fxc,SENSEI:2020dpa,DAMIC-M:2023gxo} 
and molecular targets~\cite{Arvanitaki:2017nhi,Essig:2019kfe,Blanco:2019lrf}.
Not only does the rate integrand depend on the details of the DM model (its mass, and the nature of its couplings to the SM), but it also depends on inputs from astrophysics and condensed matter or physical chemistry, all of which are generally subject to systematic uncertainties~\cite{Wu:2019nhd,Radick:2020qip,Maity:2020wic,Liu:2021avx,Buckley:2022tjy,Maity:2022enp}. 
The scattering rate in an anisotropic detector is a function of the detector's orientation with respect to the DM velocity distribution, which is itself a function of time, varying annually according to the Earth velocity around the Sun (see e.g.~\cite{Lewin:1995rx} for a review). 
Taken together, a careful analysis may need to perform the rate calculation
\begin{align}
N_\text{calcs} &= N_{\mathcal R} N_\text{DM} N_{g_\chi} N_{f_S}
\label{eq:scaling}
\end{align}
many times, where 
$N_\text{DM}$ is the number of DM particle models, specifying its mass and its SM couplings; 
$N_{g_\chi}$ counts the number of distinct lab-frame DM velocity distributions, e.g.~to account for its time dependence and its systematic uncertainties;
$N_{f_S}$ counts how many detector models are included in the analysis, e.g.~including multiple target materials, or cases where several different SM final states contribute to the same experimental observable;
and $N_{\mathcal R}$ is the number of detector orientations.
For an isotropic detector material (e.g.~unpolarized liquids or gases), the rate does not depend on the detector orientation, and $N_{\mathcal R} = 1$.

Figure~\ref{fig:shm_box_rot} shows a relatively straightforward application of VSDM: predicting the scattering rate as a function of detector orientation, for two different velocity distributions.
Both figures assume the material form factor for a particle-in-a-box from~\cite{Lillard:2023cyy},
\begin{align}
    f_s^2(\vec{q}) = \prod_{j=x,y,z} \left[ \frac{\sinc \left(\frac{|q_j L_j| - \pi(n_j-1)}{2} \right)}{1+\pi(n_j-1)/|q_j L_j|} + \frac{\sinc \left(\frac{|q_j L_j| - \pi(n_j+1)}{2} \right)}{1+\pi(n_j+1)/|q_j L_j|} \right]^2
\end{align}
where $L_j$ is the length of the side in the $j$-direction with $(L_x,L_y,L_z) = (4,7,10) \: a_0$ (where $a_0$ is the Bohr radius), and $n_j$ is the excitation mode in the $j$-direction. This figure specifically looks at the $n_x=n_y=1, n_z=2$ mode, the lowest energy excited mode with an excitation energy of $\Delta E = 4.03$ eV.

The top panel of Figure~\ref{fig:shm_box_rot} assumes that 100\% of the DM velocity distribution follows the Standard Halo Model (see~\eqref{eqn:shm}), 
while in the bottom panel a 5\% fraction of the DM is moved into a low-dispersion stream, modeled by a Gaussian with parameters
\begin{align}
    \vec{\mu} &= (v,\theta,\phi) = (238\textrm{ km/s}, \pi/4, -\pi/2) \\
    \sigma &= 23.3 \textrm{ km/s}
\end{align}
In both cases the DM mass is taken as $m_\chi = 5$ MeV, with a heavy mediator for the DM--SM scattering ($F_{\textrm{DM}} = 1$). 
Each plot shows the scattering rate $R(\theta, \phi)$ for a two-dimensional slice of the possible $\mathcal R \in SO(3)$ detector orientations, with $(\theta, \phi)$ indicating the direction of the Earth velocity $\vec v_E$ in spherical coordinates relative to the detector. For the azimuthally symmetric SHM this is a complete description of $R(\mathcal R)$. In the second example, $R$ is also a function of the rotation angle $\beta$ about the $\vec v_E$ axis, not shown in the picture: a 3d scan over $(\theta, \phi, \beta)$ would be required for a complete study of this system.

Even for this 2d subset of $SO(3)$, each plot required a scan over $10^4$ different detector rotations. With traditional methods this would ordinarily require evaluating $10^4$ numeric integrals, but {VSDM} reduces this to $10^4$ dot products involving the partial rate matrix, vastly reducing the computation time.
Repeating this analysis for other DM models ($N_\text{DM} \gg 1$), other excited states for the detector, and other parameters for the DM stream, it is easy to see how $N_\text{calcs}$ of \eqref{eq:scaling} can grow too fast for the traditional methods to handle. 

\subsection{Review: Calculating the DM--SM Scattering Rate} \label{sec:dmsmreview}

In terms of the free particle DM--SM scattering cross section, $\sigma(\vec q, \vec v)$, as a function of momentum transfer $\vec q$ and DM velocity $\vec v$ in the nonrelativistic limit; the number of SM target ``particles'' (e.g.~atoms, molecules, or unit cells of a crystal) $N_T$; 
the DM mass $m_\chi$ and local density $\rho_\chi$; and the reduced mass of the DM--SM system, $\mu_\chi$, the differential scattering rate as a function of deposited energy $E$ is:
\begin{align}
\frac{dR}{dE} \bigg|_{E, \mathcal R} &= 
N_T \frac{\rho_\chi}{m_\chi} \int\! d^3q \, d^3v \, g_\chi(\vec v, t) 
\,\delta\Big(  E + \frac{q^2}{2 m_\chi} - \vec{q} \cdot \vec{v} \Big) \frac{\sigma(\vec q, \vec v) }{4\pi \mu_\chi^2 } \mathcal R \cdot f_{S}^2(\vec q, E) .
\label{eq:rate}
\end{align}
Here $g_\chi(\vec v, t)$ is the lab frame DM velocity distribution, normalized according to
\begin{align}
\int\! d^3 v \, g_\chi(\vec v) \equiv 1.
\end{align}
For a terrestrial experiment, or any other lab that orbits the Sun, the lab frame $g_\chi(\vec v)$ is time-dependent, varying over the course of the year.
The momentum form factor $f_S^2(\vec q, E)$ is our general-purpose notation for the material response function. It has units of inverse energy. 
The rate $R$ has units of events per unit time, and it is a function of the detector orientation $\mathcal R \in SO(3)$, written in \eqref{eq:rate} as a rotation operator acting on the detector form factor. 

For transitions from the ground state ($g$) to states with discrete final energies, labeled by $E = \Delta E_s$ for some set of excited states $s$, the $g \rightarrow s$ scattering rate is:
\begin{align}
R_{g \rightarrow s} (\mathcal R) &= 
N_T \frac{\rho_\chi}{m_\chi} \int\! d^3q \, d^3v \, g_\chi(\vec v, t) 
\,\delta\Big(  \Delta E_s + \frac{q^2}{2 m_\chi} - \vec{q} \cdot \vec{v} \Big) \frac{\sigma(\vec q, \vec v) }{4\pi \mu_\chi^2 } \mathcal R \cdot f_{g\rightarrow s}^2(\vec q) ,
\label{eq:rateDiscrete}
\end{align}
with a different dimensionless material form factor $f_{g \rightarrow s}^2(\vec q)$ for each excited state $s$.

The total scattering rate $R$ is given by integrating \eqref{eq:rate} over energy, 
\begin{align}
R(\mathcal R) &= \int\! dE \frac{d R}{dE} \bigg|_{\mathcal R}, 
\end{align}
or by summing \eqref{eq:rateDiscrete} over all of the relevant excited states:
\begin{align}
R(\mathcal R) &= \sum_{s} R_{g \rightarrow s}(\mathcal R). 
\end{align}
In an idealized, perfectly efficient detector, the expected signal rate is identical to the DM--SM scattering rate $R$.
A more realistic model of the detector adjusts $R$ by some efficiency factor, which may be a function of recoil energy $E$ or the location of the scattering event inside the target material, to account for the attenuation of the signal as it propagates through the detector systems. 
VSDM does not attempt to account for these effects: it simply calculates the DM--SM scattering rates as presented in \eqref{eq:rate} and \eqref{eq:rateDiscrete}.

\paragraph{Kinematics:} 
A simple fact about the kinematics of direct detection can be gleaned from the energy-conserving $\delta$ function: 
in order to deposit energy $E$ into the detector, there must be some part of the phase space where $\vec q \cdot \vec v = E + q^2/(2 m_\chi)$. For a given value of the momentum transfer $q$, the smallest $|\vec v|$ that satisfies this constraint is:
\begin{align}
v_\text{min}(q, E) &\equiv \frac{E}{q} + \frac{q}{2 m_\chi} .
\label{eq:vmin}
\end{align}
This quantity turns out to be highly useful when simplifying the rate integrand. 
For fixed $E$ but varying $q$, the minimum possible value of $v_\text{min}$ occurs when $E/q = q/(2 m_\chi)$:
we label this special point as
\begin{align}
q_\star(E) \equiv \sqrt{ 2 m_\chi E}, 
&&
v_\star(E) \equiv \frac{q_\star}{m_\chi} = \text{min}(v_\text{min}(q, E)).
\label{def:star}
\end{align} 
This $v_\star(E)$ is related in turn to the lower limit of a detector's sensitivity to light DM masses. 
Let $E_\text{min}$ be the lowest possible energy where the detector can still detect scattering events, 
and let $v_\text{max}$ be the largest velocity where the lab frame velocity distribution $g_\chi(\vec v)$ has support. If $v_\star(E_\text{min}) > v_\text{max}$, then the scattering rate is identically zero. 
This occurs for all DM masses below some threshold, $m_\text{min}$, defined as
\begin{align}
m_\text{min}(v_\text{max}) &= \frac{2 E_\text{min} }{v_\text{max}^2} . 
\end{align}
For $m_\chi$ only slightly above this threshold, the event rate is highly suppressed: the only part of $g_\chi(\vec v)$ that can contribute to the scattering must lie within the interval $v_\star(E) \leq v \leq v_\text{min}$.

\paragraph{Rotations and Quaternions:} The rotation operator $\mathcal R \in SO(3)$ is described here as an active rotation operator that acts on the detector response function. This is equivalent to acting on the momentum variable with the operator $\mathcal R^{-1}$:
\begin{align}
\mathcal R \cdot f_S^2(\vec q, E) &= f_S^2(\mathcal R^{-1} \vec q, E).
\end{align}
Because it is the relative orientation of the detector and the DM velocity distribution that matters, in some cases it is easier to use a coordinate frame where the detector material is held fixed, and the operator $\mathcal R^{-1}$ acts on $g_\chi(\vec v)$:
\begin{align}
\mathcal R^{-1} \cdot g_\chi(\vec v) &= g_\chi(\mathcal R \vec v). 
\end{align}

VSDM and one of its dependencies, \texttt{spherical} (\texttt{SphericalFunctions.jl} in the Julia version), specify elements of $SO(3)$ using quaternions. In this language a 3d Cartesian vector $\vec u$ is represented by the imaginary quaternion $\vec u = u_x \hat i + u_y \hat j  + u_z \hat k$, 
and every unit quaternion $Q$ maps onto a rotation operator $\mathcal R$.
Any $\mathcal R \in SO(3)$ can be described as a (right-handed) rotation by an angle $\beta$ about an axis $\hat n$ (a unit vector). 
In this notation, the mapping between $Q$ and $\mathcal R$ is given by:
\begin{align}
Q(\mathcal R) &= \cos\frac{\beta}{2} + \sin\frac{\beta}{2} \hat{n}.
\label{eq:QR}
\end{align}
where $\hat{n}$ is an imaginary unit quaternion, and $\beta$ is an angle $0 \leq \beta < 2 \pi$. 
The action of $\mathcal R$ on a vector $\vec u$ is given by the quaternionic multiplication rule, $\hat i^2 = \hat j^2 = \hat k^2 = \hat i \hat j \hat k = -1$:
\begin{align}
\mathcal R \cdot \vec u &= Q^{-1} \vec u Q , 
\label{eq:RasQ}
\end{align}
and the composition of multiple rotations is given by the product of their quaternions:
\begin{align}
\mathcal R_2 \cdot \left( \mathcal R_1 \cdot \vec u \right) &= Q_2^{-1} Q_1^{-1} \vec u Q_1 Q_2 = (Q_1 Q_2)^{-1} \vec u (Q_1 Q_2) .
\end{align}
Note that $Q$ and $-Q$ encode identical rotations, 
and that for unit quaternions the inverse $Q^{-1}$ is also the complex conjugate of $Q$, 
\begin{align}
Q^\star(\mathcal R)  &=  \cos\frac{\beta}{2} - \sin\frac{\beta}{2} \hat{n} , 
&
Q^\star Q &= 1.
\end{align}
Compared to $3\times 3$ matrices, the quaternion representation is more compact and easier to interpret: i.e.~$\beta$ and $\hat n$ can be extracted directly from the real and imaginary parts of $Q$, respectively, while obtaining this information from a $3\times 3$ matrix (or a set of Euler angles) is much more tedious.

\subsection{DM Model Form Factor} \label{sec:F2DM}

For nonrelativistic DM--SM scattering, the cross section $\sigma(\vec q, \vec v)$ is a function of Galilean invariants: $\vec q$, the momentum transfer; the DM and SM spins, $\vec S_{\chi}$ and $\vec S_i$; and a velocity quantity $\vec v^\perp$~\cite{Fitzpatrick:2012ix}:
\begin{align}
\vec{v}^\perp &\equiv \vec v + \frac{\vec q}{2 \mu_\chi} ,
\end{align}
where $\mu_\chi$ is the reduced mass of the DM--SM system and $\vec v$ is the velocity of the DM particle in the lab rest frame.
Refs.~\cite{Fitzpatrick:2012ix,Catena:2022fnk} have enumerated complete sets of DM--SM couplings in the limit of nonrelativistic DM--nucleon and DM--electron scattering.
Adapting the notation of~\cite{Essig:2011nj}, the free particle cross section can be written in terms of a constant, $\bar\sigma_0$, and a velocity- and momentum-dependent form factor, $F_\text{DM}^2(\vec q, \vec v)$:
\begin{align}
\sigma(\vec q, \vec v) &= \frac{\mu_\chi^2 }{16 \pi m_\chi^2 m_\text{SM}^2 } \overline{ | \mathcal M_{\chi, \text{SM} }(\vec q, \vec v) |^2} 
\\
 &\equiv \bar\sigma_0 \cdot F_\text{DM}^2(\vec q, \vec v).
\end{align}
Here $\overline{|\mathcal M|^2}$ is the spin-averaged squared amplitude for free DM--SM scattering (i.e.~where the DM and SM incoming and outgoing states are eigenstates of momentum), and $F_\text{DM}^2$ is defined such that $F_\text{DM}^2(\vec q_\text{ref}, \vec v_\text{ref}) \equiv 1$ at some reference momentum and velocity, $\vec q_\text{ref}$ and $\vec v_\text{ref}$. 
For nuclear scattering, if $q_\text{ref} \rightarrow 0$ does not produce a well-defined $\bar\sigma_0$, then the characteristic lab-frame momentum of an incoming DM particle provides a natural scale: $q_\text{ref} = m_\chi v_0$, where $v_0$ is the local galactic rotation speed. 
In DM--electron scattering, the Bohr momentum $q_\text{ref} = \alpha m_e c$ is a common choice for $q_\text{ref}$. 

With the $F_\text{DM}^2$ notation, \eqref{eq:rate} becomes:
\begin{align}
\frac{dR}{dE} \bigg|_{E, \mathcal R} &= 
N_T \rho_\chi \bar\sigma_0 \int\! d^3q \, d^3v \, g_\chi(\vec v) 
\,\delta\Big(  E + \frac{q^2}{2 m_\chi} - \vec{q} \cdot \vec{v} \Big) \frac{F_\text{DM}^2(\vec q, \vec v) }{4\pi \mu_\chi^2 m_\chi} \mathcal R \cdot f_{S}^2(\vec q, E) .
\label{eq:rateFDM2}
\end{align}
If the spins of the DM and SM particles are unpolarized, then $F_\text{DM}^2$ is rotationally invariant, even when including the most generic set of effective operators in the nonrelativistic theory:
\begin{align}
F_\text{DM}^2(\vec q, \vec v) = F_\text{DM}^2(q, v, \vec q \cdot \vec v), 
\end{align}
with $q = |\vec q|$ and $v = |\vec v|$. 
Noting the energy-conserving $\delta$ function, instances of $\vec q \cdot \vec v$ can be replaced with $E + q^2/(2 m_\chi)$. 

For spin-independent DM--SM interactions (and also the simple $\mathcal O_4$ spin coupling of~\cite{Fitzpatrick:2012ix,Catena:2022fnk}), the form factor $F_\text{DM}^2$ is a velocity-independent function of $q$ and the scalar or vector mediator mass $m_\phi$. At tree level:  
\begin{align}
F_\text{DM}^2( q) = \left( \frac{q_\text{ref}^2 + m_\phi^2 }{q^2 + m_\phi^2} \right)^2.
\end{align}
Experimental results and projected sensitivities are often shown using the ``heavy mediator'' ($m_\phi \gg q$) and ``light mediator'' ($m_\phi \ll q$) limits of this expression, in which $F_\text{DM}^2(q) \simeq 1$ and $F_\text{DM}^2(q) \simeq (q_\text{ref} / q)^4$, respectively.

In the limits where the mediator mass is either very light or very heavy compared to $q$, the spin--momentum and spin--velocity couplings of~\cite{Fitzpatrick:2012ix,Catena:2022fnk} can all be converted into the form:
\begin{align}
F_\text{DM}^2(q, v) &= \sum_{a,b} d_{a,b}(E) \times \left( \frac{q}{q_\text{ref}} \right)^a \left( \frac{v}{c} \right)^b, 
\label{FDM2:qavb}
\end{align}
for some coefficients $d_{a,b}$, Ref.~\cite{Lillard:2023cyy} derives a number of useful analytic results for the rate calculation. The current versions of VSDM include support for arbitrary values of $a$ and $b$, including non-integer values.
Full support for form factors at finite mediator mass $m_\phi$, 
\begin{align}
F_\text{DM}^2 \propto \left( \frac{1}{q^2 + m_\phi^2 c^2} \right)^2 q^a v^b,
\end{align}
may also be added in future versions of VSDM.

\subsection{Detector Response Form Factors} \label{sec:fS2}
The material response function describes SM physics: given some momentum and energy transfer $\vec q$ and $E$, what is the probability that the detector material will undergo a transition from the ground state into a detectable final state? 
In \eqref{eq:rate}, the total material response is $N_T f_S^2(\vec q, E)$, where $f_S^2$ is the response for a single unit of SM material (e.g.~one molecule, or one atom, or the unit cell of a crystal) and $N_T$ is the total number of these units. 
As a function of the detector target mass $M_T$ and the molar mass $m_T^\text{(mol)}$ of the target atom/molecule/crystal/etc., 
\begin{align}
N_T &= N_A \times \frac{M_T}{m_T^{(\text{mol})}}, 
\label{eq:NTNA}
\end{align}
where $N_A \simeq  6.022 \cdot 10^{23}$ is the Avogadro constant.

In the simplest case, where the SM detector target is modeled as a collection of free particles (e.g.~atomic nuclei~\cite{XENON:2018voc,PandaX-II:2021nsg,LZ:2022ufs,XENON:2023sxq,DarkSide:2018ppu,XENON:2022ltv}), $f_S^2(\vec q, E)$ is simply an energy-conserving $\delta$ function:
\begin{align}
f_S^2(\vec q, E) &=   \delta\!\left( \frac{q^2}{2 m_A} + \Delta E - E \right)  
\end{align}
Here $E$ is the total energy transferred from the DM to the SM system, $m_A$ is the mass of the SM particle, and 
$\Delta E$ is the kinetic energy converted into mass or potential energy (e.g.~scattering an atomic nucleus into an excited nuclear state) in the case of inelastic scattering. The recoil energy of the SM particle is often written as $E_R \equiv q^2 / (2 m_A)$. 
The continuum of final states is labeled according to $E_R$, while $\Delta E$ may take discrete values (or it may simply be zero). 
For DM--nuclear scattering, this expression is modified to accommodate a nuclear form factor, $F^2(q)$:
\begin{align}
f_S^2(\vec q, E) &= F^2(q) \times  \delta\!\left( E_R + \Delta E - E \right) ,
\end{align}
where following the notation of~\cite{Lewin:1995rx}, $F^2(q)$ accounts for the finite size of the nucleus, $r_n$, in the limit where $q \gtrsim r_n^{-1}$. 

For DM--electron or DM--phonon scattering, $f_S^2(\vec q, E)$ depends on the wavefunctions of the SM initial and final states.
In terms of the dynamic structure factor $S(\vec q, \omega)$ of~\cite{Trickle:2019nya}, normalized using the volume of the unit cell of the crystal $V_\text{cell}$:
\begin{align}
f_S^2(\vec q, \omega) &= \frac{V_\text{cell}}{2\pi } S(\vec q, \omega).
\end{align}
In certain DM models, $S(\vec q, \omega)$ can be recast in terms of the material's dielectric function $\epsilon(\vec q, \omega)$~\cite{Hochberg:2021pkt,Knapen:2021run,Boyd:2022tcn}:
\begin{align}
f_S^2(\vec q, \omega) &= \frac{V_\text{cell}  q^2 }{4\pi^2} \frac{1}{\alpha} \text{Im}\left( \frac{-1}{\epsilon(\vec q, \omega) } \right) .
\end{align}
This result applies if the DM couples to the electron number density, and not e.g.~the spins of the SM particles. If the DM--SM interaction is mediated by a dark photon kinetically mixed with the SM photon, this formalism in terms of the energy loss function (ELF) is exact~\cite{Hochberg:2021pkt}, and so this is sometimes referred to as the dark photon model of DM--SM interactions.

For atomic and molecular systems with discrete spectra of final states $s$, with energies $\Delta E_s$,
\begin{align}
f_{S}^{2}(\vec q, E) &= \sum_{s} f_{g \rightarrow s}^2(\vec q) \, \delta(E - \Delta E_s),
\end{align}
with a different momentum form factor $f_{g \rightarrow s}^2(\vec q)$ for each excited state. 
In the case of single particle scattering from the ground state $g$ to an excited final state $s$, $f_{g \rightarrow s}^2(\vec q)$ is given by
\begin{align}
f_{g \rightarrow s}^2(\vec q) &= \left| \int\! d^3 r\, \Psi_s^\star(\vec r) e^{i \vec q \cdot \vec r} \Psi_g(\vec r) \right|^2
= \left| \int \! \frac{d^3 k}{(2\pi)^3}  \psi_s^\star(\vec k + \vec q) \psi_g(\vec k)  \right|^2 ,
\label{eq:fs2}
\end{align}
where $\Psi_{g,s}(\vec r)$ are position space wavefunctions (with normalization $\int \! d^3 r |\Psi(\vec r)|^2 = 1$),
and $\psi_{g,s}(\vec k)$ are the momentum space wavefunctions (normalized $\int \! d^3 k |\psi(\vec k)|^2 = (2\pi)^3$).
Refs.~\cite{Blanco:2019lrf,Blanco:2021hlm} describe how to calculate the momentum form factor $f_{g \rightarrow s}^2(\vec q)$ using this type of expression when the states $\Psi_{g,s}$ are multi-electron wavefunctions.

\medskip

The standard versions of the material form factors $f_S^2(\vec q, E)$ described above are sufficient for describing spin-independent DM--SM scattering, e.g.~with the interaction mediated by a new scalar particle of mass $m_\phi$. These $f_S^2$ functions are also appropriate for some, but not all, of the spin-dependent interactions of Refs.~\cite{Fitzpatrick:2012ix,Catena:2022fnk}, e.g.~the DM--SM spin coupling operator, $\mathcal O_4 = \vec S_\text{SM} \cdot \vec S_\text{DM}$ in a model of fermionic DM with a vector mediator.
To include the other possible DM--SM interactions, however, one must calculate additional material response functions. 
In the case of Ref.~\cite{Fitzpatrick:2012ix}, the full set of operators $\mathcal O_i$ couple to a total of six different types of nuclear response functions $F^2(q)$.
For DM--electron scattering, Ref.~\cite{Catena:2019gfa} introduces a vectorial form factor $\vec f_{g \rightarrow s}(\vec q)$, 
\begin{align}
\vec f_{g \rightarrow s}(\vec q) &= \int\! d^3 r\, \Psi_s^\star(\vec r) e^{i \vec q \cdot \vec r} \frac{i \nabla_{\vec r} }{m_e}  \Psi_g(\vec r) 
= \int \! \frac{d^3 k}{(2\pi)^3}  \psi_s^\star(\vec k + \vec q) \frac{\vec k}{m_e} \psi_g(\vec k) ,
\end{align}
analogous to \eqref{eq:fs2}. 
Ref.~\cite{Catena:2022fnk} enumerates four relevant combinations of $f_{g \rightarrow s}$ and $\vec f_{g \rightarrow s}$ to form the material response functions $\mathcal B_i(\vec q)$, e.g.~$\mathcal B_1(\vec q) = |f_{g \rightarrow s}(\vec q)|^2$
and $\mathcal B_3(\vec q) = (\vec f_{g \rightarrow s})^\star \cdot \vec f_{g \rightarrow s}$. 
Each $\mathcal B_i$ couples to a different ``DM response function'' $\mathcal R_i(q, v)$, which in our notation is proportional to $F_\text{DM}^2( q,  v)$. 
In terms of the $\mathcal W_i(\vec q, E)$ material response functions of~\cite{Catena:2022fnk},
\begin{align}
\bar\sigma_0 F_\text{DM}^2(q, v) f_S^2(\vec q, E) &= \frac{\mu_\chi^2}{16 m_\chi^2 m_e^2 } \frac{1}{E} \sum_{i} \text{Re}\left[ \mathcal R_i^\star(q, v) \mathcal W_i(\vec q, E) \right].
\label{eq:responseWi}
\end{align}
When using VSDM with multiple response functions $\mathcal W_i$ and operators $\mathcal O_j$, 
the product $F_\text{DM}^2 f_S^2$ should be expanded into at least one term per material response function: 
that is, 
\begin{align}
\frac{1}{E} \text{Re}\left[ \mathcal R_i^\star(q, v) \mathcal W_i(\vec q, E) \right] &\equiv \frac{16 m_\chi^2 m_e^2}{\mu_\chi^2} \bar\sigma_0^{(i)} F_{\text{DM}}^{2(i)}(q, v) f_{S( i)}^2(\vec q, E) , 
\\
\bar\sigma_0 F_\text{DM}^2(q, v) f_S^2(\vec q, E) &= \sum_i \bar\sigma_0^{(i)} F_{\text{DM}}^{2(i)}(q, v) f_{S( i)}^2(\vec q, E) .
\end{align}

\paragraph{Monomial Format in VSDM:} 
The VSDM calculation is most efficient if $F_\text{DM}^{2(i)}$ is expanded in monomials of $q$ and $v$ as in \eqref{FDM2:qavb},
\begin{align}
F_\text{DM}^{2(i)}(q, v) &= \sum_{a,b} d_{a,b}^{(i)} (E) \times \left( \frac{q}{q_\text{ref}} \right)^a \left( \frac{v}{c} \right)^b.
\label{FDM2:qavb}
\end{align}
After VSDM calculates the component of the rate generated by each $(i, a, b)$ combination, these contributions can be added together:
\begin{align}
\frac{dR}{dE} \bigg|_{E, \mathcal R} &= \frac{N_T \rho_\chi }{4\pi \mu_\chi^2 m_\chi }  \sum_{i,a,b} \bar\sigma_0^{(i)}(E)   \int\! d^3q \, d^3v \, g_\chi(\vec v) 
\,\delta\Big(  E + \frac{q^2}{2 m_\chi} - \vec{q} \cdot \vec{v} \Big)   \frac{q^a v^b}{q_\text{ref}^a c^b} \mathcal R \cdot f_{S(i)}^2(\vec q, E) ,
\end{align}
by absorbing the dimensionless coefficients $d_{a b}^{(i)}$ into the value of $\bar \sigma_0^{(i)}$ used in the exposure factor:
\begin{align}
\bar\sigma_0^{(i)}(E) &\equiv \bar\sigma_0^{(i)} \, d_{ab}^{(i)}(E) .
\end{align}

\subsection{The DM Velocity Distribution}  \label{sec:gX}

The final ingredient in the rate calculation is the local DM velocity distribution. Much of the literature uses the notation $f_\chi(\vec v)$ for this quantity. We feel, however, that there are already sufficiently many physical quantities represented by the letter $f$ or $F$, so following~\cite{Lillard:2023nyj,Lillard:2023cyy}  we use the notation $g_\chi(\vec v)$. 

VSDM will accommodate any 3d velocity distribution, as long as the DM speeds are comfortably non-relativistic. 
There are scenarios in which a subcomponent of the local dark matter could be boosted to relativistic speeds: cosmic ray upscattering~\cite{Yin:2018yjn,Bringmann:2018cvk}, for example, or models in which the DM interacts with or is produced by a hot SM source (e.g.~the Sun~\cite{An:2017ojc,Emken:2017hnp}). 
These relativistic effects are not currently included in VSDM. 


\paragraph{The Earth Velocity and \texttt{earthspeed} Package:}
Any static DM velocity distribution in the galactic rest frame becomes time-dependent when observed in the lab frame, assuming the lab is orbiting the Sun. 
Following~\cite{Lillard:2023nyj}, we have organized the rate equation \eqref{eq:rate} so that this time-dependence is accounted for in the lab frame velocity distribution $g_\chi(\vec v, \vec v_E)$,
where $\vec v_E(t)$ is the lab velocity relative to the galactic center. 
In terms of the Earth velocity with respect to the Sun, $v_\oplus(t)$, 
$\vec v_E(t)$ varies over the course of the year according to:
\begin{align}
\vec v_E(t) &\simeq \vec v_\text{LSR} + \vec v_\odot + \vec v_\oplus(t), 
\end{align}
where $\vec v_\odot$ is the peculiar velocity of the Sun relative to the LSR. On average,  $\langle | \vec v_\oplus |\rangle \simeq 29.8\,\text{km/s}$. Taking into account the angle between $\vec v_\odot$ and the plane swept out by $\vec v_\oplus(t)$, the annual variation on $|\vec v_E(t)|$ is approximately $\pm 15\,\text{km/s}$.
Refs.~\cite{Lewin:1995rx,Baxter:2021pqo} provide simplified expressions for $\vec v_E(t)$, and Refs.~\cite{Lee:2013xxa,McCabe:2013kea} correctly account for the ellipticity of the Earth's orbit. 
Using an LSR speed of 238\,km/s and the updated parameters for $\vec v_\odot$ and $\vec v_\oplus$ from~\cite{McCabe:2013kea,Baxter:2021pqo}, 
in 2024 the Earth speed reached a maximum of $v_E \simeq 266.2\,\text{km/s}$ on May 31, 11:59 UTC, and a minimum of $v_E \simeq 237.2\, \text{km/s}$ at Dec.~2, 4:07 UTC.

\begin{figure}
\centering
\includegraphics[width=0.99\textwidth]{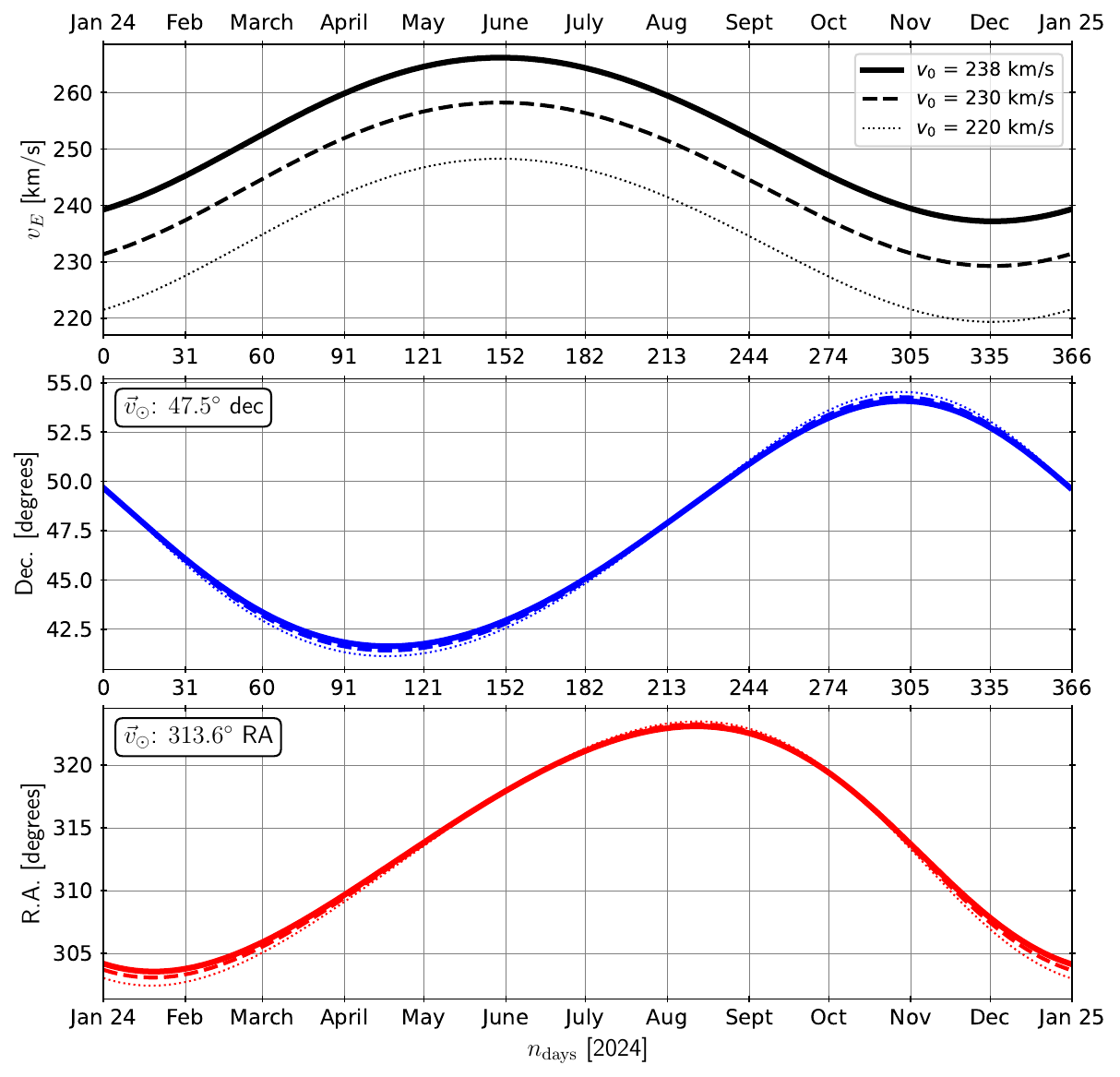}
\caption{
	The Earth velocity $\vec v_E(t)$, in the spherical coordinate system of declination (dec) and right ascension (RA) familiar to astronomers, for three values of the LSR speed: $v_0 = (220, 230, 238)$\,km/s.
	Time of year $t$ is measured in the number of days since UTC midnight on January 1, 2024. 
	The angle between the north pole and $\vec v_E$ is $\theta_N \equiv 90^\circ - \text{dec}$. Right ascension specifies the longitudinal position of $\vec v_E$ on the celestial sphere: this determines what time of day the DM wind reaches its highest and lowest points on the sky, for a lab location on or near the surface of the Earth.
}
\label{fig:SHM}
\end{figure}

A public Python package, \texttt{earthspeed}~\cite{Lillard_earthspeed_2024}, implements the precise version of $\vec v_\oplus(t)$ using the updated values of $v_\text{LSR}$ and $\vec v_\oplus$ from~\cite{Baxter:2021pqo}.
For future  reference, Figure~\ref{fig:SHM} shows how the Earth speed $|\vec v_E(t)|$ and direction of motion $\hat v_E(t)$ change from month to month. 
Several studies of anisotropic detectors and daily modulation (e.g.~\cite{Coskuner:2019odd}) define a $\hat z$ direction in the lab based on the DM wind velocity ($- \vec v_E$) at some particular time of day. For convenience to the experimentalist, Figure~\ref{fig:SHM} converts $\hat v_E(t)$ into the celestial coordinate system of right ascension (RA) and declination (dec). Identifying the $\hat z \propto \hat v_E(t = 0)$ direction is then as simple as aligning a telescope to point at a particular star (from the basement of a building, or an underground cavern).


\paragraph{The Standard Halo Model:} 
The DM velocity distribution in our neighborhood of the Milky Way is essentially unknown. For the purposes of comparing different DM experiments, however, it is convenient to approximate $g_\chi(\vec v)$ using a simplified model. In the Standard Halo Model (SHM), $g_\chi(\vec v)$ is a Maxwell--Boltzmann distribution, with a cutoff at the galactic escape velocity $v_\text{esc}$. In the galactic rest frame the SHM is isotropic, 
but in the lab frame it is
\begin{align}\label{eqn:shm}
g_\text{SHM}(\vec v, t) &= \frac{1}{N_0} \exp\left( - \frac{|\vec v + \vec v_E(t) |^2 }{v_\sigma^2 } \right) \Theta(v_\text{esc} - |\vec v + \vec v_E(t)| ),
\end{align}
where $\vec v_E(t)$ is the lab velocity (on Earth) relative to the galactic rest frame, and $N_0(v_\sigma, v_\text{esc})$ is a normalization factor. The Heaviside $\Theta$ function imposes a discontinuously sharp cutoff at the escape velocity.
The velocity dispersion, $v_\sigma$, is taken to match the local circular speed of the galaxy, i.e.~the local standard of rest (LSR). Many older references use $v_\sigma = 220\, \text{km/s}$ or $v_\sigma = 230\,\text{km/s}$~\cite{Lewin:1995rx}, but more recent measurements suggest $v_\sigma \approx 238\, \text{km/s}$~\cite{Baxter:2021pqo}.
For the escape speed, Ref.~\cite{Baxter:2021pqo} recommends $v_\text{esc} = 544\, \text{km/s}$.

While the SHM is a useful standard for predicting scattering rates in direct detection experiments, it is not supposed to be a particularly accurate representation of the as-yet unknown velocity distribution in our galaxy.
Data from the Milky Way (especially the SDSS and Gaia sky surveys) indicate that the local stellar velocity distribution contains unvirialized substructures~\cite{SDSS:2000hjo,Gaia:2018ydn} from relatively recent galactic merger events, which may well be accompanied by matching kinematic substructures in the DM velocity distribution~\cite{Necib:2018iwb,Evans:2018bqy,Bozorgnia:2019mjk}.
The SHM also neglects the gravitational effects from the Sun and Earth, 
which accelerate the incoming DM particles according to
\begin{align}
v_f^2 - v_i^2 &\simeq 2 \frac{G M_\odot}{1\, \text{AU}} + 2 \frac{G M_\oplus}{R_\oplus} \approx 2.113 \cdot 10^{-8} c^2, 
&
v_f &\simeq \sqrt{ v_i^2 + (43.58 \, \text{km/s})^2 } ,
\end{align}
with the Sun doing about 93\% of the work.
This skews the lab-frame velocity distribution towards speeds. For $v_i = 250\,\text{km/s}$ the effect is modest, increasing $v_i$ by 3.7\,km/s. For the slower part of the distribution the effect is more pronounced: an incoming $v_i = 100\,\text{km/s}$ would be boosted by 9.1\,km/s, for example, 
while the entire $0 \leq v_i \lesssim 44\,\text{km/s}$ part of the distribution shifts to $44\,\text{km/s} \lesssim v_f \lesssim 62\,\text{km/s}$. Likewise, the gravitational potential of the Sun is known to perturb the local DM density, inducing an annual variation that impacts  the predictions for annual modulation searches~\cite{Lee:2013wza}.

Additionally, the surface speed of the Earth slightly shifts the lab frame velocity distribution by an amount that decreases with latitude, from 0.465\,km/s at the equator to zero at the poles. Compared to the effects from annual modulation and the gravitational potentials of the Sun and Earth, the surface speed is a tiny perturbation, altering $\vec v_E(t)$ by no more than 0.2\,\%. However, the surface velocity is periodic over the sidereal day (23.93 hours), which does distinguish it from the annual effects. In this sense the Earth surface speed provides a minimum degree of daily modulation in materials that are highly isotropic.
Anisotropic materials can have daily modulation amplitudes that are two orders of magnitude larger than this (e.g.~\cite{Blanco:2021hlm,Blanco:2022pkt}).

\paragraph{The Integrated Velocity Distribution:} 

It is common in the literature to define an integrated version of the velocity distribution, usually notated by $\eta$, sometimes referred to as the average inverse speed. 
The usual derivations assume either that the detector response is isotropic, or that there is some velocity frame in which the DM velocity distribution is isotropic. In these cases, some or all of the angular integrals in \eqref{eq:rate} can be completed analytically. 
Refs.~\cite{Trickle:2019nya,Lillard:2023cyy} provide a more general version of this quantity, that does not rely on either assumption.
For spin-independent scattering, the quantity
\begin{align}
\eta(\vec q, E) &\equiv 2q  \int d^3 v \, g_\chi(\vec v) \, \delta\!\left(E + \frac{q^2}{2 m_\chi} - \vec q \cdot \vec v \right)
\end{align}
is sufficient for calculating the scattering rate; but for velocity-dependent $F_\text{DM}^2(q, v)$ form factors $F_\text{DM}^2 \propto v^b$,  one must also evaluate
\begin{align}
\eta^{(b)}(\vec q, E) &\equiv 2q  \int d^3 v \frac{v^b}{c^b} \, g_\chi(\vec v) \, \delta\!\left(E + \frac{q^2}{2 m_\chi} - \vec q \cdot \vec v \right) .
\end{align}
The factor of $2 q$ here ensures that the angular average of our $\eta(\vec q, E)$ matches the isotropic version, $\eta(q, E)$, that appears elsewhere in the literature. For comparison with the related quantity $g(\vec q, \omega)$ of~\cite{Trickle:2019nya}, note that $\eta(\vec q, E) = (q/\pi) g(\vec q, E)$. 
In the special case where $g_\chi(\vec v)$ is isotropic in some other velocity frame, Ref.~\cite{Lillard:2023cyy} shows how $\eta(\vec q, E)$ can be reduced to a 1d integral over speeds.
For the SHM, even this integral can be completed analytically: the result is
\begin{align}
\eta(\vec q, E) &= \frac{2\pi v_\sigma^2 }{ N_0} \left( e^{- v_-(\vec q, E)^2/v_\sigma^2 } - e^{- v_\text{esc}^2/v_\sigma^2 } \right) \Theta(v_\text{esc} - v_-(\vec q, E)),
\end{align}
where 
\begin{align}
v_-(\vec q, E) &\equiv \frac{E}{q} + \frac{q}{2 m_\chi} + \frac{\vec q \cdot \vec v_E}{q} .
\end{align}

If a closed form analytic expression can be obtained for $\eta(\vec q, E)$, then this quantity can be quite helpful for calculating the scattering rate.
Expanding $F_\text{DM}(q, v)$ in powers of $v$,
\begin{align}
F_\text{DM}^2(q, v) &= \sum_b d_b(E) \frac{v^b}{c^b} F_\text{DM}^{2(b)}(q) , 
\end{align}
the differential scattering rate becomes
\begin{align}
\frac{dR}{dE} \bigg|_{E, \mathcal R} &= \sum_b d_b(E) \frac{N_T \rho_\chi \bar\sigma_0}{8\pi \mu_\chi^2 m_\chi} \int\! \frac{d^3q}{q} \eta^{(b)}(\vec q, E)  \, F_\text{DM}^{2(b)}(q)  \, \mathcal R \cdot f_{S}^2(\vec q, E) .
\end{align}
If $\eta(\vec q, E)$ is obtained via numerical integration, however, this version of the rate calculation 
 may become even more difficult than directly evaluating \eqref{eq:rate}. 
Tabulating $\eta(\vec q, E)$ a single time would be bad enough, but this calculation needs to be repeated for every value of the DM mass and the Earth velocity.
In such a case it is almost certainly better to use the VSDM method outlined in Section~\ref{sec:vsdm} instead.

\subsection{What VSDM Does and Does Not Calculate} \label{sec:etc}

There are some physical effects that VSDM does not include, but which are straightforward for the user to implement themselves. 
Spatial or temporal variations in the local DM density $\rho_\chi$ are one such example: e.g.~if gravitational focusing effects from the Sun cause $\rho_\chi(t)$ to vary over the course of the year. As $\rho_\chi$ simply rescales the overall event rate, this type of time dependence can be accommodated by multiplying the output of VSDM by a factor of $\rho_\chi(t) / \rho_\chi(t = 0)$. 

Similarly, there may be some efficiency factors that suppress the actual number of events measured by a detector. In this case, the predicted average rate of events with $E$ in some range $E_1 \leq E \leq E_2$ would be 
\begin{align}
R(E_1 \leq E \leq E_2,  \mathcal R) &= \int_{E_1}^{E_2} \! dE \, \xi(E) \frac{dR}{dE} \bigg|_{E, \mathcal R} , 
\end{align}
where $\xi(E)$ is some efficiency factor (as a function of energy deposition $E$). For this reason and others, VSDM generally only calculates the differential rate $dR/dE$: it is left to the user to perform any integrals over energy.
The exception to this rule is for discrete final states, where
\begin{align}
f_S^2(\vec q, E) &= f_{g \rightarrow s}^2(\vec q) \delta(E - \Delta E_s) .
\end{align}
Given $f_{g \rightarrow s}^2(\vec q)$, VSDM calculates the rate $R(\mathcal R)$ directly using \eqref{eq:rateDiscrete}.

VSDM also does not calculate $f_S^2(\vec q, E)$ or $g_\chi(\vec v)$ from first principles: these inputs should generally be provided by the user. Due to the popularity (and simplicity) of the SHM, however, the Python repository \texttt{vsdm} does include some SHM-specific tools for the user's convenience. 
For analyses with continuous energy spectra, calculating $dR/dE |_{E}$ using \eqref{eq:rate}, the user should provide the form factor $f_S^2(\vec q, E)$ evaluated at the same set of $E = \{ E_j \}$ values as $dR/dE$. 
One of the intermediate steps in the VSDM calculation involves calculating angular integrals of the form 
\begin{align}
\int\! d\Omega_q\, Y_{\ell m}(\hat{q}) \, f_S^2(\vec q, E),
\end{align}
for real spherical harmonics $Y_{\ell m}$. If the user has previously tabulated $f_S^2(\vec q, E)$ on a grid of $(\vec q_i, E_j)$ values, then $f_S^2$ can be provided to VSDM as an interpolating function.
Alternatively, if $f_S^2(\vec q, E)$ is tabulated on a spherical grid in $\vec q = (|q|, \theta, \phi)$, then the necessary angular integrals can be evaluated in large batches using a discrete spherical harmonic transformation instead.
We highly recommend this approach when working with tabulated $f_S^2(\vec q, E)$: compared to numerical integration, the discrete harmonic transformation is extremely fast.

\section{Vector Space Rate Calculation} \label{sec:vsdm}
\label{sec:method}

VSDM implements the vector space integration method introduced in~\cite{Lillard:2023nyj} and described in detail in~\cite{Lillard:2023cyy}. 
In this Section we provide the key results and intermediate steps, describing how VSDM treats the units and dimensionful quantities at each stage. 
Compared to Ref.~\cite{Lillard:2023cyy}, the discussion in this section works backwards: we begin by introducing the final result for the rate in terms of the partial rate matrix $K^{(\ell)}_{mm'}$, then later in the section we go through the intermediate calculations that are designed to evaluate $K^{(\ell)}$ efficiently.

VSDM makes the following assumptions:
\begin{itemize}
\item the lab-frame $g_\chi(\vec v)$ is entirely nonrelativistic, vanishing above some cutoff $v_\text{max}$: 
\begin{align} g_\chi(|\vec v| > v_\text{max}) = 0. \end{align}
\item the cross section $\sigma(\vec q, \vec v)$ is a scalar function of $q$, $v$, and $\vec q \cdot \vec v$:  
\begin{align} F_\text{DM}^2(\vec q, \vec v) = F_\text{DM}^2(q, v, E). \end{align}
\end{itemize}

\paragraph{The Partial Rate Matrix:} 
Under these assumptions, Ref.~\cite{Lillard:2023nyj} shows that the differential scattering rate as a function of detector orientation $\mathcal R$ is entirely described by the ``partial rate matrix'' $K^{(\ell)}_{m m'}$, actually a series of matrices that transform as representations of $SO(3)$. 
Like the scattering rate itself, the partial rate matrix is a function of the DM model $(m_\chi, F^2_\text{DM})$ and the physical inputs $g_\chi(\vec v)$ and $f_S^2(\vec q, E)$.
Its differential form is defined~\cite{Lillard:2023nyj}: 
\begin{align}
\frac{dK^{(\ell)}_{m m'} }{dE} \bigg|_E &= \int\! \frac{d^3 q \,d^3 v}{q v} P_\ell\!\left( \frac{ v_\text{min}(q, E) }{v} \right) \, Y_{\ell m}(\hat v) g_\chi(\vec v) \frac{F_\text{DM}^2(q, v)}{2 \mu_\chi^2 m_\chi} Y_{\ell m'}(\hat q) f_S^2(\vec q, E),
\label{def:dKdE}
\end{align}
where $P_\ell(x)$ is the $\ell$th Legendre polynomial, $Y_{\ell m}(\theta, \phi)$ are the real spherical harmonics (see Sec.~\ref{sec:harmonics}), and $v_\text{min}$ is the  function defined in \eqref{eq:vmin}.
The velocity integral is restricted to the region where $v_\text{min} \leq v$, and the argument of $P_\ell$ is less than or equal to 1. 

Although the value of $dK/dE$ is a function of the initial orientation of $f_S^2(\vec q, E)$, it is not a function of $\mathcal R$.
Instead, the scattering rate $R(\mathcal R)$ is found by multiplying $K^{(\ell)}_{m m'}$ by the ``Wigner $G$ matrix'' of~\cite{Lillard:2023cyy}:
\begin{align}
\frac{dR}{dE} \bigg|_{E, \mathcal R} &= N_T \rho_\chi \bar\sigma_0 \sum_{\ell = 0}^\infty \sum_{m = -\ell}^{\ell} \sum_{m' = - \ell}^\ell G^{(\ell)}_{m m'}(\mathcal R) \cdot \frac{dK^{(\ell)}_{m m'} }{dE} \bigg|_E ,
\label{eq:dRdEK}
\end{align}
where $G^{(\ell)}_{m m'}(\mathcal R)$ is defined as
\begin{align}
G^{(\ell)}_{ m m'}(\mathcal R) &\equiv \int\! d\Omega_u \, Y_{\ell m}(\hat u) \, \mathcal R \cdot Y_{\ell m'}(\hat u) 
= \int\! d\Omega_u \, Y_{\ell m}(\hat u) \,Y_{\ell m'}(\mathcal R^{-1} \hat u) .
\label{def:WignerG}
\end{align}
Although \eqref{def:WignerG} is a clear way to define $G$, this integral is not the most convenient way to calculate $G(\mathcal R)$. Instead, after calculating the standard Wigner $D^{(\ell)}(\mathcal R)$ matrix using the package \texttt{spherical} (\texttt{SphericalFunctions.jl}), VSDM uses the relationship between $G^{(\ell)}$ and $D^{(\ell)}$ from~\cite{Lillard:2023cyy} to determine $G^{(\ell)}(\mathcal R)$.

VSDM works by truncating the expansion in $\ell$ at some finite $\ell_\text{max}$:
\begin{align}
\frac{dR}{dE} \bigg|_{E, \mathcal R} &\simeq \sum_{\ell= 0}^{\ell_\text{max}} \frac{dR_{(\ell)}}{dE} \bigg|_{E, \mathcal R}, 
\\
\frac{dR_{(\ell)}}{dE} \bigg|_{E, \mathcal R} &=  N_T \rho_\chi \bar\sigma_0 \sum_{m, m' = -\ell}^{\ell} G^{(\ell)}_{m m'}(\mathcal R) \cdot \frac{dK^{(\ell)}_{m m'} }{dE} \bigg|_E .
\end{align}
For each value of $\ell$, the sum over $m, m'$ is equivalent to taking the dot product of two vectors of length $(2\ell + 1)^2$. 
Including all $\ell \leq \ell_\text{max}$, the total number of terms in \eqref{eq:dRdEK} grows as $4\ell_\text{max}^3/3$ in the limit of large $\ell_\text{max}$.

Given a particular precision goal, the appropriate value for $\ell_\text{max}$ can be estimated \textit{a priori} by inspecting the sizes of $g_{\ell m}(v)$ and $f_{\ell m'}^2(q, E)$ input functions.
To verify that $\ell_\text{max}$ is large enough, however,  the user should inspect the relative magnitudes of $dR_{(\ell)} / dE$ as a function of $\ell$ at the end of the calculation,
to verify that this series does converge to within the desired precision. 

For discrete final states, $f_S^2(\vec q, E) = f_{g \rightarrow s}^2(\vec q) \delta(E - \Delta E)$, the differential form of $K$ is replaced by:
\begin{align}
K_{m m'}^{(\ell)}(\Delta E) &= \int\! \frac{d^3 q \,d^3 v}{q v} P_\ell\!\left( \frac{ v_\text{min}(q, \Delta E) }{v} \right) \, Y_{\ell m}(\hat v) g_\chi(\vec v) \frac{F_\text{DM}^2(q, v)}{2 \mu_\chi^2 m_\chi} Y_{\ell m'}(\hat q) f_{g \rightarrow s}^2(\vec q),
\label{def:KdeltaE} ,
\\
R(\mathcal R, \Delta E) &= N_T \rho_\chi \bar\sigma_0 \sum_{\ell = 0}^\infty \sum_{m = -\ell}^{\ell} \sum_{m' = - \ell}^\ell G^{(\ell)}_{m m'}(\mathcal R) \cdot K^{(\ell)}_{m m'}(\Delta E) .
\end{align}

\paragraph{Result:}

The dot product version of \eqref{eq:dRdEK} needs to be evaluated for each of the $N_\text{DM} N_{\mathcal R} N_{g_\chi} N_{f_S}$ combinations of model parameters and physical inputs described in \eqref{eq:scaling}.
With typical values of $\ell_\text{max} \sim 10^1$, each rate calculation might take some number of microseconds to evaluate. Using an earlier version of the Python implementation \texttt{vsdm}, for example, Ref.~\cite{Lillard:2023cyy} found that the evaluation time in seconds was approximately $T_\text{eval} \sim \ell_\text{max}^3 \times 10^{-9}\,\text{s}$. 
Given $dK^{(\ell)}_{m m'}/dE$ for each combination of models, a typical personal computer using \texttt{vsdm} could calculate the DM scattering rate for billions of combinations of $\mathcal R$, $g_\chi$, $f_S^2$, $m_\chi$ and $F_\text{DM}$ per hour.
The Julia version is faster.

\medskip 

Now that the importance of the partial rate matrix has been established, the remainder of this section describes the steps VSDM takes to calculate it. 
Sections~\ref{sec:harmonics} and~\ref{sec:wavelets} begin by defining a basis of orthogonal 3d functions, in terms of real spherical harmonics and the spherical Haar wavelets introduced in Ref.~\cite{Lillard:2023cyy}. 
A user familiar with~\cite{Lillard:2023cyy} can skip directly to Section~\ref{sec:basis}, which 
specifies the conventions that we use to keep track of the dimensionful quantities.

\subsection{Spherical Harmonics} \label{sec:harmonics}

VSDM uses real spherical harmonics for its rate calculation, because the input functions $g_\chi(\vec v)$ and $f_S^2(\vec q, E)$ are real-valued.
As in~\cite{Lillard:2023cyy},  for polar and azimuthal angles $\theta, \phi$, $Y_{\ell m}(\theta, \phi)$ is defined explicitly as:
\begin{align}
Y_{\ell m}(\theta, \phi) &=
\left\{ \begin{array}{l c c}
\sqrt{2}  (-1)^m    \sqrt{ \dfrac{2\ell + 1}{4\pi} \dfrac{(\ell - |m|)!}{(\ell+|m|)!} } P_\ell^{|m|} (\cos\theta) \sin(|m|\phi)
&& \text{for } m < 0,
\\[\bigskipamount]
\sqrt{ \dfrac{2\ell + 1}{4\pi} } P_\ell(\cos\theta)  && \text{for } m = 0,
\\[\bigskipamount]
\sqrt{2}   (-1)^m   \sqrt{ \dfrac{2\ell + 1}{4\pi} \dfrac{(\ell - m)!}{(\ell+m)!} } P_\ell^m (\cos\theta) \cos(m\phi)
&& \text{for } m > 0,
\end{array}  
\right.
\end{align}
where $P_\ell^m(x)$ on $-1 \leq x \leq 1$ are the associated Legendre polynomials, 
\begin{align}
P_\ell^m(x) &\equiv \frac{(-1)^m}{2^\ell \ell!}  (1 - x^2)^{m/2} \frac{d^{\ell+m}}{dx^{\ell+m}}  (x^2 - 1)^\ell. 
\end{align}
To avoid numerical instabilities at large $\ell$, VSDM uses the efficient iterative method outlined in Appendix~\ref{sec:legendre} to calculate $\sqrt{(\ell - m)! / (\ell + m )!} P_{\ell}^m(x)$ directly. 

\paragraph{Properties:} 
Under central inversion, $\hat u \rightarrow -\hat u$, i.e.~$\theta \rightarrow \pi - \theta$ and $\phi \rightarrow \phi \pm \pi$, the harmonics transform as 
\begin{align}
Y_{\ell m}(- \hat u) = Y_{\ell m}(\pi - \theta, \phi \pm \pi ) = (-1)^{\ell} Y_{\ell m}(\theta, \phi),
\end{align}
while under reflections along the $x$ axis ($\hat x \rightarrow - \hat x$, or $\phi \rightarrow -\phi$), 
\begin{align}
Y_{\ell m}(\theta, - \phi) =  (-1)^{m} Y_{\ell m}(\theta, \phi).
\end{align}
Under rotations, a spherical harmonic $Y_{\ell m}$ mixes with the other $Y_{\ell m'}$ harmonics with matching $\ell$:
\begin{align}
\mathcal R \cdot Y_{\ell m}(\hat u) &= \sum_{m'} G_{m' m}^{(\ell)}(\mathcal R) \,Y_{\ell m'}(\hat u).
\label{eq:rotG}
\end{align}
For the trivial rotation $\mathcal R = \mathbbm{1}$, the identity operator, $G^{(\ell)}$ is just the Kronecker $\delta$ function:
\begin{align}
G^{(\ell)}_{m m'}(\mathbbm{1}) &= \delta_{m m'}. 
\end{align} 
For the relationship between $Y_{\ell m}$ and the complex spherical harmonics $Y_\ell^m$, we refer the reader to Ref.~\cite{Lillard:2023cyy}. Likewise, this reference also provides the explicit expression for $G_{mm'}^{(\ell)}(\mathcal R)$ as a function of $D_{m m'}^{(\ell)}(\mathcal R)$. 


The spherical harmonics provide an orthonormal basis for real-valued functions on the sphere $S^2$: their inner product is given by
\begin{align}
\int\! d\Omega \, Y_{\ell m}(\theta, \phi) Y_{\ell', m'}(\theta, \phi) &= \delta_\ell^{\ell '} \delta_{m}^{m'} . 
\end{align}
A function $f(\vec u)$ of a 3d variable $\vec u$ can be written as a series expansion in spherical harmonic modes:
\begin{align}
f(\vec u) &= \sum_{\ell = 0}^\infty \sum_{m = - \ell}^\ell   f_{\ell m}(u)  Y_{\ell m}(\hat u),  
\end{align}
defining the $(\ell, m)$ radial function $f_{\ell m}(u)$ as 
\begin{align}
 f_{\ell m}(u) &\equiv  \int\! d\Omega_u \,Y_{\ell m}(\hat u) \, f(\vec u).
\end{align}

\paragraph{Revisiting the Partial Rate Matrix:} 

With this notation, the partial rate matrix \eqref{def:dKdE} simplifies to a series of 2d integrals on $q$ and $v$:
\begin{align}
\frac{dK^{(\ell)}_{m m'} }{dE} \bigg|_E &= \int_0^\infty\! q d q \int_{v_\text{min}(q, E)}^{v_\text{max}} v d v \,  P_\ell\!\left( \frac{ v_\text{min}(q, E) }{v} \right) \, 
g_{\ell m}(v)  \cdot \frac{F_\text{DM}^2(q, v) }{2 \mu_\chi^2 m_\chi} \cdot f_{\ell m'}^2(q, E) ,
\label{eq:dK2d}
\end{align}
where
\begin{align}
g_{\ell m}(v) &\equiv \int\! d\Omega_v \, Y_{\ell m}(\hat v) g_\chi(\vec v), 
&
f^2_{\ell m'}(q, E) &\equiv \int\! d\Omega_q \, Y_{\ell m'}(\hat q) f_S^2(\vec q, E). 
\end{align}

If the input functions $g_\chi(\vec v)$ and $f_S^2(\vec q, E)$ allow  \eqref{eq:dK2d} to be evaluated analytically, then our work is done: 
$dR/dE$ can be found directly from  \eqref{eq:dRdEK}. 
If the number of $g_\chi$ and $f_S^2$ models is small, then it might be feasible to integrate \eqref{eq:dK2d} numerically. That said,  following \eqref{eq:scaling}, there are
\begin{align}
N_\text{DM} N_{g_\chi} N_{f_S} \times \left( \frac{4}{3} \ell_\text{max}^3 + \mathcal O(\ell_\text{max}^2) \right) 
\end{align}
many of these 2d integrals to evaluate. 
To solve this problem, Section~\ref{sec:wavelets} introduces the spherical wavelets, replacing $g_{\ell m}(v)$ and $f_{\ell m'}^2(q, E)$ with a set of simpler basis functions, 
for which \eqref{eq:dK2d} has a closed-form solution.

\subsection{Spherical Haar Wavelets} \label{sec:wavelets}

In this section we introduce the last major component of VSDM: a basis of orthogonal functions in $v = |\vec v|$ and $q = |\vec q|$.
As far as the integration method of~\cite{Lillard:2023nyj} is concerned, nearly any type of basis function would work. 
In VSDM, we use  the spherical wavelets derived in~\cite{Lillard:2023cyy}. Their major advantages are:
\begin{itemize}
\item spherical Haar wavelets are piecewise-constant, making it easy to integrate \eqref{eq:dK2d}; 
\item the wavelet expansion of 1d functions converge predictably and generically; 
\item given tabulated values of $f(x)$, there are efficient discrete transformations for calculating the coefficients of the wavelet expansion.
\end{itemize}
Together, these properties of wavelets greatly reduce the number of integrals that need to be calculated.

\paragraph{Definition:}
The $d$-dimensional spherical Haar wavelets defined in~\cite{Lillard:2023cyy} form a set of orthogonal basis functions on $x \in [0, 1]$, satisfying
\begin{align}
\int_0^1 \! x^{d-1} dx \, h_{n}(x) h_{n'}(x) &= \delta_{n n'}  .
\end{align} 
For all applications in this paper, we specialize to $d=3$. 
As the radial index $n =0,1,2,\ldots$ becomes larger, the $h_n(x)$ functions become narrower, with $h_n(x) = 0$ for $x$ outside of an $n$-dependent region. 
The $n=0$ and $n=1$ wavelets have support over the full region, $0 \leq u \leq 1$, starting with:
\begin{align}
h_{n=0}(0 \leq x \leq 1) &= \sqrt{3} .
\end{align}
All other wavelets ($n \geq 1$) are organized by a ``generation'' label, $\lambda = 0, 1, 2, \ldots$, and a position label,  $\mu = 0, 1, \ldots, 2^{\lambda} - 1$:
\begin{align}
n &= 2^\lambda + \mu.
\end{align}
The generation label $\lambda$ controls the width of the wavelet: with each generation, the width of the wavelet decreases by a factor of 2.  
Within each generation, the wavelets are non-overlapping, with their relative positions within the interval $[0, 1]$ determined by $\mu$. 

Precisely:
\begin{align}
{h}_{n} (x) = 
\left\{
\begin{array}{r c l}
+A_{n} && x_1(n) \leq x <  x_2(n) ,
\\[4pt]
-B_{n} &&  x_2(n)  < x \leq x_3(n) ,
\\[4pt]
0 && \text{otherwise} ,
\end{array}
\right.
\label{def:sphericalHaar}
\end{align}
where
\begin{align}
x_1(n) = 2^{-\lambda} \mu,
&&
x_2(n) = 2^{-\lambda} (\mu + \tfrac{1}{2} ), 
&&
x_3(n) = 2^{-\lambda} (\mu + 1) ,
\label{def:x123}
\end{align}
and 
\begin{align}
A_{n} &= \sqrt{ \frac{3}{x_3^3 - x_1^3} \frac{x_3^3 - x_2^3 }{x_2^3 - x_1^3} }
&
B_{n} &= \sqrt{ \frac{3}{x_3^3 - x_1^3} \frac{x_2^3 - x_1^3}{x_3^3 - x_2^3 } } .
\label{haar:AB}
\end{align}
For the $n=0$ wavelet, $x_1(0) \equiv 0$ and $x_2(0) \equiv 1$. Although there is no second region $[x_2, x_3]$, we define 
\begin{align}
A_0 &\equiv \sqrt{3}, 
&
B_0 &\equiv 0
\end{align}
for uniformity.

For functions of velocity or momentum, we take $x = v / v_\text{max}$ or $x = q/q_\text{max}$, for velocity and momentum cutoffs $v_\text{max}$ and $q_\text{max}$, respectively.

\paragraph{Setting $q_\text{max}$:} 
This is one possible disadvantage of the spherical Haar wavelets: because they are defined on a finite interval, $x \in [0, 1]$, we must impose some cutoff on the momentum $q$. 
We have already established that $g_\chi$ is only nonzero over a finite range, $0 \leq v \leq v_\text{max}$, so for functions of velocity this is no problem;   $f_S^2(\vec q, E)$, on the other hand, does not necessarily vanish identically at large $q$. 

In the context of direct detection, there is a strict upper limit on which values of $q$ contribute to the scattering rate.
Thanks to the lab-frame velocity cutoff $|\vec v| \leq v_\text{max}$ and the nature of $v_\text{min}(q, E)$, the rate integrand \eqref{eq:dK2d} vanishes for 
\begin{align}
q \geq 2 m_\chi v_\text{max}, 
\end{align}
because at these $q$ the value of $v_\text{min}$ is always larger than $v_\text{max}$:
\begin{align}
v_\text{min}(q = 2 m_\chi v_\text{max}) > v_\text{max}. 
\end{align}
So, the momentum integral of \eqref{eq:dK2d} can be cut off at $q_\text{max} = 2 m_\chi v_\text{max}$ without any approximation.

The form factor $f_S^2(\vec q, E)$ may also have a natural cutoff, some $q_\text{max}$ at which $f_S^2$ vanishes or is at least negligibly small. 
In DM--electron scattering, for example, the Bohr momentum 
\begin{align}
q_\text{Bohr} &\equiv \alpha m_e c \simeq 3.729\,\text{keV}
\end{align}
is the characteristic scale for the momentum transfer, and $f_S^2(\vec q, E)$ can become quite small in the $q \gg \alpha m_e$ limit. In~\cite{Blanco:2019lrf}, for example, $f_{g \rightarrow s}^2(\vec q) \propto (q_\text{Bohr} / q)^7$ at high momenta.
For such cases, setting $q_\text{max} \sim 10 q_\text{Bohr}$ can capture all the relevant dynamics with a negligible loss of precision, even when this $q_\text{max}$ is orders of magnitude smaller than $2 m_\chi v_\text{max}$.

Whatever the physical system, before proceeding with VSDM, the user should identify the physically relevant range of momenta $0 \leq q \leq q_\text{max}$ that should be included in the rate integral.

\subsection{Basis Functions In VSDM} \label{sec:basis}

In VSDM, functions of $\vec v$ and $\vec q$ are expanded in a series of 3d wavelet-harmonic basis functions, 
\begin{align}
\phi_{n \ell m}(\vec u) &\equiv h_n\!\left( \frac{u}{u_\text{max}} \right) Y_{\ell m}(\hat u).
\end{align}
At this stage of the calculation it is more convenient to work with the bra-ket notation of Ref.~\cite{Lillard:2023cyy}, 
with $g_\chi(\vec v) \rightarrow \ket{g_\chi}$, $f_S^2(\vec q, E) \rightarrow \ket{f_S^2}$. 
Basis functions of velocity or momentum are written in terms of their indices,  $\phi_{n \ell m}(\vec u) \rightarrow \ket{n \ell m}$. For real-valued functions, there is no distinction between ``bra'' and ``ket'' vectors. 
In this notation, 
\begin{align}
g_\chi(\vec v) &= \sum_{n = 0}^\infty \sum_{\ell = 0}^{\infty} \sum_{m = -\ell}^\ell \langle n \ell m | g_\chi \rangle \phi_{n \ell m}(\vec v), 
&
f_S^2(\vec q, E) &= \sum_{n \ell m} \langle n \ell m | f_S^2 \rangle \phi_{n \ell m}(\vec q), 
\end{align}
where the inner products $\langle f | g \rangle$ are defined following the Ref.~\cite{Lillard:2023cyy} convention:
\begin{align}
\langle f | g \rangle &\equiv \int\! \frac{d^3 u}{u_\text{max}^3} f(\vec u) \cdot g(\vec u) ,
\end{align}
where $\vec u$ in this expression is either $\vec v$ or $\vec q$. 
Note that $\langle n \ell m | f_S^2 \rangle$ is a function of $E$, while $\langle n \ell m | g_\chi \rangle$ is a constant.
With these conventions,  the $\ket{n \ell m}$ are dimensionless and orthonormal,
\begin{align}
\langle n \ell m | n' \ell' m' \rangle &= \delta_{n n'} \delta_{\ell \ell'} \delta_{m m'}, 
\end{align}
and the coefficients $\langle n \ell m | f \rangle$ have the same units as the function $f(\vec u)$. 

Tensor operators $\hat{\mathcal O}$ map vectors from one space onto another, e.g.~velocity space onto momentum space, or vice versa. 
In a given basis, the coefficients of a generic $\hat{\mathcal O}(\vec v, \vec q)$ are:
\begin{align}
\langle n \ell m | \hat{\mathcal O} | n' \ell' m' \rangle &\equiv \int\! \frac{d^3 v}{v_\text{max}^3} \frac{d^3 q}{q_\text{max}^3} \phi_{n \ell m}(\vec v) \, \hat{\mathcal O}(\vec v, \vec q)  \, \phi_{n' \ell' m'}(\vec q) .
\end{align}
After shuffling around some factors of $v_\text{max}$ and $q_\text{max}$, the scattering rate $R$ from \eqref{eq:rateFDM2} can be written in this form: 
\begin{align}
\frac{dR}{dE} \bigg|_{E, \mathcal R} &= 
\frac{N_T \rho_\chi \bar\sigma_0}{4\pi \mu_\chi^2 m_\chi} \left( q_\text{max} v_\text{max} \right)^3 
\left\langle g_\chi \left| F_\text{DM}^2(\vec q, \vec v) \,\delta\Big(  E + \frac{q^2}{2 m_\chi} - \vec{q} \cdot \vec{v} \Big) \right| \mathcal R \cdot f_S^2 \right\rangle .
\label{eq:dRdEM}
\end{align}

\subsection{The Kinematic Scattering Matrix} \label{sec:kinematicI}
Ref.~\cite{Lillard:2023cyy} shows that the tensor operator $F_\text{DM}^2 \delta(E \ldots - \vec q \cdot \vec v) $ is diagonal in the basis of spherical harmonics: i.e.
\begin{align}
\left\langle n \ell m \left| \frac{F_\text{DM}^2( q,  v)}{4 \pi \mu_\chi^2 m_\chi} \delta\Big(  E + \frac{q^2}{2 m_\chi} - \vec{q} \cdot \vec{v} \Big) \right| n' \ell' m' \right\rangle
&\equiv \delta_{\ell\ell'} \delta_{m m'}  \times \frac{v_\text{max}^2 }{q_\text{max}^4} \mathcal I^{(\ell)}_{n n'}(E) ,
\end{align}
where $\mathcal I^{(\ell)}$ is the \emph{kinematic scattering matrix}. In the wavelet basis, it is:
\begin{align}
\mathcal I^{(\ell)}_{n n'}(E) &\equiv \frac{q_\text{max}^3 / v_\text{max}^3 }{2 m_\chi \mu_\chi^2 }\int_0^{q_\text{max}} \!\frac{qdq}{q_\text{max}^2} \, h_{n'}\!\left( \frac{q}{q_\text{max}} \right) \int_{v_\text{min}(q, E) }^{v_\text{max}} \! \frac{vdv}{v_\text{max}^2} P_\ell\! \left( \frac{v_\text{min}(q,E) }{v} \right) h_n\!\left( \frac{v}{v_\text{max}} \right) \, F_\text{DM}^2(q, v)  .
\label{def:mathcalI}
\end{align}
A factor of $q_\text{max}^4 / v_\text{max}^2$ has been absorbed into $\mathcal I$ to make it dimensionless. 
More importantly, the factors of $m_\chi \mu_\chi^2$ are included as well. 
This way, $\mathcal I^{(\ell)}$ contains all of the terms that depend on the DM particle model, $m_\chi$ and $F_\text{DM}^2$. 

In terms of $\mathcal I^{(\ell)}$, the scattering rate \eqref{eq:dRdEM} is:
\begin{align}
\frac{dR}{dE} \bigg|_{E, \mathcal R} &= N_T \rho_\chi \bar\sigma_0 \frac{v_\text{max}^5}{q_\text{max} }   
\sum_{n, n'} \sum_{\ell m}  \langle g_\chi | n \ell m \rangle \, \mathcal I^{(\ell)}_{n n'}(m_\chi, F_\text{DM}^2; E)  \langle n' \ell m | \mathcal R | f_S^2 \rangle.
\label{eq:dRdEmI}
\end{align}
The $m_\chi$ dependence of $\mathcal I$ enters through its dependence on $v_\text{min}(q, E)$.

\paragraph{In VSDM:} 

In the current version of VSDM, if the user wishes to evaluate an $F_\text{DM}^2$ that is not of the form $F_\text{DM}^2 \propto q^a v^b$, they must provide the solution to \eqref{def:mathcalI} themselves. 
Otherwise, for $F_\text{DM}^2$ given by \eqref{FDM2:qavb}, VSDM provides the contribution to $\mathcal I^{(\ell)}$ from each combination of $(a, b)$:
\begin{align}
F_\text{DM}^2(q, v) &= \sum_{a b} d_{ab}(E) (q / q_\text{ref})^a (v / c)^b, 
\label{eq:FDM2dab}
\\
\mathcal I^{(\ell)}_{n n'}(E) &= \sum_{ab} d_{ab}(E) \times \mathcal I^{(\ell)}_{n n'}(m_\chi, a, b; E), 
\end{align}
where $ \mathcal I^{(\ell)}_{n n'}(a,b; E)$ is defined as in \eqref{def:mathcalI} but with $F_\text{DM}^2 \rightarrow (q/q_\text{ref})^a (v / c)^b$.

For any pair of velocity and momentum basis functions, there are up to four regions that contribute to the integral, 
within which $h_n(v) h_{n'}(q)$ is constant. 
Defining $q_i(n) = x_i(n) q_\text{max}$ and $v_i(n) = x_i(n) v_\text{max}$, each of the intervals $[v_1, v_3]$ and $[q_1, q_3]$ is subdivided at $v_2$ and $q_2$, respectively, and within each region:
\begin{align}
h_n(v/v_\text{max})  \, h_{n'}(q/q_\text{max}) &= \left\{ A_n A_{n'},  - A_n B_{n'}, -B_n A_{n'}, + B_n B_{n'} \right\} ,
\end{align}
where $A_n$, $B_n$ are as defined in \eqref{haar:AB}. 
Referring to these regions as $AA$, $AB$, $BA$, and $BB$, we define $\mathcal I^{(\ell)}$ in terms of the convenient $I_\star$ defined in~\cite{Lillard:2023cyy}:
\begin{align}
\mathcal I^{(\ell)}_{n n'}(m_\chi, \alpha, \beta; E) &= \sum_{AA\ldots BB} \left( h_n h_{n'} \right) \times \frac{q_\text{max} q_\star^2 v_\star^2 }{2 m_\chi \mu_\chi^2 v_\text{max}^5} \times  I_{\alpha, \beta}^{(\ell)}(E, [q_{ab}, v_{ab} ]),
\end{align}
where $v_\star$ and $q_\star$ are as defined in \eqref{def:star}. 
In each term $AA$, $AB$, etc., the limits of integration $u_{a, b}$ are determined by $n$ and $n'$.
For the $A$-type intervals, $[u_a, u_b] = u_\text{max} \times [x_1(n), x_2(n)]$, while for $B$-type intervals $[u_a, u_b] = u_\text{max} \times [x_2(n) , x_3(n) ] $. 
The integral function $I_{a,b}^{(\ell)}$ is defined as:
\begin{align}
 I_{a,b}^{(\ell)}(E, [q_{ab}, v_{ab} ]) &\equiv  \int_{q_a}^{q_b}\! \frac{qdq}{q_\star^2} \left( \frac{q}{q_\star} \right)^a \int_{v_a}^{v_b} \! \frac{ v dv}{v_\star^2} \left( \frac{v}{v_\star} \right)^{b} P_\ell\! \left( \frac{v_\text{min} }{v} \right) 
 \Theta(v - v_\text{min}(q, E)) ,
 \label{eq:mIstar}
\end{align}
which has a closed form solution derived in Appendix~B of Ref.~\cite{Lillard:2023cyy},  for integer or non-integer values of $a$ and $b$. 



\subsection{The Partial Rate Matrix} 

By changing the order of the sums in \eqref{eq:dRdEmI}, and applying \eqref{eq:rotG} to evaluate $\mathcal R \ket{ f_S^2}$, 
the kinematic scattering matrix can be used to calculate $K^{(\ell)}_{m m'}$ or $dK^{(\ell)}_{mm'}/dE$ from the wavelet-harmonic representations of $\ket{g_\chi}$ and $\ket{f_S^2 }$.
Defining the \emph{reduced partial rate matrix} as
\begin{align}
\mathcal K^{(\ell)}_{m m'}(g_\chi, f_S^2, F_\text{DM}^2; E) &\equiv  v_\text{max}^3 \sum_{n n'} \langle g_\chi | n \ell m \rangle \, \mathcal I^{(\ell)}(F_\text{DM}^2; E) \, \langle n' \ell m' | f_S^2 \rangle ,
\label{eq:KfromI}
\end{align}
and assuming that $F_\text{DM}^2 \rightarrow \sum_{ab} (a,b)$ is given by \eqref{eq:FDM2dab}, 
\begin{align}
\frac{dR}{dE}\bigg|_{E, \mathcal R} &= N_T \rho_\chi \frac{v_\text{max}^2}{q_\text{max} } \sum_{\alpha, \beta} \bar\sigma_0 d_{\alpha, \beta}(E) \sum_{\ell m m'} \mathcal K^{(\ell)}_{m m'}(g_\chi, f_S^2, (m_\chi, \alpha, \beta); E) \cdot G^{(\ell)}_{m m'}(\mathcal R) ,
\end{align}
where $G_{mm'}^{(\ell)}$ is the Wigner $G$ matrix defined in \eqref{def:WignerG}.
The factor of $v_\text{max}^3$ is included in $\mathcal K$ to make it nearly dimensionless: that is, it has the same units as $f_S^2$. 

The $\mathcal K$ notation accommodates either continuous or discrete final states. In the former case, $\mathcal K$ has units of inverse energy, like $f_S^2(\vec q, E)$, and 
\begin{align}
\mathcal K^{(\ell)}_{m m'}(E)  &= \frac{q_\text{max}}{v_\text{max}^2} \frac{d K^{(\ell)}_{m m'} }{dE} \bigg|_E,
\end{align}
in the limit where the sums over $n, n'$ are complete.
For discrete final states, 
we simply define $\mathcal K$ in terms of $f_{g \rightarrow s}^2(\vec q)$ rather than $f_S^2(\vec q, E)$:
\begin{align}
\mathcal K^{(\ell)}_{m m'}(g_\chi, f_{g \rightarrow s}^2, F_\text{DM}^2; \Delta E) &\equiv  v_\text{max}^3 \sum_{n n'} \langle g_\chi | n \ell m \rangle \, \mathcal I^{(\ell)}(F_\text{DM}^2; \Delta E) \, \langle n' \ell m' | f_{g \rightarrow s}^2 \rangle .
\end{align}
In this case $\mathcal K_{m m'}^{(\ell)}$ is dimensionless, and equal to
\begin{align}
\mathcal K^{(\ell)}_{m m'}(\Delta E)  &= \frac{q_\text{max}}{v_\text{max}^2} K^{(\ell)}_{m m'}(\Delta E) .
\end{align}

For convenience, 
the remaining dimensionful constants can be combined into a dimensionless, basis-dependent quantity $k_0$ proportional to $\bar\sigma_0$ and the experimental exposure time $T_\text{exp}$: 
\begin{align}
\label{eqn:k0}
k_0 &\equiv N_T T_\text{exp} \bar\sigma_0 \rho_\chi \frac{v_\text{max}^2}{q_\text{max} }.
\end{align}
This way the total (differential) rate is given by 
\begin{align}
\frac{dR}{dE}\bigg|_{E, \mathcal R} &=  \frac{k_0}{T_\text{exp}} \sum_{ab} d_{ab}(E)  \sum_{\ell m m'} \mathcal K^{(\ell)}_{m m'}(g_\chi, f_S^2, (m_\chi, a, b);  E) \cdot G^{(\ell)}_{m m'}(\mathcal R) .
\label{eq:dRdEfromKG}
\end{align}
For discrete final states \eqref{eq:rateDiscrete}, the total rate is given directly by a sum over final states $s$:
\begin{align}
R(\mathcal R) &= \frac{k_0}{T_\text{exp}} \sum_{s}  \sum_{ab} d_{ab}(\Delta E_s)  \sum_{\ell m m'} \mathcal K^{(\ell)}_{m m'}(g_\chi, f_{g \rightarrow s}^2, (m_\chi, a, b); \Delta E_s) \cdot G^{(\ell)}_{m m'}(\mathcal R).
\label{eq:RfromKG}
\end{align}


Both the Python and Julia implementations store $\mathcal K^{(\ell)}_{mm'}$ and $G^{(\ell)}_{mm'}$ as 1d vectors 
\begin{align}
\vec G_i &= G^{(\ell)}_{m m'} , 
&
\vec{\mathcal K}_i &= \mathcal K^{(\ell)}_{m m'} ,
\end{align}
of length 
\begin{align}
\text{len}(\vec{\mathcal K}) = \text{len}(\vec G) 
&= \frac{(\ell_\text{max} + 1)(2\ell_\text{max} + 1)(2\ell_\text{max} + 3) }{3},
\end{align}
with $i$ indexed by
\begin{align}
\texttt{index}(\ell, m, m') &= \frac{\ell}{3} (4 \ell^2 - 1) + (2\ell + 1)(\ell + m) + (\ell + m') 
\end{align}
for $\texttt{index} = 0, 1, 2, \ldots, \text{len}(G) -1$.

\subsection{Summary} \label{sec:summary}

A VSDM calculation proceeds in the following steps:
\begin{enumerate}
\item The user selects basis function parameters $v_\text{max}$ and $q_\text{max}$ that are appropriate for the $g_\chi$ and $f_S^2$ functions in question.  
The user should also determine which values to use for continuous parameters $m_\chi$, the excitation energy (for $dR/dE$), and possibly the time of year (if including annual variation in $g_\chi(t)$). It is not necessary to specify the list of $\mathcal R \in SO(3)$ detector orientations at this time.
\item \label{item:two} The functions $g_\chi$ and $f_S^2$ (or $f_{g \rightarrow s}^2$) are projected onto the wavelet-harmonic basis, saving the coefficients $\langle n \ell m | g_\chi \rangle$ and $\langle n \ell m | f_S^2(E) \rangle$ for every version of $g_\chi$ and $f_S^2(E)$. At this stage, the user should determine the largest values of $n_\text{max}$ and $\ell_\text{max}$ that are actually needed in order to represent the functions $\ket{g_\chi}$ and $\ket{f_S^2}$ within the desired precision.
\item \label{item:three} The kinematic scattering matrix $\mathcal I^{(\ell)}_{n n'}(m_\chi, F_\text{DM}^2; E)$ is evaluated for every combination of $(m_\chi, F_\text{DM}^2)$ and $E$ that appear in the problem, up to some predetermined values of $\ell_\text{max}$ and $n_\text{max}^{(q, v)}$. 
\item For each combination of $g_\chi$, $f_S^2(E)$, $F_\text{DM}^2$, and $m_\chi$, evaluate the reduced partial rate matrix $\mathcal K$.
\end{enumerate}
 Once the partial rate matrices are known, it is very fast to calculate the scattering rate as a function of detector orientation: after evaluating $G^{(\ell)}_{mm'} (\mathcal R)$ for every $\mathcal R$, the scattering rate is given by the dot product between the vectors $\vec G$ and $\vec K$: 
\begin{align}
\label{eqn:contrate}
\frac{dR}{dE}\bigg|_{E, \mathcal R} &= \frac{k_0}{T_\text{exp}}   \sum_{ab} d_{ab}( E) \times \vec{\mathcal K}  \cdot \vec G(\mathcal R) .
\end{align} 
For a transition to a single discrete final state $g \rightarrow s$, the expression for $R(\mathcal R)$ is identical:
\begin{align}
\label{eqn:rate}
R_s(\mathcal R) &= \frac{k_0}{T_\text{exp}}   \sum_{ab} d_{ab}( E) \times \vec{\mathcal K}  \cdot \vec G(\mathcal R)  .
\end{align} 

Compared to numerical integration of \eqref{eq:rate}, evaluating the $\vec{\mathcal K} \cdot \vec G$ dot product is extremely fast.
Section~\ref{sec:performance} quantifies the improvement. 
Step~\ref{item:two} is often the slowest step in this pipeline, at least if  $\langle n \ell m | g_\chi \rangle$ and $\langle n \ell m | f_S^2 \rangle$ are found by numeric integration. However, if $g_\chi(\vec v)$ or $f_S^2(\vec q, E)$ have already been tabulated, then the wavelet coefficients can be obtained more quickly using discrete transformations. Section~\ref{sec:shortcuts} describes this method, as well as the analytic shortcut from Ref.~\cite{Lillard:2023cyy} that can be used for functions (e.g.~$g_\chi$) that are defined as the sum of multiple Gaussians.

\subsection{Faster Wavelet-Harmonic Transformations} \label{sec:shortcuts}
If a function of a 3d variable $f(\vec u)$ is tabulated on a regular grid of points in spherical coordinates, $\vec u = (u, \theta, \phi)$, 
then the user can rapidly evaluate $\langle n \ell m | f \rangle$ without using the numeric integration methods included in VSDM. 
Bypassing the numeric integration step can drastically reduce the total VSDM evaluation time. The only nonstandard step in the procedure is the wavelet transformation for $d > 1$ spherical wavelets, which we provide in this section.

For calculating $\langle f | n \ell m \rangle$ with discretized 3d functions $f(\vec u)$, we recommend a two-step process. 
First, for a sequence of spherical shells  $|\vec u | = u_i$, evaluate 
\begin{align}
f_{\ell m}(u) &\equiv \langle \ell m | f \rangle = \int\! d\Omega\, Y_{\ell m}(\hat u) \, f(\vec u)  
\end{align}
for all of the relevant $\ell m$ with $\ell \leq \ell_\text{max}$. 
This can be done by using a discrete spherical harmonic transform, for which there are many pre-existing packages and typically requires tabulating $f(u_i, \theta, \phi)$ on a grid of $(\theta, \phi)$. The sampling density of the grid determines the largest value of $\ell$ that can be accurately recovered.

Next, for each $f_{\ell m}(u)$ function (evaluated at $u = u_i$ for all $u_i$ on the radial grid), evaluate 
\begin{align}
\langle n \ell m | f \rangle &= \langle n | f_{\ell m} \rangle = \int\! u^2 du\, h_n(u/u_\text{max}) \, f_{\ell m}(u)  
\end{align}
for each $n \leq n_\text{max}$, with $n_\text{max}$ determined by the total number of $u_i$ points. If the values of $u_i$ are sufficiently finely spaced, then $\langle n | f_{\ell m} \rangle$ can be found using the discrete wavelet transformation: otherwise, a finer grid can be generated from $f_{\ell m}(u_i)$ by interpolation.

It is not necessary to use the same grid of $(\theta, \phi)$ points for each $u_i$. Near the origin $u \sim 0$, for example, the higher-$\ell$ moments are expected to vanish, and there is no need for a particularly fine angular grid. In spherical harmonics, the Taylor series of $f(\vec u)$ has the form
\begin{align}
f(\vec u) \approx \sum_{\ell m} c_{\ell m} u^\ell Y_{\ell m}(\hat u), 
\end{align}
so the expected size of $f_{\ell m}(u_i)$ scales as $u_i^\ell$ for small $u_i$ (assuming that $f$ and its derivatives are finite at the origin).


Unless it is essentially effortless to evaluate $f(\vec u)$ on a fine grid of points, we recommend the following procedure:
\begin{enumerate}
\item identify a list of spherical shells $|\vec u| = u_i$ to evaluate $f_{\ell m}(u_i)$. This grid can be irregularly spaced, to increase the sampling density in the places where $f(\vec u)$ varies most quickly.
\item \label{step:ui} For each $u_i$:
\begin{enumerate} 
\item evaluate $f(u_i, \theta, \phi)$ on a grid of $(\theta, \phi)$ points, with the angular grid spacing chosen according to the chosen spherical harmonic transform algorithm and the desired $\ell_\text{max}$ for this $i$ 
\item find $f_{\ell m}(u_i)$ from the tabulated $f(\vec u)$ via the discrete spherical harmonic transformation.
\end{enumerate}
\item In preparation for the wavelet transformation, evaluate each $f_{\ell m}$ on a finer grid of $2^{L}$ regularly-spaced values of $u = \bar u_i$, interpolating between the evaluated $f_{\ell m}(u_i)$ if $\bar u_i \not\in \{ u_i \}$. 
\item Apply the discrete spherical wavelet transformation (defined below) to obtain $\langle n  \ell m | f  \rangle$ from the tabulated values of $f_{\ell m}(\bar u_i)$, for wavelet generations up to $\lambda_\text{max} = L - 1$.
\end{enumerate}
If it is easy to evaluate $f(\vec u)$ on a denser grid of points, then the interpolation step can be skipped by sampling $f(\vec u)$ on the same $|\vec u| = \bar u_i$ spherical shells used by the wavelet transformation. 

If interpolation is necessary, it is much faster to evaluate $f_{\ell m}(u_i)$ first and interpolate second (to get $f_{\ell m}(\bar u_i)$), rather than interpolating first (to get $f(\bar u_i, \theta_j, \phi_k)$ for all $\bar u_i$)  and getting $f_{\ell m}(\bar u_i)$ directly from the spherical harmonic transformation on the interpolated data.

\paragraph{Discrete Spherical Wavelet Transformation:}
Here we generalize the discrete Haar transformation to 
define the discrete spherical wavelet transformation for $d \geq 2$. 

A 1d function $f(x)$, for $x \in [0, 1]$, is 
discretized on a regularly spaced grid, 
\begin{align}
f(x_i \leq x < x_{i + 1} ) &\approx \bar f_i, 
& 
x_i &= i \times \Delta , 
&
\Delta &\equiv 2^{-L}, 
\end{align}
where $x_0, x_1, \ldots, x_{i_\text{max}}, x_{i_\text{max} + 1}$ mark the boundaries between the different intervals, $\Delta$ is the interval width, $i = 0, 1, \ldots, 2^{L} - 1$, and $\bar f_i$ 
is the volume-weighted average of $f$ in the $i$th interval: 
\begin{align}
\bar f_i &\equiv \int_{x_i}^{x_{i+1}}\! x^{d-1} dx\, f(x).
\label{def:barfi}
\end{align}
This is equivalent to approximating $f(x)$ by a sum of orthogonal ``tophat'' basis functions $\ket{\Theta_i}$, 
\begin{align}
\Theta_i(x) &\equiv \left\{ \begin{array}{r l} 1 & \text{if } x_i < x < x_{i + 1} \\
0 & \text{otherwise}.
\end{array}
\right.
\end{align}
vanishing everywhere outside the $i$th interval. For this discussion it is simpler not to normalize  $\ket{\Theta}$, so we can write 
\begin{align}
f(x) &\approx \sum_{i} \bar f_i \ket{\Theta_i} .
\end{align}
The error in $f(x) - \sum_i \bar f_i \ket{\Theta_i}$ is proportional to $\Delta$ and the first derivative of $f$, 
\begin{align}
f(x) - \sum_i \bar f_i \ket{\Theta_i} \lesssim f'(x) \Delta.
\end{align}

Rather than finding $\bar f_i$ from the integral \eqref{def:barfi}, it can be sufficiently well approximated by sampling $f(x)$ at a particular point, $\bar x_i$: for arbitrary $d$, 
\begin{align}
\bar x_i &\equiv \frac{d}{d+1} \frac{x_{i + 1}^{d+1} - x_{i}^{d+1} }{x_{i+1}^d - x_{i}^d} 
\simeq x_i + \frac{1}{2} \Delta + \frac{d-1}{12} \frac{\Delta^2}{x_i} + \mathcal O(\Delta^3) . 
\end{align}
For the 3d spherical wavelets, $d = 3$. 
With this choice, $f(\bar x_i) \simeq \bar f_i$ to second order in the small $\Delta$ expansion: 
\begin{align}
f(\bar x_i) - \bar f_i &\sim  \Delta^2 \cdot f''(\bar x_i)  .
\end{align}
At any other value of $\bar x_i$, the correction to the approximation above would have been proportional to $\Delta \cdot f'(x_i)$, 
but $\bar x_i$ is chosen so that this linear correction cancels itself out. 

After evaluating the function $f$ at each $\bar x_i$ for $i = 0, 1, \ldots, i_\text{max}$, the coefficients of $f$ in the tophat basis are approximated by:
\begin{align}
f_i &= f(\bar x_i) \simeq \langle f  | \Theta \rangle,  
&
f(x) &\simeq \sum_i f_i \ket{\Theta_i} .
\end{align}
By applying a basis transformation, these coefficients can be used to find the wavelet coefficients. Writing the spherical wavelet basis functions as $\ket{n}$, 
the coefficients of the basis transformation $\langle n | \Theta_i \rangle$ are given by:
\begin{align}
\langle n | \Theta_i \rangle &= \int_{x_i}^{x_{i + 1} } \! x^2 dx \, h_n(x), 
\end{align}
which for the 3d spherical Haar wavelets simplifies to:
\begin{align}
\langle n < 2^{\lambda_\text{max}} | \Theta_i \rangle &= \frac{\Delta }{3} \left( 3 x_i^2 + 3 x_i \Delta + \Delta^2 \right) \times \left\{ \begin{array}{r l }
A_n & \text{if } x_1(n) < \bar x_i < x_2(n) 
\\
- B_n & \text{if } x_2(n) < \bar x_i < x_3(n) 
\\
0 & \text{otherwise}  ,
\end{array} \right.
\\
\langle n \geq  2^{\lambda_\text{max}} | \Theta_i \rangle &= 0, 
\end{align}
for the $x_{1,2,3}$ defined in \eqref{def:x123}, 
and $x_i = \Delta \cdot i$. $A_n$ and $B_n$ are defined in \eqref{haar:AB}. 
Because $x_i$ and $x_{i+1}$ are aligned with the wavelet grid, $\langle n | i \rangle$ vanishes for most pairings of $n$ and $i$. 
In particular, the wavelet expansion terminates at finite $n \leq i_\text{max}$: for higher-frequency wavelets $n > i_\text{max}$, $\ket{\phi_i}$ is constant over the full range of support of the wavelet, and so $\langle n | \phi_i \rangle = 0$ by the same orthogonality condition that sets $\langle n | 0 \rangle$ for the $h_0$ wavelet. 
For $n \leq i_\text{max}$, the tophat function $\phi_i$ is narrower than any of the wavelets, and it overlaps with exactly one wavelet from each generation.
Consequently, there are only $(i_\text{max} + 1) \cdot (\log_2(i_\text{max} + 1) + 1)$ many nonzero terms, rather than $(i_\text{max} + 1)^2$. 

With this result, the  wavelet transformation of $f(x)$ can be read off from:
\begin{align}
f(x) \simeq 
\sum_{i=0}^{i_\text{max}} f_i \ket{ \Theta_i } &= \sum_{i=0}^{i_\text{max}} f_i  \sum_{n=0}^{i_\text{max}} \langle n | \Theta_i \rangle \ket{n} ,
\end{align}
and the $n$th wavelet coefficient $\langle f | n \rangle$ is:
\begin{align}
\langle n | f \rangle &\simeq \sum_i f(\bar x_i) \langle n | \Theta_i \rangle, 
\end{align}
where the sum over $i$ can be restricted to only those $i$ for which $x_1(n) < \bar x_i < x_3(n)$. 
In terms of the $n \rightarrow (\lambda, \mu)$ wavelet indexing, there are $2^{L - \lambda}$ many terms in this sum: as few as 2 for the narrowest wavelets, at $2^{\lambda_\text{max}} \leq n \leq 2^{\lambda_\text{max}} - 1$, or as many as $2^L$ for the $n =0, 1$ wavelets.

\paragraph{Discrete Spherical Harmonic Transformation:}
A function $f(\theta, \phi)$ is defined on the $S^2$ sphere, with $0 \leq \theta \leq \pi$ and $0 \leq \phi < 2\pi$. 
While there exist methods for extracting the spherical harmonic coefficients $\langle \ell m | f \rangle$ from irregular sets of $(\theta_i, \phi_j)$ values, 
many discrete transformations use regular grids in $(\theta, \phi)$, e.g.~an equiangular $N \times (2N-1)$ rectangular grid in $\theta, \phi$ with 
\begin{align}
\theta_i &= \frac{i + \frac{1}{2} }{N} \pi , 
&
\phi_j &= \frac{j}{(2N-1)} \times 2\pi 
\end{align}
for $i = 0, 1, \ldots, N-1$ and $j = 0, 1, \ldots, 2(N - 1)$.
In this scheme, the maximum $\ell$ for a given $N$ is $\ell_{\textrm{max}} = N-1$, thus the function $f(u_i, \theta, \phi)$ would need to be evaluated $M$ times for a desired $\lmax$, where $M = 2 \lmax^2 + 3 \lmax + 1$. Generally, the number of function evaluations needed for a particular $\lmax$ will scale with $\lmax^2$, but the exact grid that $f(u_i, \theta, \phi)$ should be evaluated on will depend directly on the choice of algorithm for performing the discrete spherical harmonic transform.


\section{Julia Implementation: VectorSpaceDarkMatter.jl}
\label{sec:julia}
\texttt{VectorSpaceDarkMatter.jl} is the Julia implementation of the wavelet-harmonic integration method.
The Julia programming language~\cite{JuliaLang} was chosen mostly for its just-in-time compilation, which allows for an even larger speedup over the Python implementation. 
Julia also offers a few additional benefits, including multiple dispatch, which allows for a more simplified set of functions, and built-in parallelization, allowing for additional speedup proportional to the number of threads.

In order to calculate the rate, there are two main steps: projecting velocity distribution(s) and material form factor(s) onto a basis, and then calculating the partial rate matrix $\mathcal{K}$. This section outlines the steps that one needs to take in order to do these calculations. 
An example notebook is available at \href{https://github.com/ariaradick/VectorSpaceDarkMatter.jl/blob/main/examples/VSDM\_Example.ipynb}{https://github.com/ariaradick/VectorSpaceDarkMatter.jl}.

\subsection{Projecting a Function}
\label{subsec:juproj}

The first step in projecting a function is choosing a basis for the radial part. There are two options included in \texttt{VectorSpaceDarkMatter.jl}, spherical Haar wavelets and spherical tophat functions. These are described in more detail in Section~\ref{sec:wavelets} and Appendix~\ref{sec:tophats} respectively. We can initialize a basis by running the appropriate constructor function,
\begin{minted}[bgcolor=pink]{Julia}
Wavelet(umax)
Tophat(xi, umax)
\end{minted}
where \texttt{umax} is the (dimensionful) maximum value of the magnitude of the velocity (momentum) in the chosen velocity distribution (material form factor), and \texttt{xi} is a list of boundaries for the tophat function on the range [0,1]. Both of these are subtypes of the abstract type \texttt{RadialBasis}. Importantly, \texttt{umax} should be defined for each function in natural units (units of $c$ for the velocity, and eV for the momentum) for consistency with the rate calculation. It is possible to omit the \texttt{umax} option from either, which results in a default value of \texttt{umax=1.0}.

Next, if the function $f(\vec u)$ has rotational or reflection symmetries, it can speed up the projection greatly if they are specified when defining the function. In order to specify these, we include a type
\begin{minted}[bgcolor=pink]{Julia}
f_uSph(f::Function; z_even=false, phi_even=false, 
       phi_cyclic=1, phi_symmetric=false, center_Z2=false)
\end{minted}
The options are
\begin{itemize}
  \item \texttt{z\_even} : if $f(x,y,z) = f(x,y,-z)$. Implies $\langle \ell m | f \rangle = 0$ if $(\ell+m)$ is odd.
  \item \texttt{center\_Z2} : if $f(\vec{u}) = f(-\vec{u})$ is symmetric under central inversion. Implies $\langle \ell m | f \rangle = 0$ if $\ell$ is odd.
  \item \texttt{phi\_even} : if $f(u,\theta,\phi) = f(u,\theta,-\phi)$. Implies $\langle \ell m | f \rangle = 0$ if $m$ is negative.
  \item \texttt{phi\_cyclic} : $f(u,\theta,\phi) = f(u,\theta,\phi + 2\pi/\alpha)$ for integer $\alpha = \texttt{phi\_cyclic}$. Implies $\langle \ell m | f \rangle = 0$ unless $m$ is divisible by $\alpha$.
  \item \texttt{phi\_symmetric} : if $f(u,\theta)$ is independent of $\phi$. Takes precedence over other $\phi$ options. Sets $\langle \ell m | f \rangle = 0$ for all $m \neq 0$.
\end{itemize}

If one wishes to project a Gaussian function, there are special routines implemented to handle these efficiently. Primarily, we have included a \texttt{GaussianF} type, which stores the relevant parameters of the Gaussian,
\begin{minted}[bgcolor=pink]{Julia}
GaussianF(c::Float64, uSph::Vector{Float64}, sigma::Float64)
\end{minted}
where \texttt{uSph} is the spherical vector indicating the center of the Gaussian, \texttt{uSph} = $(u, \theta, \phi)$, \texttt{sigma} is the standard deviation, and \texttt{c} is the overall amplitude. \texttt{uSph} and \texttt{sigma} should be in natural units. Instances of this type can be called as $g(u,\theta,\phi)$ to return the value of the Gaussian at the specified point.

In order to calculate the coefficients, one simply needs to use
\begin{minted}[bgcolor=pink]{Julia}
ProjectF(f, nl_max::Tuple{Int,Int}, radial_basis::RadialBasis; 
         dict=false, use_measurements=false, 
         integ_method=:cubature, integ_params=(;))
\end{minted}
where \texttt{f} is the function to be projected, which can be a \texttt{f\_uSph}, \texttt{GaussianF}, or a plain function. \texttt{nl\_max} $= (n_{\textrm{max}}, \ell_{\textrm{max}})$ is a tuple containing the number of coefficients in the radial basis, $n$, and the number of $\ell$ modes in the spherical harmonics -- all non-zero $m$ will be calculated for each $\ell$. \texttt{radial\_basis} is either an instance of \texttt{Wavelet} or \texttt{Tophat}. \texttt{use\_measurements} will return the numerical integration error along with the values of the coefficients stored in a \texttt{Measurements} object from the \texttt{Measurements.jl} package~\cite{Measurements.jl-2016}. \texttt{integ\_method} chooses between three options: \texttt{:cubature}, which uses the integration method \texttt{hcubature} from the package \texttt{HCubature.jl}~\cite{HCubature}, and \texttt{:vegas} or \texttt{:vegasmc}, both options for integration from \texttt{MCIntegration.jl}~\cite{MCIntegration}. \texttt{integ\_params} is a named tuple containing keyword arguments for the chosen integration method, see the documentation for each package for available options.

If \texttt{dict} is false (default), \texttt{ProjectF} will return a \texttt{ProjectedF} object that stores the values of the coefficients in a matrix,
\begin{minted}[bgcolor=pink]{Julia}
struct ProjectedF{A, B<:RadialBasis}
    fnlm::Matrix{A}
    lm::Vector{Tuple{Int,Int}}
    radial_basis::B
end
\end{minted}
where \texttt{lm} stores the corresponding $(\ell,m)$ values as a vector and \texttt{radial\_basis} is a copy of the supplied radial basis object. An instance of this \texttt{ProjectedF} can be called as $p(u,\theta,\phi)$ and use the values of the coefficients to evaluate the function at the specified point.

If \texttt{dict} is true, \texttt{ProjectF} will return an \texttt{FCoeffs} object that stores the values of the coefficients in a dictionary,
\begin{minted}[bgcolor=pink]{Julia}
struct FCoeffs{A, B<:RadialBasis}
    fnlm::Dict{Tuple{Int,Int,Int}, A}
    radial_basis::B
end
\end{minted}
where the \texttt{fnlm} values are stored as $(n, \ell, m) \Rightarrow f_{n \ell m}$ and again, \texttt{radial\_basis} is a copy of the supplied radial basis object. Note that an \texttt{FCoeffs} object \emph{cannot} be called as a function, and \emph{cannot} be used in functions calculating the rate. The main advantage to an \texttt{FCoeffs} is the ability to update individual coefficients via the function
\begin{minted}[bgcolor=pink]{Julia}
update!(fc::FCoeffs, f, nlm; integ_method=:cubature,
        integ_params=(;))
\end{minted}
which updates the \texttt{nlm} $= (n, \ell, m)$ coefficient of \texttt{fc} with the coefficient obtained by projecting \texttt{f} onto the stored radial basis in \texttt{fc}. \texttt{nlm} can also be a vector of tuples and this will update each coefficient. There are also convenience functions that convert between \texttt{ProjectedF} and \texttt{FCoeffs} objects,
\begin{minted}[bgcolor=pink]{Julia}
FCoeffs(pf::ProjectedF)
ProjectedF(fcoeffs::FCoeffs)
\end{minted}

In order to check the accuracy of the projection, we include a function that calculates the $L^2$ norm,
\begin{minted}[bgcolor=pink]{Julia}
f2_norm(f[, a, b])
\end{minted}
where \texttt{f} can be a function, \texttt{ProjectedF}, \texttt{FCoeffs}, \texttt{GaussianF}, or \texttt{Vector\{GaussianF\}}. In each case, the appropriate norm will be returned:
\begin{align}
  L^2[f] = \int_a^b d^3 u \; f^2(\textbf{u}) = \left( \sum_{n \ell m} f_{n \ell m} \right)^2
\end{align}
\texttt{a} and \texttt{b} represent the bounds in spherical coordinates when integrating a function. Using \texttt{f2\_norm} with a Gaussian will return the analytic integral over all space, unless the bounds \texttt{a} and \texttt{b} are specified. It should be noted that accuracy in the $L^2$ norm is not necessarily equivalent to accuracy in the rate. \emph{E.g.}, if one has a material with high $\ell$ support, but a velocity distribution that falls off with $\ell$, it may be the case that the rate can be reproduced with high accuracy but the $L^2$ norm for the material can be off by 50\%.
So, $L^2$ accuracy is a sufficient, but not necessary, condition for convergence in the VSDM rate calculation.

Finally, as it can be computationally expensive to calculate coefficients, we include the ability to save and load coefficients to and from CSV files via
\begin{minted}[bgcolor=pink]{Julia}
writeFnlm(outfile, pf)
readFnlm(infile[, radial_basis]; dict=false, use_err=true)
\end{minted}
where \texttt{pf} is either a \texttt{ProjectedF} or \texttt{FCoeffs}. In \texttt{readFnlm} the \texttt{radial\_basis} argument is optional if using \texttt{Wavelet} for the radial basis as \texttt{writeFnlm} saves this information along with the coefficients. If the coefficients were calculated with tophats, it \emph{is} necessary to specify the basis. \texttt{use\_err} is equivalent to \texttt{use\_measurements}, choosing whether or not to load the uncertainty values (if present).

\subsection{Calculating the Rate}
\label{subsec:jurate}

It is necessary to supply some parameters that specify the DM mass $m_\chi$, form factor $F_{\textrm{DM}}$, SM target mass $m_{\textrm{SM}}$, and the difference in energy between the initial and final SM states $\Delta E$. We store these parameters in a \texttt{ModelDMSM} object,
\begin{minted}[bgcolor=pink]{Julia}
struct ModelDMSM
    a::Int
    b::Int
    mX::Float64
    mSM::Float64
    deltaE::Float64
end
\end{minted}
where $a$ and $b$ are parameters of the dark matter form factor such that $F_{\textrm{DM}}^2 \sim v^a (q/q_{\textrm{ref}})^b$ (Eqn.~\ref{FDM2:qavb}) and the other three parameters should be in units of eV. $q_\textrm{ref}$ is taken to be $q_\textrm{ref} = \alpha m_e$ by default (users can change this by editing the relevant variable in the \texttt{units.jl} file and installing the package locally). One can construct a \texttt{ModelDMSM} either by supplying all five arguments, or by providing \texttt{fdm\_n} $= n$ where $F_{\textrm{DM}}^2 \sim (\alpha m_e / q)^{2n}$ as an alternative way of defining $F_{\textrm{DM}}$,
\begin{minted}[bgcolor=pink]{Julia}
ModelDMSM(a, b, mX, mSM, deltaE)
ModelDMSM(fdm_n, mX, mSM, deltaE)
\end{minted}

Given that our primary concern is anisotropic detectors, the last important factor is the relative rotation between the material and the velocity distribution.
This can be specified using a \texttt{Quaternion} or \texttt{Rotor} object from the \texttt{Quaternionic.jl} package~\cite{Boyle_Quaternionic_jl_2023}.
If one wishes to use Euler angles or another way of specifying the rotation, \texttt{Quaternionic.jl} provides the conversion functions (see \href{https://moble.github.io/Quaternionic.jl/stable/manual/#Quaternionic.from\_euler\_angles-Tuple{Any,\%20Any,\%20Any}}{their documentation}).

Calculating the rate once all of these ingredients have been assembled is fairly simple,
\begin{minted}[bgcolor=pink]{Julia}
rate(R, model::ModelDMSM, pfv::ProjectedF, pfq::ProjectedF; 
     ell_max=nothing, use_measurements=false)
\end{minted}
where \texttt{R} is the rotation and \texttt{ell\_max} specifies the maximum $\ell$ used when both $\ell_{\textrm{max}}$ for \texttt{pfv} and \texttt{pfq} are larger than \texttt{ell\_max}. To scan over rotations it is possible to pass an array of quaternions or rotors for \texttt{R}. For continuous final state energies this returns $dR/dE$ from Eqn.~\ref{eqn:contrate} while for discrete final state energies this returns $R$ from Eqn.~\ref{eqn:rate}. Note that this will \emph{not} include the constant prefactor $k_0 / T_{\textrm{exp}}$.

The only part of the computation that takes a significant amount of time is calculating the reduced partial rate matrix $\mathcal{K}$, specifically the intermediate step of calculating the kinematic scattering matrix $\mathcal{I}$. It is possible to pre-compute $\mathcal{I}$ with
\begin{minted}[bgcolor=pink]{Julia}
kinematic_I(lnvnq_max, model::ModelDMSM, 
            v_basis::RadialBasis, q_basis::RadialBasis)
\end{minted}
with \texttt{lnvnq\_max} $= (\lmax, n^v_{\textrm{max}}, n^q_{\textrm{max}})$. This is useful if one has many velocity distributions and/or material form factors projected with the same $v_{\textrm{max}}$, $q_{\textrm{max}}$, and $\Delta E$. From $\mathcal{I}$ it is straightforward to obtain $\mathcal{K}$,
\begin{minted}[bgcolor=pink]{Julia}
get_mcalK(mcI, pfv::ProjectedF, pfq::ProjectedF; 
          ell_max=nothing, use_measurements=true)
\end{minted}
where \texttt{mcI} is the precomputed $\mathcal{I}$ from \texttt{kinematic\_I}.
It is also possible to calculate $\mathcal{K}$ from a \texttt{ModelDMSM} instead, which will calculate $\mathcal{I}$ as an intermediate step,
\begin{minted}[bgcolor=pink]{Julia}
get_mcalK(model::ModelDMSM, pfv::ProjectedF, pfq::ProjectedF; 
          ell_max=nothing, use_measurements=true)
\end{minted}
Both versions will return the values in a flattened vector in an \texttt{McalK} object,
\begin{minted}[bgcolor=pink]{Julia}
struct McalK{A<:Union{Measurement,Float64}}
    K::Vector{A}
    ell_max::Int64
    vmax::Float64
    qmax::Float64
end
\end{minted}
and these values can be saved and loaded via
\begin{minted}[bgcolor=pink]{Julia}
writeK(outfile, mcK::McalK)
readK(infile[, vmax, qmax]; 
      use_err=true)
\end{minted}
where the basis maximums can optionally be specified when reading the $\mathcal{K}$ matrix. The dimensionful, basis-independent partial rate matrix $K$ (or $dK/dE$ in the case of continuous final states) only differs from this $\mathcal{K}$ by a factor of $v_{\textrm{max}}^2 / q_{\textrm{max}}$ and can be computed with
\begin{minted}[bgcolor=pink]{Julia}
partial_rate(K::McalK)
\end{minted}
An \texttt{McalK} object can be passed to the \texttt{rate} function as
\begin{minted}[bgcolor=pink]{Julia}
rate(R, mcK::McalK)
\end{minted}
where \texttt{R} is the rotation, equivalent to the case when not providing a pre-computed $\mathcal{K}$ matrix.

The constants in Eqn~\ref{eqn:rate} can be incorporated via an \texttt{ExposureFactor} struct,
\begin{minted}[bgcolor=pink]{Julia}
struct ExposureFactor
    N_T::Float64
    sigma0::Float64
    rhoX::Float64
    total::Float64
end
expf = ExposureFactor(N_T, sigma0, rhoX)
\end{minted}
where \texttt{N\_T} is the number of individual targets, \texttt{N\_T} $= M_T / m_{\textrm{cell}}$ with $M_T$ the target mass and $m_{\textrm{cell}}$ the mass of the unit cell, \texttt{sigma0} is the reference cross-section in cm\textsuperscript{2}, \texttt{rhoX}~$=\rho_\chi$ is the local DM density in GeV / cm\textsuperscript{3}. \texttt{total} is computed based on the given inputs and is used to calculate the rate; it is equal to \texttt{total}~$= k_0 q_{\textrm{max}} / (T_{\textrm{exp}} v_{\textrm{max}}^2)$, where the constant $k_0$ is defined in Eqn.~\ref{eqn:k0}. This can be passed as the second argument in any of the rate functions,
\begin{minted}[bgcolor=pink]{Julia}
rate(R, exp::ExposureFactor, model::ModelDMSM, pfv::ProjectedF, 
     pfq::ProjectedF; ell_max=nothing, use_measurements=false,
     t_unit='s')
rate(R, exp::ExposureFactor, mcK::McalK; t_unit='s')
\end{minted}
which will return the rate for the given parameters in the specified units according to \texttt{t\_unit}, which can either be \texttt{`s'} or \texttt{`y'} for seconds or years.

Finally, to get the total event count for a specified parameter point, we provide \texttt{N\_events},
\begin{minted}[bgcolor=pink]{Julia}
N_events(R, T_exp_s, exp::ExposureFactor, model::ModelDMSM, 
         pfv::ProjectedF, pfq::ProjectedF; ell_max=nothing, 
         use_measurements=false)
N_events(R, T_exp_s, exp::ExposureFactor, mcK::McalK)
\end{minted}
which is similar to rate called with a provided \texttt{ExposureFactor}, but also includes the total detector run time \texttt{T\_exp\_s}, which should be in units of seconds.

\section{Python Implementation: VSDM}
\label{sec:python}

The Python implementation is available on the Python Package Index (PyPI) as \texttt{vsdm}, and on Github at \href{https://github.com/blillard/vsdm}{\texttt{https://github.com/blillard/vsdm}}. It can be installed via \texttt{pip}:
\begin{align*}
\texttt{pip install vsdm} 
\end{align*}
Beyond the core package, the Github repository includes several example notebooks for the $\langle f | n \ell m \rangle$ and rate calculations, 
as well as an efficient program for evaluating $\langle g_\chi | n \ell m \rangle$ for Standard Halo Model velocity distributions, \texttt{SHM\_gX.py}. 

A rate calculation proceeds in two stages: first, the functions $g_\chi$ and $f_S^2$ are projected onto velocity and momentum vector spaces; then, the scattering rate as a function of detector orientation is found by evaluating the partial rate matrix $K^{(\ell)}_{m m'}$. 
In the Python implementation, a few fundamental \texttt{class} objects handle each stage of the calculation: 
\begin{itemize}
\item \texttt{Fnlm}: calculates and saves the coefficients $\langle f | n \ell m \rangle$ for functions $f(\vec u)$ of a 3d variable $\vec u$
\item \texttt{McalI}: calculates and saves the kinematic scattering matrices $\mathcal I^{(\ell)}(E)$ for each DM model $(m_\chi, F_\text{DM}^2)$ and energy transfer $E$
\item \texttt{McalK}: calculates the dimensionless partial rate matrices $\mathcal K^{(\ell)}(E)$, for combinations of $\langle g_\chi | n \ell m \rangle$, $\langle f_S^2(E) | n \ell m \rangle$, and $\mathcal I^{(\ell)}(E)$. 
\end{itemize}
\texttt{McalK} includes several functions (class methods) for calculating the rate $R \propto G^{(\ell)}_{m m'} \mathcal K^{(\ell)}_{m m'}$ from \eqref{eq:dRdEK}, with or without the exposure factor $k_0$.

Below, we provide a brief description of the contents of the \texttt{vsdm} package.
Sections~\ref{vsdm:projection} and~\ref{vsdm:rate} describe the usage of each object and its variants in detail. 
Appendix~\ref{appx:PyExample} summarizes the workflow for a \texttt{vsdm}, showing in a single page how to perform all steps in a standard computation. More detailed demonstrations are available on the Github repository, in the \href{https://github.com/blillard/vsdm/tree/main/tools}{\texttt{tools}} directory.

\paragraph{Constants and Basic Functions:} 
\begin{itemize}
\item \texttt{units.py}: defines the units for velocity and energy used in the internal calculations.\footnote{All built-in choices use $\hbar = 1$, but the fundamental unit for velocity can be changed from its default of $c=1$ by editing \texttt{VUNIT\_c}.} This file also defines convenient dimensionful parameters $\texttt{km\_s} =1\, \text{km/s}$  and $\texttt{g\_c} = c$ for velocities; $\texttt{qBohr} = \alpha m_e c$ for momenta; and \texttt{eV}, \texttt{keV}, \texttt{MeV}, etc.~for energies. This file also defines the $q_\text{ref} = \texttt{q0\_fdm}$ and $v_\text{ref} = \texttt{v0\_fdm}$ parameters that appear in \eqref{FDM2:qavb}. By default, $q_\text{ref} = \texttt{qBohr}$ and $v_\text{ref} = c$.
\item \texttt{utilities.py}: includes several general-purpose functions, including our implementation of spherical harmonics and associated Legendre polynomials; the exposure factor $\texttt{g\_k0} = k_0$; 
	the DM scattering form factor functions $\texttt{fdm2\_n(n,q)} =   (q_\text{ref} / q)^{2n}$ and $\texttt{fdm2\_ab(a,b,q,v)} = (q / q_\text{ref})^a (v/c)^b$;
	transformations between Cartesian and spherical coordinate systems; 
	the numerical integration function \texttt{NIntegrate}; 
	and the interpolation objects \texttt{Interpolator1d} and \texttt{Interpolator3d}, for functions of $u$ or $\vec u$. 
\item \texttt{portfolio.py}: defines a consistent format for saving and reading HDF5 files.
\end{itemize}

\paragraph{Projection and Basis Functions:}
\begin{itemize}
\item \texttt{basis.py}: defines the \texttt{Basis} and \texttt{Basis1d} classes, for generic $\ket{n \ell m }$ and $\ket{n}$ basis functions (respectively). Used by \texttt{Fnlm}: see Section~\ref{vsdm:projection} for detail. 
\item \texttt{haar.py}: contains the wavelet-specific functions for Haar and spherical Haar wavelets, including the wavelet transformation (\texttt{haar\_transform}) and its inverse transformation (\texttt{haar\_inverse}) for arbitrary $d$ dimensional spherical wavelets, reverting to the standard Haar wavelets at $d = 1$. Used by \texttt{Basis} for the wavelet-harmonic $\ket{n \ell m }$ functions.
\item \texttt{projection.py}: includes the \texttt{Fnlm} class, which stores, manipulates and saves the $\langle f | n \ell m \rangle$ coefficients; and \texttt{EvaluateFnlm}, which automatically calculates $\langle f | n \ell m \rangle$ for a predetermined list of $(n, \ell, m)$ coefficients. 
\item \texttt{Gaussian.py}: implements an analytic shortcut for calculating $\langle f | n \ell m \rangle$ for functions $f(\vec u)$ that are provided as the sum of several Gaussians.  
\item \texttt{adaptive.py}: applies a wavelet-specific adaptive technique in \texttt{WaveletFnlm}. This method estimates the values of $\langle f | n' \ell m \rangle$ coefficients from the $n \approx n'/2$ previously evaluated coefficients $\langle f | n \ell m \rangle$, and calculates additional $\langle f | n' \ell m \rangle$ from numerical integration until the wavelet-harmonic expansion converges to within a specified absolute accuracy goal. 
\end{itemize}

\paragraph{Rate Calculation and Rotations:} 
\begin{itemize}
\item \texttt{analytic.py}: provides the $\mathcal I^{(\ell)}$ integrals for piecewise-constant radial basis functions, following \eqref{eq:mIstar}. 
\item \texttt{matrixcalc.py}: defines the \texttt{McalI} object, which evaluates the kinematic scattering matrix $\mathcal I^{(\ell)}$ for arbitrary radial basis functions. For wavelets and tophat basis functions, \texttt{McalI} uses the \texttt{analytic.py} results for the \eqref{eq:mIstar} integral. \texttt{McalI} can also be used with arbitrary radial basis functions, by integrating \eqref{def:mathcalI} numerically. 
\item \texttt{wigner.py}: evaluates the Wigner $G$ matrix for real spherical harmonics in \texttt{WignerG}
\item \texttt{ratecalc.py}: defines the \texttt{McalK} object, which calculates and stores the partial rate matrices $\mathcal K^{(\ell)}$. The class \texttt{McalK} includes methods for calculating the scattering rate as a function of detector orientation $\mathcal R \in SO(3)$, for rotation operators $\texttt{R} = \mathcal R$ provided in the quaternionic representation of \eqref{eq:QR}.
\end{itemize}

In this section we use the \texttt{vsdm} built-in units, e.g.~$\texttt{km\_s} = 1 \, \text{km/s}$, but otherwise we keep the \texttt{vsdm} prefix for functions and classes from the \texttt{vsdm} namespace, as if the package has been loaded via: 
\begin{minted}[bgcolor=pink]{Python}
import vsdm
from vsdm.units import *
\end{minted}

\subsection{Defining a Function} 
\label{vsdm:definition}

As the first step of the calculation, the user must provide the input distributions $g_\chi(\vec v)$ and $f_S^2(\vec q, E)$ as functions of 3d velocity $\vec v$ and momentum $\vec q$, in spherical coordinates: $\vec u = (u, \theta, \phi)$, where $\theta$ is the polar angle and $u = |\vec u|$. 
After defining a function $f(\vec u)$, the symmetries of $f(\vec u)$ can be specified by adding decorations to the Python function. Just as in the Julia version, the options are:
\begin{itemize}
  \item \texttt{z\_even = True} : if $f(x,y,z) = f(x,y,-z)$. Sets $\langle \ell m | f \rangle = 0$ for odd $(\ell+m)$.
  \item \texttt{phi\_even = True} : if $f(u,\theta,\phi) = f(u,\theta,-\phi)$. Sets $\langle \ell m | f \rangle = 0$ for negative $m$.
  \item \texttt{phi\_cyclic = n} : if $f(u,\theta,\phi) = f(u,\theta,\phi + 2\pi/n)$. Sets $\langle \ell m | f \rangle = 0$ unless $m$ is divisible by \texttt{n}. 
  \item \texttt{phi\_symmetric} : if $f(u,\theta)$ is independent of $\phi$. Sets $\langle \ell m | f \rangle = 0$ for all $m \neq 0$.
  \item \texttt{center\_Z2} : if $\vec f(\vec u) = \vec f(- \vec u)$ is symmetric under central inversion. Sets $\langle \ell m | f \rangle = 0$ for odd $\ell$.
\end{itemize}
For example:
\begin{minted}[bgcolor=pink]{Python}
def f_uSph(uSph):
	...
	return f_u
f_uSph.z_even = True 
f_uSph.phi_even = True 
f_uSph.phi_cyclic = 2 
f_uSph.phi_symmetric = False
f_uSph.center_Z2 = True
\end{minted}
defines a function that has the symmetries of a rectangular box. It is symmetric under the reflections $z \rightarrow -z$ and $\phi \rightarrow -\phi$ (a.k.a.~$y \rightarrow -y$), and also has a $Z_2$ rotational symmetry about the $z$ axis. 

The combination of \texttt{z\_even} and \texttt{phi\_cyclic = 2} means that $f$ is also center-symmetric. VSDM would set \texttt{center\_Z2 = True} in this case automatically, but for clarity we have added it explicitly to \texttt{f\_uSph}'s list of discrete symmetries. 
By default, VSDM assumes that $f$ has no symmetries, so the line \texttt{f\_uSph.phi\_symmetric = False} can be omitted without consequence.

If $f(\vec u)$ can be written as a sum of Gaussian functions, \texttt{vsdm} uses some analytic shortcuts to simplify the evaluation of $\langle f | n \ell m \rangle$. We use the normalization: 
\begin{align}
f(\vec u) &= \sum_i c_i  \frac{1 }{\sigma_i^3 (2 \pi )^{3/2} } \exp\left( - \frac{|\vec u - \vec u_i|^2}{2 \sigma_i^2} \right), 
\end{align}
with each term denoting a Gaussian of width $\sigma_i$, centered at $\vec u_i$, with $L^1$ norm $c_i$. 
In \texttt{vsdm}, a function of this type can be expressed as a list of $\{ c_i, \vec u_i, \sigma_i \}$ parameters, with $\vec u_i$ in spherical coordinates: 
\begin{minted}[bgcolor=pink]{Python}
v_1 = (250*km_s, 0, 0)
sigma_1 = 238*km_s / math.sqrt(2) 
v_2 = (320*km_s, 1.2*math.pi, 4.0) 
sigma_2 = 188*km_s / math.sqrt(2) 
gvec_list = [(0.8, v_1, sigma_1), (0.2, v_2, sigma_2)]
gX = vsdm.GaussianF(gvec_list)
\end{minted}
Each entry in \texttt{gvec\_list} can be provided as a tuple, list, or \texttt{numpy} array. 
This excerpt defines a Gaussian-type function, \texttt{gX}, with \texttt{len(gvec\_list) = 2} many Gaussian contributions. 

Just like \texttt{f\_uSph} above, the function \texttt{gX} can be decorated with the optional attributes \texttt{z\_even}, \texttt{center\_Z2}, etc. 
In this example, even though $f(\vec u)$ is symmetric under reflections along the $\vec v_1 \times \vec v_2$ axis, the plane of symmetry is not aligned with the $\vec u$ coordinate system, so there are no sets of $\langle f | \ell m \rangle$ coefficients that systematically vanish. 

\subsection{Projecting a Function} 
\label{vsdm:projection}

Once $g_\chi(\vec v)$ and $f_S^2(\vec q, E)$ are defined, we must specify the velocity and momentum basis functions $\ket{n \ell m}$, 
\begin{align}
\ket{n \ell m} &= r_n(u) \cdot Y_{\ell m}(\theta, \phi).
\end{align}
The built-in options for $r_n$ include $d$-dimensional spherical Haar wavelets, and similarly normalized spherical ``tophat'' basis functions, for $x \equiv u/u_\text{max}$ defined on $0 \leq x \leq 1$. 
Both of these options are piecewise-constant, allowing $\mathcal I$ to be calculated analytically.  It is possible, however, for the user to add their own radial basis functions $r_n^{(\ell)}$ to the \texttt{class Basis1d}: see for example the \texttt{laguerre} option included in \texttt{Basis1d.\_r\_n\_x}. We do not recommend using any set of basis functions that forces $\mathcal I^{(\ell)}$ to be evaluated via numerical integration.

In \texttt{vsdm}, the basis functions are specified by a dictionary object, that at a minimum defines the function type (wavelet or tophat) and the value of $u_\text{max}$. For example, 
\begin{minted}[bgcolor=pink]{Python}
basisV = dict(uMax=820*km_s, type='wavelet')
\end{minted}
identifies a wavelet basis with $v_\text{max} = 820\, \text{km/s}$. This dictionary can be used to create a \texttt{Basis} or \texttt{Fnlm} type class instance.

Given a function $f(\vec u)$ and a set of basis functions, any of the classes \texttt{Fnlm}, \texttt{EvaluateFnlm}, and \texttt{WaveletFnlm} can be used to calculate the coefficients $\langle f | n \ell m \rangle$:
\begin{itemize}
\item  \texttt{EvaluateFnlm} converts a function $f(\vec u)$ into $\langle f | n \ell m \rangle$ coefficients for arbitrary basis functions $\phi_{n \ell m}(\vec u)$ using numerical integration.
\item  \texttt{WaveletFnlm} uses properties of Haar wavelets to enhance and reduce evaluation time of \texttt{EvaluateFnlm}. 
\item \texttt{Fnlm} is the underlying class for both methods, and can write and read the $\langle f | n \ell m \rangle$ coefficients to/from CSV and HDF5 files.  
\end{itemize}
Each class has a \texttt{self.f\_nlm} dictionary, which saves the value of $\langle f | n \ell m \rangle$ with the key \texttt{(n, l, m)}: 
\begin{align}
\texttt{self.f\_nlm[(n, l, m)]} = \langle f | n \ell m \rangle.
\end{align}
When calculating the partial rate matrix, or saving the coefficients to an HDF5 file, \texttt{Fnlm} converts this dictionary into a more convenient 2d array, \texttt{self.f\_lm\_n}, 
discussed in subsection~\ref{subsec:arrayF}.

\subsubsection{Integration} 

The actual evaluation of $\langle f | n \ell m \rangle$ inner products is implemented in the class \texttt{Basis}, which is a parent class of \texttt{Fnlm}, via the \texttt{Basis.getFnlm()} method: 
\begin{align}
\texttt{Basis.getFnlm(f\_uSph, nlm, integ\_params)} = \int\! \frac{d^3 u}{u_0^3} \phi_\texttt{nlm}(\vec u) \, \texttt{f\_uSph}(\vec u) , 
\end{align}
with the dictionary \texttt{integ\_params} providing the instructions for evaluating the integral numerically. 
Here $\phi_{n \ell m} = \ket{n \ell m}$ and $u_0 = \texttt{Basis.u0}$, which for wavelet and tophat basis functions is equivalent to $u_\text{max}$. 

There are two numerical integration methods implemented in \texttt{vsdm.NIntegrate}: a Gaussian quadrature method, from \texttt{scipy.integrate}; and the adaptive Monte-Carlo method, \texttt{vegas}. 
The dictionary \texttt{integ\_params} contains an item \texttt{`method'}, which is set to either \texttt{`gquad'} or \texttt{`vegas'}, respectively.
In the former case: 
\begin{minted}[bgcolor=pink]{Python}
integ_params = dict(method='gquad', rtol_f=1e-5, atol_f=1e-8)
\end{minted}
tells \texttt{NIntegrate} to use quadrature, with a relative accuracy goal of \texttt{rtol = rtol\_f} and an absolute accuracy goal of \texttt{atol = atol\_f}. 

Alternatively, with \texttt{method=`vegas'}, the dictionary must specify the number of iterations, \texttt{nitn}, and the number of integrand evaluations per iteration, \texttt{n\_eval}: 
\begin{minted}[bgcolor=pink]{Python}
integ_params = dict(method='vegas', nitn=10, neval=1e4, 
                    nitn_init=3, neval_init=1e3)
\end{minted}
This instructs \texttt{vegas} to use $10^4$ evaluations for each of 10 iterations of the adaptive method. The optional \texttt{\_init} parameters add a lower-resolution initialization step, allowing \texttt{vegas} to adjust its sampling density to match the integrand before performing the main \texttt{nitn, neval} integration.   

For low-dimensional integrals, including the 3d $\langle f | n \ell m \rangle$ inner products, we find that quadrature is typically faster.
With either method, the dictionary can include an optional \texttt{integ\_params[`verbose'] = True} entry, which tells \texttt{NIntegrate} to print certain intermediate results. 

The output of \texttt{NIntegrate} includes an estimate of the uncertainty from the numerical integration. For \texttt{vegas} this is the standard deviation from the Monte Carlo integration; for \texttt{scipy.integrate}, the uncertainty is calculated from the \texttt{rtol} and \texttt{atol} parameters.
In both cases, we use the \texttt{gvar} Python package to combine the mean value and standard deviation into a single Gaussian variable.
The \texttt{gvar} package adjusts the $1\sigma$ widths as needed when adding or multiplying \texttt{gvar} variables, making it a convenient way to propagate the uncertainty from the numerical integration.
The mean value and standard deviation can be extracted from a \texttt{gvar} object via \texttt{.mean} and \texttt{.sdev}:
\begin{minted}[bgcolor=pink]{Python}
result = vsdm.NIntegrate(integrand, volume, integ_params)
value = result.mean
error = result.sdev
\end{minted}
Here \texttt{value} and \texttt{error} are float-valued, and \texttt{result} is an object of type \texttt{\_gvarcore.GVar}. 
Arithmetic with \texttt{gvar} objects can be substantially slower, so \texttt{Fnlm} includes a boolean parameter \texttt{use\_gvar} that can be set to \texttt{False} to replace \texttt{result} with \texttt{result.mean} for the relevant calculations. 

Note that the \texttt{gquad} integration method sometimes remains accurate to around one part in $10^8$ even when \texttt{rtol} is substantially larger than $10^{-8}$. In these cases, the reported error (\texttt{result.sdev} from \texttt{NIntegrate}) is substantially overestimated.
For this reason, the \texttt{use\_gvar} option is more useful when using the Monte Carlo integrator \texttt{vegas}. 

\subsubsection{Evaluation} 

The $\langle f | n \ell m \rangle$ calculation can be carried out using \texttt{Fnlm}, by running \texttt{getFnlm()} for each value of \texttt{nlm}: 
\begin{minted}[bgcolor=pink]{Python}
gXnlm = vsdm.Fnlm(basisV, f_type='gX', use_gvar=bool)
for nlm in nlmlist:
    gXnlm.updateFnlm(f_uSph, nlm, integ_params, 
                     csvsave_name='filename.csv')
\end{minted}
given some \texttt{integ\_params} and some list of $(n, \ell, m)$ indices.
The \texttt{updateFnlm} function automatically adds $\langle f | n \ell m \rangle$ to \texttt{self.f\_nlm}, either as a \texttt{gvar} variable or as a float, according to whether \texttt{use\_gvar} is \texttt{True} or \texttt{False}.  
Here we have included an optional label \texttt{f\_type}, to indicate that this function is a $g_\chi(\vec v)$ velocity distribution, 
and we provided a CSV file destination (\texttt{filename.csv}) for saving the $\langle f | n \ell m \rangle$ coefficients. 
The \texttt{f\_type} is only relevant when saving the results to HDF5 files.

If a file name is provided for \texttt{csvsave\_name}, then \texttt{updateFnlm} automatically saves each new value of $\langle f | n \ell m \rangle$ to the end of the CSV file immediately after evaluation. This  option is particularly convenient whenever the $\langle f | n \ell m \rangle$ integration time is relatively slow, as the coefficients are saved even if \texttt{Fnlm} is interrupted before it has processed the full \texttt{nlmlist}. 

\medskip 

\texttt{EvaluateFnlm} and \texttt{WaveletFnlm} both inherit the class methods of \texttt{Fnlm}, 
and can evaluate batches of $(n, \ell, m)$ coefficients upon initialization. 
With \texttt{EvaluateFnlm}, this is done by supplying a list \texttt{nlmlist} of $(n, \ell, m)$ tuples: 
\begin{minted}[bgcolor=pink]{Python}
gXnlm = vsdm.EvaluateFnlm(basisV, f_uSph, integ_params, 
                          nlmlist=[(0,0,0), (1,0,0), ...], 
                          csvsave_name='filename.csv', 
                          f_type='gX', use_gvar=bool)
\end{minted}
This example initializes an \texttt{Fnlm} type class instance called \texttt{gXnlm}, and calculates $\langle f | n \ell m \rangle$ in series for all of the $(n, \ell, m)$ coefficients supplied by \texttt{nlmlist}. Here  $f(\vec u)$ given by \texttt{f\_uSph}, and the basis functions $\phi_{n \ell m}(\vec u)$ are defined according to the \texttt{basisV} dictionary. 

Additional $n \ell m $ coefficients can be calculated (and added to \texttt{self.f\_nlm}) using the class method \texttt{Fnlm.updateFnlm()}. In this example, running
\begin{minted}[bgcolor=pink]{Python}
gXnlm.updateFnlm(f_uSph, (0,14,2), integ_params, 
                 csvsave_name='filename.csv')
\end{minted}
would calculate $\langle f | 0, 14, 2 \rangle$, add its value to \texttt{gXnlm.f\_nlm}, and append the result to the end of the CSV file \texttt{filename.csv}. The values of \texttt{integ\_params} (and even \texttt{f\_uSph}) can be changed. Running \texttt{updateFnlm} for an \texttt{nlm} already present in \texttt{self.f\_nlm} will overwrite the older value. 

\paragraph{WaveletFnlm:} 
When using wavelet basis functions, we recommend using \texttt{WaveletFnlm} instead of \texttt{EvaluateFnlm} whenever high precision in $\langle f | n \ell m \rangle$ is needed (e.g.~better than $10^{-8}$ relative precision).
\texttt{WaveletFnlm} calculates the $\langle f | n \ell m \rangle$ coefficients in batches of fixed $(\ell, m)$, for all $n < 2^p$ for some power $p$. 
Rather than integrating over $0 \leq x \leq 1$ for $n=0$ and $n = 1$, and $0 \leq x \leq 0.5$ for $n=2$ (etc.), \texttt{WaveletFnlm} calculates the $\langle f | N \ell m \rangle$ coefficients for $2^p$ many tophat basis functions $r_N(x)$, each of width $2^{-p}$, and applies a discrete transformation to convert $\langle f | N \ell m \rangle$ into wavelet coefficients $\langle f | n \ell m \rangle$. This can reduce the number of integrand evaluations by a factor of $p$ without negatively impacting the integration accuracy. \texttt{EvaluateFnlm}, on the other hand, calculates the integral $\langle f | n \ell m \rangle$ directly using wavelet functions, requiring relatively more integrand evaluations to reach the same precision as \texttt{WaveletFnlm}.

In the sample implementation of \texttt{WaveletFnlm} below, the $\langle f | n \ell m \rangle$ batch calculation is performed in a separate line: 
\begin{minted}[bgcolor=pink]{Python}
gXnlm = vsdm.WaveletFnlm(basisV, f_uSph, integ_params, f_type='gX', 
                         csvsave_name='filename.csv')
gXnlm.initialize_lm((0,0), 9)
\end{minted}
The \texttt{initialize\_lm(lm, power2)} function evaluates all $\langle f | n \ell m \rangle$ coefficients with $n < 2^9$ and $\texttt{lm} = (0,0)$. 
If \texttt{self.csvsave\_name} is not \texttt{None}, then \texttt{WaveletFnlm} will also save the results to the specified CSV file. 
\texttt{WaveletFnlm} can be parallelized simply by assigning different $(\ell, m)$ modes to different processors. 

The complete wavelet-harmonic projection can be performed (in series) with a single line, by supplying a non-empty dictionary \texttt{power2\_lm} when initializing \texttt{WaveletFnlm}: for example,
\begin{minted}[bgcolor=pink]{Python}
lm_p = {(0,0):10, (2,0):9, (2,2):9, (4,0):8, ...}
gXnlm = vsdm.WaveletFnlm(basisV, f_uSph, integ_params, power2_lm=lm_p)
\end{minted}
sets $p=10$ for $(\ell, m) = (0,0)$, $p=9$ for $\ell = 2$, and so on. 
One expects the magnitudes of $\langle f | n \ell m \rangle$ to decrease as $\ell$ becomes large, so the user may wish to keep somewhat larger values of $p$ at small $\ell$. 

\texttt{WaveletFnlm} also includes a \texttt{refine\_lm} function, that implements the polynomial wavelet extrapolation method from~\cite{Lillard:2023cyy} to predict the values of higher-$n$ wavelet coefficients from the set of previously calculated $\langle f | n \ell m \rangle$. The main reason to use \texttt{refine\_lm} is for functions $f(\vec u)$ that may have large, isolated fluctuations at small scales in $u$: \texttt{refine\_lm} subdivides this region of $u$ with progressively narrower basis functions, until the wavelet expansion converges.

If all values of $\langle f | n \ell m \rangle$ for $n = 2^{p-1}, 2^{p-1}+ 1, \ldots, 2^{p} - 1$  match the predictions from the ref.~\cite{Lillard:2023cyy} wavelet extrapolation method to within a tolerance set by \texttt{atol\_fnlm}, then there is nothing for \texttt{refine\_lm} to do.
Otherwise, for each $n$ where this difference is larger than \texttt{atol\_fnlm}, the \texttt{refine\_lm} function evaluates the descendant wavelets $n \rightarrow 2n$ and $n \rightarrow 2n + 1$ via numerical integration, and checks again to see if the wavelet extrapolation method has converged. 
The process continues until all $\langle f | n \ell m \rangle$ coefficients match the predictions from the polynomial extrapolation, with the maximum number of wavelet subdivisions set by \texttt{max\_depth}. For example,
\begin{minted}[bgcolor=pink]{Python}
gXnlm.refine_lm((0,0), max_depth=3)
\end{minted}
applies this process to the $(\ell, m) = (0, 0)$ harmonic mode, allowing up to three stages of wavelet subdivision. 
This optional part of the calculation can also be completed upon initialization of \texttt{WaveletFnlm}, by setting \texttt{refine\_at\_init = True} and supplying values for \texttt{max\_depth} and the absolute and relative tolerances \texttt{atol\_fnlm} and \texttt{epsilon}.
By default, \texttt{refine\_lm} applies the cubic order wavelet extrapolation, but the polynomial order can be changed by setting \texttt{p\_order} to e.g.~1 for linear extrapolation.

\subsubsection{Array Format for $\langle f | n \ell m \rangle$} \label{subsec:arrayF}

The dictionary format for \texttt{Fnlm.f\_nlm} is convenient when adding $\langle f | n \ell m \rangle$ coefficients one at a time, particularly for sparse lists. 
However, when calculating the partial rate matrix, it is much faster to convert this data into an array, so that the double sum $\sum_{n, n'}$ in \eqref{eq:KfromI} can be implemented via matrix multiplication. 
\texttt{Fnlm} organizes these coefficients into a 2d array, \texttt{Fnlm.f\_lm\_n}, where each row corresponds to a different $(\ell, m)$, and each column corresponds to $n$.
The mapping between the row label and $(\ell, m)$ is stored in \texttt{Fnlm.lm\_index}.

To populate the \texttt{self.f\_lm\_n} array with values from \texttt{self.f\_nlm}, simply run the line:
\begin{minted}[bgcolor=pink]{Python}
gXnlm._makeFarray(use_gvar=bool)
\end{minted}
with the option to set \texttt{use\_gvar = False} for this array even if \texttt{self.use\_gvar = True} for the class instance.  
For large sets of coefficients (e.g.~$\texttt{len(nlmlist)} \gtrsim 10^5$) this step may take several seconds, 
so \texttt{vsdm} usually waits to calculate \texttt{f\_lm\_n} until it is needed, e.g.~when saving an HDF5 file or before calculating the partial rate matrix.

\subsubsection{Reading and Writing CSV Files}

There are two built-in file formats for saving $\langle f | n \ell m \rangle$ coefficients from \texttt{Fnlm}: CSV and HDF5. 
The CSV format is simpler: each row saves the values of 
\begin{minted}[bgcolor=pink]{Python}
n, l, m, fnlm.mean, fnlm.sdev
\end{minted}
for each value of \texttt{fnlm} in the \texttt{Fnlm.f\_nlm} dictionary. The indices $n$, $\ell$, and $m$ are not necessarily in any sort of order. 
Every \texttt{Fnlm} type object has a class method \texttt{writeFnlm\_csv()}, 
\begin{minted}[bgcolor=pink]{Python}
Fnlm.writeFnlm_csv(csvfile, nlmlist=None)
\end{minted}
which when called with \texttt{nlmlist=None} will save all values in \texttt{self.f\_nlm} to the CSV file. Alternatively, supplying a nonempty list for $\texttt{nlmlist} = \{ (n, \ell, m) \}$ will save only those values of $\langle f | n \ell m \rangle$.
Rather than overwriting any existing lines of \texttt{csvfile}, \texttt{writeFnlm\_csv()} appends the new values to the end of the file.

Once a CSV file has been created with this format, it can be used to populate a new \texttt{Fnlm} object. 
For example: 
\begin{minted}[bgcolor=pink]{Python}
gXnlm = vsdm.Fnlm(basisV, f_type='gX', use_gvar=bool)
gXnlm.importFnlm_csv(filename.csv, use_gvar=bool)
\end{minted}
The second line reads the values in the CSV file \texttt{filename.csv}, and adds each $\langle f | n \ell m \rangle$ to the \texttt{self.f\_nlm} dictionary. 
If the file includes multiple rows with the same value of $\texttt{nlm} = (n, \ell, m)$, then the last entry (the most recent) is the one ultimately saved to \texttt{f\_nlm[nlm]}.

\subsubsection{Reading and Writing HDF5 Files}

The CSV format is human-readable and highly convenient when working with one or two functions at a time, but it becomes increasingly challenging to organize larger sets of $g_\chi$ or $f_S^2$ distributions. 
In this scenario the HDF5 format is preferable: all of the data can be saved to the same file, and each dataset within the file can store the relevant metadata needed to interpret the tabulated coefficients, e.g.~the description of the $\ket{n \ell m}$ basis functions, and any parameter values $\vartheta$ that went in to the definition of $f(\vec u, \vartheta)$.

By running the function \texttt{Fnlm.writeFnlm}, 
\begin{minted}[bgcolor=pink]{Python}
gXnlm.writeFnlm(filename.hdf5, modelName, use_gvar=bool)
\end{minted}
the array \texttt{self.f\_lm\_n} and the $\ell m$ index \texttt{self.lm\_index} are both saved as datasets within the HDF5 subgroup \texttt{typeName/modelName}, where \texttt{typeName} is given by the function's \texttt{f\_type} (e.g.~`\texttt{gX}' for the \texttt{gXnlm} example), and \texttt{modelName} is a name string chosen by the user. 
Unless otherwise specified, the \texttt{f\_lm\_n} dataset name is `\texttt{fnlm}', and the $(\ell, m)$ index dataset is saved with the name `\texttt{lm\_index}'. 
The default dataset names \texttt{fnlm}, \texttt{lm\_index} are defined in \texttt{portfolio.py}. 

The \texttt{modelName} can be organized with additional subgroups: for example, setting 
\begin{minted}[bgcolor=pink]{Python}
modelName = 'SHM/vEsc_544/vE_250'
\end{minted}
would generate a subgroup \texttt{SHM} under the group \texttt{gX}, with nested subgroups \texttt{vEsc\_544} and \texttt{vE\_250}, 
e.g.~for an SHM velocity distribution with escape velocity 544\,km/s and Earth speed 250\,km/s.

Every HDF5 dataset comes with a \texttt{.attrs} dictionary to store auxiliary data. By default, \texttt{writeFnlm} includes all entries in its \texttt{self.basis} dictionary, such as the \texttt{type} and the values of \texttt{uMax} and/or \texttt{u0}. 
The easiest way to feed additional data to \texttt{.attrs} is to add them as items in \texttt{self.basis} before running \texttt{writeFnlm}.
For the \texttt{gXnlm} example above, one might add:
\begin{minted}[bgcolor=pink]{Python}
gXnlm.basis['vEsc'] = 544.*km_s
gXnlm.basis['vCirc'] = 238.*km_s
gXnlm.basis['vE'] = 250.*km_s
gXnlm.writeFnlm(filename.hdf5, 'SHM/vE_Feb15', alt_type=None)
\end{minted}
to ensure that the SHM parameter values are saved with the data. 
The optional \texttt{alt\_type} argument can be used to save the data under a different \texttt{typeName = alt\_type}, rather than the default value \texttt{typeName = self.f\_type}. This capability is included for cases such as the one encountered in \eqref{eq:responseWi}, where an analysis includes multiple types of response functions for the same material.

To import the $\langle f | n \ell m \rangle$ coefficients from an HDF5 file, the user only needs to specify the \texttt{modelName} and the file name: 
\begin{minted}[bgcolor=pink]{Python}
gXnlm = vsdm.Fnlm(basisV, f_type='gX', use_gvar=bool)
dataFnlm, basis_info = gXnlm.importFnlm(filename.hdf5, modelName)
\end{minted}
By default, \texttt{importFnlm} looks for the dataset `\texttt{fnlm}' under \texttt{f\_type/modelName}, but if a string is provided for \texttt{alt\_type} then \texttt{Fnlm} will import the coefficients from \texttt{alt\_type/modelName} instead. 

The class method \texttt{self.importFnlm} does not automatically populate \texttt{self.basis} with the values from the HDF5 dataset's \texttt{.attrs} dictionary. Similarly, running \texttt{importFnlm} fills \texttt{self.f\_nlm} with the saved $\langle f | n \ell m \rangle$ coefficients, but not the array \texttt{self.f\_lm\_n}. 
However, both types of information can be recovered from the \texttt{return} value of \texttt{importFnlm}, as demonstrated above. 
Here, the first item \texttt{dataFnlm} is the \texttt{f\_lm\_n} array from the dataset, and \texttt{basis\_info} is the \texttt{.attrs} dictionary.

\paragraph{Automatic Dataset Naming:} 
The function \texttt{Fnlm.writeFnlm()} uses a naming scheme to avoid overwriting existing datasets. If a dataset by the name \texttt{fnlm} already exists in the subgroup \texttt{typeName/modelName}, then \texttt{writeFnlm} saves the  newest dataset under the modified names \texttt{fnlm\_\_1}, \texttt{fnlm\_\_2}, etc., appending a double underscore and an integer to the original dataset name.

The same procedure is used for the \texttt{lm\_index} dataset. 
Each \texttt{lm\_index.attrs} dictionary (\texttt{lm\_index\_\_1.attrs}, etc.) saves the name of its matching \texttt{fnlm} dataset, under the key \texttt{dname}. 
So, the $\langle f | n \ell m \rangle$ dataset matching a given \texttt{lm\_index} can be recovered via:
\begin{minted}[bgcolor=pink]{Python}
dsetname = lm_index__2.attrs['dname']
\end{minted}
If all edits to the HDF5 file have been made using the \texttt{Portfolio} class, then one expects \texttt{dsetname = fnlm\_\_2} to match the suffix of \texttt{lm\_index}. 

To import datasets that are not named ``\texttt{fnlm}'' and ``\texttt{lm\_index},'' run \texttt{importFnlm} with the dataset names specified via the \texttt{d\_fnlm} and \texttt{lm\_ix} options:
\begin{minted}[bgcolor=pink]{Python}
gXnlm.importFnlm(filename.hdf5, modelName, 
                 d_fnlm='fnlm__2', lm_ix='lm_index__2')
\end{minted}

\subsection{Calculating A Rate} \label{vsdm:rate}

Following \eqref{eq:RfromKG} and \eqref{eq:dRdEfromKG}, 
the scattering rate $R$, or the differential scattering rate $dR/dE$, can be rapidly evaluated from the partial rate matrix $\mathcal K$ defined in \eqref{eq:KfromI}. 
The precursor to $\mathcal K$ is the kinematic scattering matrix $\mathcal I^{(\ell)}_{n n'}(E, m_\chi, \ldots)$ of Section~\ref{sec:kinematicI}, which must be evaluated for every DM particle model $(m_\chi, F_\text{DM}^2)$. 
In \texttt{vsdm} this part of the calculation is handled by the class \texttt{McalI} from \texttt{matrixcalc.py}. 

Section~\ref{vsdm:KinematicI} calculates $\mathcal I$, and describes how it can be saved in the HDF5 format. 
Section~\ref{vsdm:PartialRate} shows how to calculate the partial rate matrix $\mathcal K$ from $\mathcal I$ and the vector space representations of $g_\chi(\vec v)$ and $f_S^2(\vec q, E)$. 
After assembling the Wigner $G$ matrices for each rotation $\mathcal R \in SO(3)$ in Section~\ref{subsec:wigner}, 
Section~\ref{subsec:rate} evaluates the scattering rate (or differential rate) for every detector orientation.

\subsubsection{Kinematic Scattering Matrix $\mathcal I$} \label{vsdm:KinematicI}

Like the \texttt{Fnlm} classes, \texttt{McalI} is a basis-dependent object, and it defines the velocity and momentum basis functions using the same \texttt{basisV} and \texttt{basisQ} dictionaries. Unlike \texttt{Fnlm}, \texttt{McalI} depends on the details of the DM particle model, via the DM mass $m_\chi$ and the DM--SM scattering form factor $F_\text{DM}^2(q, v)$. 
This information, and the energy transfer $E$, are provided to \texttt{McalI} via another dictionary, \texttt{modelDMSM}, so that an \texttt{McalI} class instance is defined via: 
\begin{minted}[bgcolor=pink]{Python}
modelDMSM = dict(mX=..., fdm=(a,b), DeltaE=..., mSM=...)
mI = vsdm.McalI(basisV, basisQ, modelDMSM, mI_shape=(...),
                do_mcalI=bool, center_Z2=bool, use_gvar=bool) 
\end{minted}
The DM--SM scattering model depends on four parameters:
\begin{itemize}
\item \texttt{mX}: the DM particle mass, e.g. \texttt{1e9*eV} or \texttt{15*MeV} (using \texttt{units.py})
\item \texttt{fdm}: can be a tuple \texttt{(a, b)}, specifying the form factor parameters $a,b$ for $F_\text{DM}^2 \propto q^a v^b$, following \eqref{FDM2:qavb}. Or, can be an integer or float, \texttt{fdm = n}, in which case $a = -2n$ and $b=0$.
\item \texttt{DeltaE}: the energy transfer from DM to the SM system, provided in \texttt{eV}, \texttt{keV}, etc.
\item \texttt{mSM}: this is the mass of the SM target particle,  e.g.~$m_e$ for DM--$e^-$ scattering, so that the reduced mass of the DM--SM system is $\mu_{\chi \text{SM}} = m_\text{SM} m_\chi / (m_\text{SM} + m_\chi )$. 
\end{itemize}
So, the first two arguments of \texttt{McalI} define the basis for the scattering matrix, and the third argument provides the physical parameters for the scattering process. 

The values of $\mathcal I^{(\ell)}_{n_v, n_q}$ are saved as a 3d array, 
\begin{align}
\texttt{McalI.mcalI[l,nv,nq]} &= \mathcal I^{(\ell)}_{n_v, n_q}(\Delta E, m_\chi, F_\text{DM}^2),
\end{align}
with the same normalization as \eqref{def:mathcalI}. 
The fourth argument in the \texttt{McalI} initialization, the tuple \texttt{mI\_shape}, sets the size of this array, so that it includes all values $\ell = 0, 1, \ldots, \ell_\text{max}$, where $\ell_\text{max} = \texttt{mI\_shape}[0] - 1$; likewise for the radial indices $n_v$ and $n_q$.

When $\ell_\text{max}$ and $n_\text{max}^{(v, q)}$ are large, \texttt{McalI} can be one of the most time-consuming parts of the calculation, and so the class is structured to allow for parallel computation of its components. 
Initializing \texttt{McalI} with \texttt{do\_mcalI = True} will instruct it to calculate all elements of $\mathcal I^{(\ell)}_{n n'}$ in series; with \texttt{do\_mcalI = False}, the \texttt{McalI.mcalI} array is initialized with each value equal to zero. Then, different sets of $(\ell, n_v, n_q)$ indices can be assigned to multiple processors, and evaluated using the \texttt{McalI.updateIlvq} class method.

As long as both \texttt{basisV} and \texttt{basisQ} are wavelet or tophat basis functions, $\mathcal I^{(\ell)}_{n_v, n_q}$ can be integrated analytically. Each coefficient can be evaluated by running:
\begin{minted}[bgcolor=pink]{Python}
mI.updateIlvq((l,nv,nq), analytic=bool, integ_params={'verbose':bool})
\end{minted}
Running \texttt{McalI} with the default values \texttt{analytic=True} and  \texttt{integ\_params=None} will replace \texttt{integ\_params} with the dictionary \texttt{dict(verbose=False)}.
Setting \texttt{verbose=True} causes \texttt{McalI} to print the intermediate integration results even for the analytic calculation.

For non-piecewise-constant basis functions, \texttt{McalI} must integrate each coefficient numerically (\texttt{analytic=False}). Tn this case, \texttt{integ\_params} holds the parameters needed for the numerical integrator, with the same options as \texttt{Fnlm}. Note that compared to the analytic calculation,  numerical integration for $\mathcal I$ will be extremely slow.

Once $\mathcal I^{(\ell)}$ has been calculated, it can be saved to an HDF5 file for future use, using the class method
\begin{minted}[bgcolor=pink]{Python}
mI.writeMcalI(hdf5file, modelName, alt_type=None)
\end{minted}
Here \texttt{hdf5file} can be the same file that saves the \texttt{Fnlm} results for the $g_\chi$ and $f_S^2$ models. By default, \texttt{McalI} arrays are stored under the type name \texttt{mcalI}, but the type name can be edited by passing a string to the \texttt{alt\_type} argument. 
In either case, the dataset is saved under a subgroup name given by the \texttt{modelName} argument.
The details of the \texttt{basisV} and \texttt{basisQ} basis functions, and \texttt{modelDMSM}, are saved to the \texttt{.attrs} dictionary.
When running \texttt{McalI} with \texttt{use\_gvar = True}, the central values and error bars are saved to two separate datasets, with names ending with ``\texttt{\_mean}'' and ``\texttt{\_sdev},'' respectively.

To read in the values of $\mathcal I^{(\ell)}_{n n'}$ from an HDF5 file, simply run:
\begin{minted}[bgcolor=pink]{Python}
mI.importMcalI(hdf5file, modelName, d_pair=['Ilvq_mean'])
\end{minted}
providing an additional \texttt{alt\_type} keyword argument if needed. 
When using the recommended wavelet or tophat basis functions, $\mathcal I$ is found analytically, so there is no \texttt{Ilvq\_sdev} dataset with error bars. However, if $\mathcal I$ is found using numerical integration, then the error bars can be imported by setting:
\begin{minted}[bgcolor=pink]{Python}
d_pair=['Ilvq_mean', 'Ilvq_sdev']
\end{minted}
when running \texttt{importMcalI}.

\subsubsection{The Reduced Partial Rate Matrix $\mathcal K$} \label{vsdm:PartialRate}

Once $\langle g_\chi | n \ell m \rangle \rightarrow \texttt{gV}$, $\langle f_S^2 | n' \ell m' \rangle\rightarrow \texttt{fsQ}$, and $\mathcal I^{(\ell)}_{n n'} \rightarrow \texttt{mI}$ are all known, the partial rate matrix $\mathcal K$ can be calculated directly using the \texttt{RateCalc} class:
\begin{minted}[bgcolor=pink]{Python}
mK = vsdm.RateCalc(gV, fsQ, mI, ellMax=None, lmod=1)
\end{minted}
\texttt{RateCalc} automatically finds $\mathcal K$ from \texttt{gV}, \texttt{fsQ} and \texttt{mI} upon initialization. 
By default, the cutoff $\ell_\text{max}$ is taken to be the smallest value from the three physics inputs \texttt{gV, fsQ, mI}, but \texttt{RateCalc} can be initialized with a different value via the optional \texttt{ellMax} argument.

For convenience, the parent class of \texttt{RateCalc}, \texttt{McalK}, keeps track of two versions of the partial rate matrix: the ``dimensionless'' $\mathcal K$ defined in \eqref{eq:KfromI}, 
\begin{align}
\texttt{mK.vecK[indexLMvMq]} &= \mathcal K^{(\ell)}_{m_v, m_q},
\end{align}
which has the same units as the momentum space form factor $f_S^2$; 
and the original partial rate matrix $K^{(\ell)}_{m m'}$ or $dK^{(\ell)}_{m m'}/ dE$, defined in \eqref{def:dKdE}:
\begin{align}
\texttt{mK.PartialRate[indexLMvMq]} &= \frac{v_\text{max}^2 }{q_\text{max}} \mathcal K^{(\ell)}_{m_v, m_q}.
\end{align}
Recall from \eqref{def:dKdE} that $\texttt{PartialRate} = K$ is basis-independent, but not dimensionless, while $\texttt{vecK} = \mathcal K$ depends on the value of $v_\text{max}$ and $q_\text{max}$. For discrete final states, $\ket{f_S^2}$ is dimensionless, and so is \texttt{vecK}; for continuum final states, $\ket{f_S^2}$ and $\texttt{vecK}$ have units of inverse energy. 

Both objects are stored as 1d arrays, indexed according to the function \texttt{Gindex(l, mv, mq, lmod=1)} from \texttt{wigner.py}: i.e.
\begin{align}
\texttt{indexLMvMq} &= \texttt{vsdm.Gindex}(\ell, m_v, m_q). 
\end{align}
The \texttt{lmod} option in \texttt{Gindex} is used to skip all values of $\ell$ that are not divisible by \texttt{lmod}. The default \texttt{lmod = 1} includes all $\ell = 0, 1, 2, \ldots$, but for systems where either $f_S^2$ or $g_\chi$ is center-symmetric, setting \texttt{lmod = 2} will skip all odd values of $\ell$. 
With \texttt{lmod = 1}, the entries of $\mathcal K^{(\ell)}_{m_v, m_q}$ are taken in the order:
\begin{align}
(\ell, m_v, m_q) &= (0, 0, 0), (1, -1, -1), (1, -1, 0), \ldots, (1, 1, 1), (2, -2, -2), \ldots, 
\end{align}
while with \texttt{lmod = 2} the odd $\ell$ (but not odd $m$) entries are skipped:
\begin{align}
(\ell, m_v, m_q) &= (0, 0, 0), (2, -2, -2), (2, -2, -1), \ldots, (2, 2, 2), (4, -4, -4), \ldots.
\end{align}

\paragraph{Saving $K$ or $\mathcal K$ to CSV/HDF5:}
The partial rate matrices contain all of the information needed to evaluate $dR/dE$ or $R$ for arbitrary detector orientations. This data is also quite concise: the number of coefficients, \texttt{len(PartialRate)}, simply scales as $\frac{4}{3}\ell_\text{max}^3 / \texttt{lmod}$, which is typically orders of magnitude smaller than $\mathcal I^{(\ell)}_{n n'}$. 

Because each $\mathcal K$ depends on the values of $E$ and $m_\chi$, as well as the versions of $g_\chi(\vec v)$, $F_\text{DM}^2(q, v)$ and $f_S^2(\vec q, E)$ under consideration, we leave it to the reader to save and organize their partial rate matrices according to their own preferences. 

When reading in $\mathcal K$ or $K$ from tabulated \texttt{vecK} or \texttt{PartialRate} data, it is simplest to use \texttt{McalK}, the parent class of \texttt{RateCalc}. For example:
\begin{minted}[bgcolor=pink]{Python}
mK = vsdm.McalK(ellMax, lmod=lmod)
mK.PartialRate = ... # import from CSV/HDF5 data with matching lmod
# alternatively, if vecK was tabulated instead:
mK.vecK = ... # import from CSV/HDF5
mK.PartialRate = VMAX**2 / QMAX * mK.vecK
\end{minted}
The first line initializes an \texttt{McalK} object with empty \texttt{vecK} and \texttt{PartialRate} vectors, which are then filled in by hand in the subsequent lines.

Unlike \texttt{PartialRate}, 
$\texttt{vecK} = \mathcal K$ depends on the values of $v_\text{max}$ and $q_\text{max}$, which must be saved independently when storing \texttt{vecK} data. 
On the other hand, \texttt{PartialRate} has dimensions of 
\begin{align}
[\texttt{PartialRate}] \propto \frac{\text{velocity}}{\text{mass}} \cdot [f_S^2] = \frac{\text{velocity}^3}{\text{energy}} \cdot \left\{ \begin{array}{c l} 1 & \text{discrete} \\ \text{energy}^{-1} & \text{continuous} \end{array}
\right.
,
\end{align}
so in this case the user must ensure consistency in their unit conventions. By default, \texttt{vsdm} represents velocities in units of $c$, and energies in units of electronvolts. These defaults can only be changed by editing \texttt{VUNIT\_c} and \texttt{EUNIT\_eV} in \texttt{units.py}.

\subsubsection{The Wigner $G$ Matrix} \label{subsec:wigner}

Converting $K$ into a scattering rate requires one final ingredient: the Wigner $G$ matrix \eqref{def:WignerG}, which is a function of detector orientation $\mathcal R \in SO(3)$. 
We represent rotation operators using the standard quaternion convention, \eqref{eq:RasQ}, and calculate $G^{(\ell)}_{m m'}$ from the Wigner $D^{(\ell)}$ matrix for complex spherical harmonics, following~\cite{Lillard:2023cyy}. 

The class \texttt{WignerG} saves the values of each $G^{(\ell)}_{m m'}(\mathcal R)$ as a 1d array, with the same indexing as \texttt{vecK} and \texttt{PartialRate}. 
If the list of rotation operators is known ahead of time, then \texttt{WignerG} can be initialized to evaluate $G$ for this list:
\begin{minted}[bgcolor=pink]{Python}
wG = vsdm.WignerG(ellMax, rotations=[Q1, Q2, Q3, ...], lmod=lmod)
\end{minted}
for quaternions \texttt{Q1}, \texttt{Q2}, etc., and for some values of \texttt{ellMax} and \texttt{lmod}. 

\texttt{WignerG} saves the values of $G(\mathcal R)$ to a 2d array, \texttt{G\_array}. It can also be used to evaluate additional values of $G(\mathcal R)$ for $\mathcal R$ not included in the initial \texttt{rotations} list, via the class method \texttt{G()}:
\begin{minted}[bgcolor=pink]{Python}
newG_R = wG.G(R, save=bool)
\end{minted}
If \texttt{save==True}, then the newest \texttt{G} vector is also appended to the \texttt{G\_array} list, and the corresponding value of \texttt{R} is appended to the list \texttt{wG.rotations} of saved rotation quaternions.

\subsubsection{Calculating Rates} \label{subsec:rate}

The partial rate class \texttt{McalK} includes several methods for evaluating the scattering rate for lists of $\mathcal R$ rotations.
The simplest of these options is the function \texttt{mu\_R}, which calculates the dot product
\begin{align}
\texttt{mu\_R}(\mathcal R) &= \sum_{\ell, m, m'} G^{(\ell)}_{m m'}(\mathcal R) \cdot \mathcal K^{(\ell)}_{m m'}
\end{align}
for every $\mathcal R$ in the \texttt{rotations} list. 
This is quite fast: it is as simple as multiplying the 2d array \texttt{wG.G\_array} by the 1d vector \texttt{mK.vecK}. 
The output, a list of $\mu(\mathcal R)$, is proportional to the rate $R$ or $dR/dE$ without the exposure factor $k_0$ from \eqref{eqn:k0}. 

To include the exposure factor, one can multiply \texttt{mu\_R} by the function \texttt{vsdm.g\_k0}. For discrete final states, this returns the total number of expected events, based on the target exposure (in $\text{kg}\,\text{yr}$), cross section $\bar\sigma$, and local DM density $\rho_\chi$:
\begin{align}
N_\text{events} &= \texttt{g\_k0} \cdot \texttt{mu\_R}, 
\\
\texttt{g\_k0} &= k_0 = \frac{M_T}{m_\text{cell}^{(\text{mol})}} N_A T_\text{exp} \bar\sigma_0 \rho_\chi \frac{v_\text{max}^2}{q_\text{max} }
\end{align}
Here $N_A\,M_T/m_\text{cell}^{(\text{mol})}$ is the number of unit cells in the detector, written in terms of the target mass $M_T$ and the molar mass of the unit cell, $m_\text{cell}^{(\text{mol})}$, as in \eqref{eq:NTNA}.
For continuum final states, the product \texttt{g\_k0 * mu\_R} is the number of expected events per unit energy, i.e.
\begin{align}
T \cdot \frac{dR}{dE}\bigg|_E &= \texttt{g\_k0} \cdot \texttt{mu\_R}(E), 
\end{align}
for exposure time $T$. 

For either case, \texttt{McalK} provides a convenient class method, \texttt{Nevents}, that evaluates the total expected number of events for each rotation in \texttt{wG}:
\begin{align}
\texttt{mK.Nevents(wG)} &= \left\{ 
\begin{array}{cl} 
    T \cdot R(\mathcal R) & \text{discrete} 
    \\
    T \cdot \frac{dR}{dE} \big|_{E, \mathcal R} & \text{continuous} 
\end{array}
\right.
\end{align}
The exposure factor is determined by the values supplied in the optional arguments of \texttt{Nevents}: \texttt{exp\_kgyr}, $M_T T$ in units of kg\,yr; \texttt{mCell\_g}, the molar cell mass in grams; \texttt{sigma0\_cm2}, the cross section $\bar\sigma_0 d_{\alpha \beta}$ in $\text{cm}^2$; and \texttt{rhoX\_GeVcm3}, the local DM density in $\text{GeV}/\text{cm}^3$.

\section{Performance} \label{sec:performance}


In this section we provide some specific results for the evaluation times of both the Python and Julia versions of \texttt{VSDM} to compare to the full multi-dimensional numeric integration. The total evaluation time for a set of velocity distributions, material form factors, dark matter models (masses and form factors), and rotations is given by
\begin{align}
    T_{\textrm{tot}} = N_{g_\chi} N_v T_{\textrm{proj}} + N_{f_S} N_q T_{\textrm{proj}} +  N_\mathcal{I} T_\mathcal{I} + N_{g_\chi} N_{f_S} N_{\textrm{DM}} \left( N_\mathcal{K} T_\mathcal{K} + N_{\mathcal{R}} T_{\textrm{tr}} \right) ,
\end{align}
where
\begin{itemize}
    \item $N_{g_\chi}$ is the number of different velocity distributions
    \item $N_{f_S}$ is the number of different material form factors
    \item $N_v$ and $N_q$ are the total number of coefficients computed for each velocity distribution and material form factor respectively
    \item $T_{\textrm{proj}}$ is the average time to project a single coefficient
    \item $T_\mathcal{I}$ is the average time per coefficient to compute the $\mathcal{I}^{(\ell)}_{n,n'}$ matrix
    \item $T_\mathcal{K}$ is the average time per coefficient to compute a reduced partial rate matrix $\mathcal{K}^{(\ell)}_{m,m'}$ coefficient from a pre-computed $\mathcal{I}$
    \item $N_{\textrm{DM}}$ is the number of dark matter models
    \item $N_\mathcal{R}$ is the number of rotational configurations
    \item $T_{\textrm{tr}}$ is the average time to compute the final dot product between the Wigner $G$ matrix and the partial rate matrix $K$.
\end{itemize}
$N_v$ ($N_q$) is determined by the number of radial coefficients $N_n^{(v)}$ ($N_n^{(q)}$), $\lmax$, and the symmetries of the given function to be projected. For a function with \emph{no} symmetries the total number of coefficients is $N_x = N_n^{(x)} (\lmax+1)^2$. If \texttt{z\_even = true} or \texttt{phi\_even = true} this number is approximately halved and if \texttt{phi\_symmetric = true} divide the total number of coefficients by a factor of $\lmax+1$. $N_{\mathcal{I}}$ is equal to the number of coefficients in $\mathcal{I}$ times the number of $\mathcal{I}$ matrices that need to be computed, $N_\mathcal{I} = N_{\textrm{DM}} N_{\Delta E} N_n^{(v)} N_n^{(q)} \ell_{\textrm{max}}$ where $N_{\Delta E}$ is the number of different excitation energies. $N_{\mathcal{K}}$ depends on the total number of coefficients over both the velocity distribution and material form factor, $N_\mathcal{K} = N_n^{(v)} N_n^{(q)} \ell_{\textrm{max}}^3$.

\renewcommand{\arraystretch}{1.25}
\begin{table}[h]
    \centering
    \begin{tabular}{|c||c|c|c|c|}
    \hline
     Language & $T_{\textrm{proj}}$ & $T_\mathcal{I}$ & $T_\mathcal{K}$ & $T_{\textrm{tr}} / \lmax^3$ \\
     \hline\hline
     Python & 0.74 & $1.4 \times 10^{-4}$ & $1.1 \times 10^{-10}$ & $9.8 \times 10^{-10}$ \\
     Julia & $7.0 \times 10^{-3}$ & $9.1 \times 10^{-7}$ & $3.4 \times 10^{-11}$ & $1.6 \times 10^{-10}$ \\
     \hline
    \end{tabular}
     \caption{All listed times are in seconds. $T_{\textrm{proj}}$ was computed using the example material form factor from~\cite{Lillard:2023cyy} with $n_{\textrm{max}} = 2^8-1$, $\lmax = 4$. $T_{\mathcal{I}}$ and $T_\mathcal{K}$ were computed with same $n_{\textrm{max}}$, $\lmax = 10$, for $m_\chi = 100$~MeV, $\Delta E = 4.03$~eV, and $F_{\textrm{DM}} = 1$. $T_{\textrm{tr}}$ is also computed for $\lmax = 10$ and only depends on $\lmax$.}
    \label{tab:timing}
   \end{table}

Table~\ref{tab:timing} shows a concrete example of evaluation times on a Ryzen R5 3600 processor (with hyperthreading disabled in the Julia version). For reference, it takes $\sim 30$ sec to numerically calculate the full integral rate for a single parameter point using the \texttt{MCIntegration.jl} package in Julia, with options \texttt{solver=:vegas}, \texttt{niter}=$10$, and \texttt{neval}$=10^6$ (everything else default). In Python, using the package \texttt{vegas} it takes $\sim 100$ sec to perform the same integral (with \texttt{nitn}$=10$, \texttt{neval}$=10^6$). Both packages end up with about $\mathcal{O}(0.1\%)$ error with these options.

If we take $N_{g_\chi} = N_{f_S} = 10$, $N_{\Delta E} = 10$, $N_n^{(v)} = N_n^{(q)} = 2^8$, $\lmax = 10$, $N_\textrm{DM} = 60$, and $N_{\mathcal{R}} = 10^4$ (a reasonable number of parameter points) and assume that neither the velocity distribution nor the material form factor have any symmetries, the \emph{first time} we calculate everything the total evaluation times would be:
\begin{align*}
    \textrm{Monte Carlo (Julia): } &\sim 1.8 \times 10^{9} \textrm{ sec} \: \simeq 57\textrm{ yr} \\
    \textrm{Python: } &\sim 5.1 \times 10^{5} \textrm{ sec} \: \simeq 142\textrm{ hr} \\
    \textrm{Julia: } &\sim 4.7 \times 10^{3} \textrm{ sec} \: \simeq 79\textrm{ min} .
\end{align*}
On subsequent analyses, however, we can make use of the fact that we have already computed the projection coefficients, and load them from a file. For this number of coefficients and models, the total time to load projected coefficients and assemble them into a matrix is about 4 sec (92 sec) for Julia (Python) (coefficients stored on a WD SN770 drive). This decreases the evaluation times for \texttt{VSDM},
\begin{align*}
    \textrm{Python: } &\sim 5.5 \times 10^{4} \textrm{ sec} \: \simeq 15.3\textrm{ hr} \\
    \textrm{Julia: } &\sim 3.8 \times 10^{2} \textrm{ sec} \: \simeq 6.3\textrm{ min} .
\end{align*}
The majority of the computation time here is taken up by $\mathcal{I}$. There are two options for saving this: 1. you can directly save $\mathcal{I}$, although it is a large object ($N_n^{(v)} \times N_n^{(q)} \times (\lmax+1)$), or 2. you can save $\mathcal{K}$, which is in general much smaller. It is useful to save $\mathcal{I}$ when you have many different velocity distributions and/or material form factors that all use the same basis (and $\Delta E$), but if you do not anticipate needing to add more of these, saving $\mathcal{K}$ will save a lot of space. If we have saved all of the $\mathcal{K}$ matrices to a file the $R(\mathcal R)$ evaluation time is even faster,
\begin{align*}
    \textrm{Python: } &\sim 59\textrm{ sec} \\
    \textrm{Julia: } &\sim 9.6\textrm{ sec} ,
\end{align*}
where in both cases it would take about 20 sec to load all of the $\mathcal{K}$ matrices.
Calculating the scattering rate from a tabulated partial rate matrix is faster by factors of
\begin{align*}
    \textrm{Python: } &\sim 3.0 \times 10^{7} \\
    \textrm{Julia: } &\sim 1.9 \times 10^{8},
\end{align*}
compared to brute-force numerical integration.

\section{Conclusion}
\label{sec:conclusion}

VSDM is an invaluable tool for researchers in dark matter direct detection. By projecting the DM velocity distributions and the material form factors onto their respective bases of wavelet-harmonic functions, the three-dimensional input functions from astrophysics, condensed matter and/or physical chemistry are represented concisely in a format that is easily shared between researchers. VSDM easily assembles the partial rate matrix for DM--SM scattering from this data. For an anisotropic detector, VSDM can evaluate the orientation-dependent scattering rate in fractions of a microsecond for each point in parameter space.

Even an isotropic analysis can benefit from using VSDM, thanks to the factorization inherent in the VSDM rate calculation. For large sets of input distributions, i.e.~$N_{g_\chi} \gg 1$ velocity distributions and $N_{f_S} \gg 1$ response functions, the total number of rate evaluations can become large enough that even low-dimensional numerical integration becomes inconveniently slow. With VSDM, one only has to calculate the $\ell = m = 0$ modes of the velocity and momentum distributions, $\langle g_\chi | n 0 0 \rangle$ and $\langle f_S^2 | n' 0 0 \rangle$, and the $\ell_\text{max} = 0$ kinematic scattering matrix $\mathcal I^{(0)}_{n n'}$. The angular average rate, proportional to $\mathcal K^{(0)}_{0,0}$, is then efficiently obtained from \eqref{eq:KfromI} for each combination of models.

\section*{Acknowledgments}
The work of B.L.~was supported by the U.S. Department of Energy under Grant Number DE-SC0011640. The work of A.R.~was supported in part by NSF CAREER grant PHY-1944826.
We thank Pankaj Munbodh for providing helpful comments on the manuscript, and Pierce Giffin and P.K.\ for their comments on an earlier version of the Python implementation \texttt{vsdm}. 

\begin{appendix}
\numberwithin{equation}{section}

\section{Associated Legendre Polynomials and Spherical Harmonics}
\label{sec:legendre}

The associated Legendre polynomials $P_\ell^m(x)$ appear frequently in physics. They are solutions to the angular components of the 3d Laplace equation, i.e.:
\begin{align}
\frac{d}{dx} \left( (1 - x^2) \frac{d}{dx} P_\ell^m(x) \right)  + \left( \ell(\ell + 1) - \frac{m^2}{1 - x^2} \right) P_\ell^m(x) &= 0, 
\label{ddx:Plm}
\end{align}
and they are a crucial component in defining the spherical harmonics. 

Standard computational libraries (e.g.~SciPy for Python) include routines for evaluating spherical harmonics and the associated Legendre polynomials $P_\ell^m$, but these tend to fail for $\ell, m > 10^2$. 
In this Appendix, we outline the recursive method that VSDM uses to evaluate $Y_{\ell m}$. It retains its accuracy up to $\ell, m \leq 1800$, which is more than sufficient for our needs.

\subsection{Definitions}
We define the associated Legendre polynomial of order $\ell$ and degree $m$ via its Rodriguez formula, 
\begin{align}
P_\ell^m (x) &\equiv \frac{(-1)^{\ell+m} }{2^\ell \ell!} (1 - x^2)^{m / 2} \frac{d^{\ell + m}}{dx^{\ell + m } } (1 - x^2)^\ell,
\label{def:Plm}
\end{align}
including the Condon--Shortley phase $(-1)^m$. At $m = 0$, $P_\ell^0 = P_\ell$ is the regular (i.e. not associated) Legendre polynomial, 
\begin{align}
P_\ell(x) &\equiv \frac{(-1)^{\ell} }{2^\ell \ell!}  \frac{d^{\ell }}{dx^{\ell  } } (1 - x^2)^\ell .
\end{align}
\eqref{def:Plm} is valid for negative $m$, but these negative-order polynomials can also be derived from positive-order $P_{\ell}^{|m|}$ as follows:
\begin{align}
P_\ell^{-|m|} &=  (-1)^m \frac{(\ell - |m|)!}{(\ell + |m|)!} P_\ell^{|m|}(x) .
\end{align}

We define the complex spherical harmonics ($Y_\ell^m$) as:
\begin{align}
Y_\ell^m(\theta, \phi) &\equiv  \sqrt{ \frac{2\ell + 1}{4\pi} \frac{(\ell - m)!}{(\ell+m)!} } P_\ell^m (\cos\theta) e^{i m \phi} ,
\end{align}
while for the real spherical harmonics ($Y_{\ell m}$), we take: 
\begin{align}
Y_{\ell m}(\theta, \phi) &\equiv
\left\{ \begin{array}{l c c}
\sqrt{2}  (-1)^m    \sqrt{ \dfrac{2\ell + 1}{4\pi} \dfrac{(\ell - |m|)!}{(\ell+|m|)!} } P_\ell^{|m|} (\cos\theta) \sin(|m|\varphi)
&& \text{for } m < 0,
\\[\bigskipamount]
\sqrt{ \dfrac{2\ell + 1}{4\pi} } P_\ell(\cos\theta)  && \text{for } m = 0,
\\[\bigskipamount]
\sqrt{2}   (-1)^m   \sqrt{ \dfrac{2\ell + 1}{4\pi} \dfrac{(\ell - m)!}{(\ell+m)!} } P_\ell^m (\cos\theta) \cos(m\varphi)
&& \text{for } m > 0,
\end{array}  
\right.
\end{align}
so that the negative-$m$ real harmonics are odd under $\varphi \rightarrow - \varphi$, and the positive-$m$ harmonics are even.

\subsection{Difficulty At Large $\ell$} 
The associated Legendre polynomial $P_\ell^m$ is a polynomial of degree $\ell$, with coefficients that can be determined analytically. For example, 
\begin{align}
P_\ell^m(x) &=  (-1)^m 2^\ell (1 - x^2)^{m/2} \sum_{k= m}^\ell \frac{k! }{(k- m)!} x^{k - m} \left( \begin{array}{c} \ell \\ k \end{array} \right) \left( \begin{array}{c} \frac{1}{2}(\ell + k - 1) \\ \ell \end{array} \right) ,
\label{series:plm} 
\end{align}
with the ``$n$ choose $k$'' symbols defined for non-integer $n$ as
\begin{align}
\left( \begin{array}{c} n \\ k \end{array} \right) &\equiv \frac{\Gamma(n + 1)}{\Gamma(k + 1) \, \Gamma(n - k + 1) } .
\end{align}
At $m = \ell$ this yields a particularly simple expression, 
\begin{align}
P_\ell^\ell(x) &= (-1)^\ell (2 \ell - 1)!! \, (1 - x^2)^{\ell/2}, 
\label{eq:Pellell}
\end{align}
where $!!$ is the double factorial, $n!! = (n)(n - 2)(n-4) \ldots (3)(1)$ for odd $n$. 

Attempting to evaluate $P_\ell^m$ directly from \eqref{series:plm} quickly leads to underflow and overflow errors when $\ell$ becomes large. For example, Stirling's approximation gives the asymptotic form for the double factorial as
\begin{align}
(2 \ell - 1)!! &\approx \sqrt{ \frac{4\ell - 4}{2\ell - 1} } \left[ \frac{(2\ell - 1)^2 }{2e (\ell -1) } \right]^\ell, 
\end{align}
which exceeds $10^{100}$ for $\ell > 60$. This large factor must cancel against similarly exponentially small factors from $\sqrt{(\ell - m)! / (\ell + m )!}$, $x^{k - m}$, and $(1 - x^2)^{m / 2}$ to yield $\mathcal O(1)$ values for $Y_\ell^m(\theta, \phi)$, and for $\ell \sim 10^2$ this cannot be done accurately from the polynomial form of \eqref{series:plm}. 

A great many recursion relations for $P_\ell^m$ and $P_{\ell \pm 1}^{m \pm 1}$ can be derived from \eqref{ddx:Plm}. Two relatively simple recursion relations keep either $\ell$ or $m$ fixed: we refer to these as fixed-degree or fixed-order recursion relations, respectively. 
SciPy (\texttt{v1.15.0}) appears to use the fixed-degree recursion relations: to find $P_\ell^m(x)$, it first evaluates $P_\ell^\ell(x)$ via \eqref{eq:Pellell}, then it employs the appropriate decreasing-order recursion relation to evaluate $P_\ell^{\ell -1}$, $P_\ell^{\ell - 2}$, \ldots, $P_\ell^m$. 
As a result, SciPy is only able to calculate $Y_\ell^m$ for $\ell$ and $m$ up to about $\ell \approx m \approx 85$. 
This may be insufficient for our purposes.

Instead, we employ a fixed-order recursion relation, beginning with $P_m^m$ and evaluating $P_{m+1}^m$, $P_{m + 2}^m$, \ldots, $P_\ell^m$ by increasing the degree. This is much more numerically stable, permitting the calculation of $Y_\ell^m$ up to $\ell \sim 1800$ with an absolute accuracy of $10^{-10}$ or better, using standard 64-bit floating precision.

\subsection{Recursion Relations} 

The associated Legendre polynomial satisfies the modified Bonnet recurrence formula: 
\begin{align}
(\ell - m) P_\ell^m &= x ( 2\ell -1 ) P_{\ell - 1}^m - (\ell + m - 1) P_{\ell - 2}^m ,
\label{recur:fixm}
\end{align}
suppressing the $x$ dependence of $P_\ell^m(x)$.
Given $P_m^m$ and $P_{m + 1}^m$, where
\begin{align}
P_{m  +1}^m &= x ( 2m + 1) P_m^m, 
\end{align}
\eqref{recur:fixm} provides the means to evaluate $P_\ell^m$ for $\ell \geq m  +2$. 
This is the \textit{Numerical Recipes}~\cite{10.5555/148286} recommendation for evaluating $P_\ell^m$.

For VSDM, the reason to calculate $P_\ell^m$ is to subsequently calculate $Y_{\ell m}$ or $Y_\ell^m$, so we define the ``reduced'' (or partially normalized) associated Legendre polynomials $\mathcal P_\ell^m$, 
\begin{align}
\mathcal P_\ell^m(x) &\equiv (-1)^m \sqrt{ \frac{(\ell - m )!}{(\ell + m)!} } P_\ell^m (x) ,
\end{align}
to absorb the exponentially small prefactor.
Including the factor of $(-1)^m$ removes the Condon--Shortley phase, so as to simplify the evaluation of the real spherical harmonics:
\begin{align}
Y_{\ell m}(\theta, \phi) &= \sqrt{ \frac{2 \ell + 1}{4\pi} } \mathcal P_\ell^{|m|}(\cos\theta) \times \left\{
\begin{array}{l c l}
\sqrt{2} \sin( |m| \phi) && m < 0,
\\
1 && m=0,
\\
\sqrt{2} \cos(m \phi) && m > 0.
\end{array}
\right.
\end{align}

Adapting \eqref{recur:fixm} for $\mathcal P_\ell^m$, we find:
\begin{align}
\sqrt{\ell^2 - m^2} \mathcal P_\ell^m &= (2 \ell - 1) x \mathcal P_{\ell - 1}^m - \sqrt{ ( \ell - 1)^2 - m^2 } \mathcal P_{\ell - 2}^m ,
\end{align}
with the initial values calculated via
\begin{align}
\mathcal P_m^m &= (1 - x^2)^{m /2} \prod_{j = 0}^{m - 1} \sqrt{ 1 - \frac{1/2}{1 + j}  } , 
\label{eq:Pmm}
&
\mathcal P_{m + 1}^m &= \sqrt{2 m + 1} \, x\,  \mathcal P_m^m. 
\end{align}
The form of the product above is the reason why we use $\mathcal P_\ell^m$ rather than $P_\ell^m$: 
each term in this product is approximately $1$, even when $m$ is absurdly large. 

The analogous fixed-degree recursion (apparently used by SciPy) is less stable. While $\mathcal P_\ell^\ell$ can be calculated precisely using \eqref{eq:Pmm}, the decreasing-order recursion relation, 
\begin{align}
\mathcal P_\ell^m(x) &= \frac{ (m + 1) \alpha_x \mathcal P_\ell^{m+1}(x) - \sqrt{ (\ell - m - 1)(\ell + m + 2) } \mathcal P_\ell^{m+2}(x) }{ \sqrt{(\ell - m)(\ell + m + 1) } } ,
\label{eq:descend}
\end{align}
depends on a coefficient 
\begin{align}
\alpha_x &\equiv \frac{x}{\sqrt{1 - x^2}} 
\end{align}
that becomes much larger than $\mathcal O(1)$ as $x \rightarrow \pm 1$.
Consequently, this routine for calculating $\mathcal P_\ell^m$ and $Y_\ell^m$ becomes unreliable at significantly smaller values of $\ell_\text{max}$, compared to the fixed-order recursion. 

\subsection{Accuracy At Large $\ell$}

At large $m$, the term $(1 - x^2)^{m /2}$ in \eqref{eq:Pmm} becomes exponentially small everywhere except about $x \approx 0$. 
With the fixed-order recursion relation, this is one limiting factor in our ability to accurately resolve $Y_{\ell m}(\theta, \phi)$: if  $(1 - x^2)^{m /2} = |\sin^m\theta|$ is too small, then our algorithm sets $Y_\ell^m \rightarrow 0$ for those values of $\theta$. This is related to a physical effect, however: the large $m$ spherical harmonics vanish as $P_\ell^m(\cos \theta) \propto \sin^m \theta$ near the $\theta \approx 0$ poles. The $m \approx \ell$ harmonics have support only near the equator, $\theta \approx \pi/2$, where $x = \cos \theta \approx 0$.
Our fixed-order recursive method is stable to extremely large values of $\ell$ and $m$ in this regime: for example, taking $\ell = 3 \times 10^9$ and $m = 1.5 \times 10^9$, VSDM finds
\begin{align}
Y_{\,3\,000\,000\,000}^{1\,500\,000\,000}(\theta = 1.570, \phi = 0) &=  -0.335\,001\,108 ,
\end{align}
which matches the result from Mathematica~12 to better than one part in $10^{9}$. 

Away from the equator, however, the $m \ll \ell$ harmonics become inaccurate at $\ell > 1800$. 
For example, 
\begin{align}
Y_{2000}^{500}(2.91, 0) &\approx 3.90715 \cdot 10^{-6}, 
\end{align}
but at $\theta = 2.92$ the result drops abruptly (and unphysically) to zero. 
So, in applications that require a precision of better than $10^{-6}$, our $\ell \gtrsim 2000$ spherical harmonics are not {guaranteed} to be accurate.

At $\ell = 1800$, discontinuities of this type still occur, but only once $Y_\ell^m$ itself is quite small. For example, we find
\begin{align}
Y_{1800}^{700}(2.78, 0) &\simeq 8.039\,439 \cdot 10^{-10}, 
&
Y_{1800}^{700}(2.79, 0) \rightarrow 0,
\end{align}
losing precision above above $\theta > 2.785$ (where $Y_\ell^m(\theta, 0) \lesssim 1.1 \cdot 10^{-11}$). This value of $m\sim 700$ roughly maximizes the error for the $\ell = 1800$ harmonics.
Because this error remains smaller than $10^{-10}$  at all values of $\theta$, we say that our spherical harmonic evaluation remains precise up to $\ell_\text{max} = 1800$.

For $\ell \leq 1800$, we find smooth results for the full domain of $|x| \leq 1$, for all $| m | \leq \ell$, 
even in the polar regions where $Y_\ell^m$ is exponentially small. Taking $\ell = 10^3$ for a final demonstration,
\begin{align}
Y_{1000}^{100}(3.1, 0) &= 9.603\,018\,299\,15 \times 10^{-28}, 
\\
Y_{1000}^{500}(2.8, 0) &= 8.399\,092\,757\,04 \times 10^{-52}\, 
\\
Y_{1000}^{997}(1.9, 0) &= 2.583\,063\,136\,80 \times 10^{-21} .
\end{align}
These values of $\ell$ are an order of magnitude larger than the largest values of $\ell_\text{max}$ that we expect to need for VSDM.

In the Python package \texttt{vsdm}, the spherical harmonics and partially normalized associated Legendre polynomials are given by: 
\begin{align}
\texttt{plm\_norm}(\ell, m, x) &= \mathcal P_\ell^m(x), 
\\
\texttt{ylm\_cx}(\ell, m, \theta, \phi) &= Y_\ell^m(\theta, \phi), 
\\
\texttt{ylm\_real}(\ell, m, \theta, \phi) &= Y_{\ell m} (\theta, \phi),
\end{align}
with $\theta$ and $\phi$ in radians.

\section{Spherical Wavelet and Tophat Basis Functions}
\label{sec:tophats}

The $d$-dimensional spherical Haar wavelets and spherical tophat functions are both normalized according to
\begin{align}
\int_0^1 \! x^{d-1} dx \, r_{n}(x) r_{n'}(x) &= \delta_{n n'}  
\end{align} 
for $x \in [0, 1]$. 
The $d=3$ wavelet functions $r_n(x) = h_n(x)$ are already provided in Section~\ref{sec:wavelets}. For arbitrary values of $d$, the wavelet base of support remains the same as in \eqref{def:x123}, but the values of $A_n$ and $B_n$ change:
\begin{align}
{h}_{n} (x) = 
\left\{
\begin{array}{r c l}
+A_{n} && x_1(n) \leq x <  x_2(n) ,
\\[4pt]
-B_{n} &&  x_2(n)  < x \leq x_3(n) ,
\\[4pt]
0 && \text{otherwise} ,
\end{array}
\right.
\end{align}
where
\begin{align}
x_1(n) = 2^{-\lambda} \mu,
&&
x_2(n) = 2^{-\lambda} (\mu + \tfrac{1}{2} ), 
&&
x_3(n) = 2^{-\lambda} (\mu + 1) ,
\end{align}
and 
\begin{align}
A_{n} &= \sqrt{ \frac{d}{x_3^d - x_1^d} \frac{x_3^d - x_2^d }{x_2^d - x_1^d} }
&
B_{n} &= \sqrt{ \frac{d}{x_3^d - x_1^d} \frac{x_2^d - x_1^d}{x_3^d - x_2^d } } .
\end{align}
For the $n=0$ wavelet,  
\begin{align}
A_0 &\equiv \sqrt{d}, 
&
B_0 &\equiv 0.
\end{align}
The $d=1$ case returns the standard Haar wavelets.  
In the Python implementation, \texttt{vsdm}, the value of $d$ can be altered from its default ($d = 3$) by adding a \texttt{dim} entry to the basis dictionary: e.g.,
\begin{minted}[bgcolor=pink]{Python}
basisV = dict(uMax=820*km_s, type='wavelet', dim=1)
\end{minted}

\paragraph{Tophat Functions:}

Some of the intermediate steps of the calculation are faster when evaluated using piecewise-constant tophat basis functions, and then applying a discrete wavelet transformation to obtain the wavelet coefficients.  (See Section~\ref{sec:shortcuts}.)
Unlike the wavelets, the tophat basis functions are allowed to have varying widths. 
A list of boundary points $x_0, x_1, x_2, x_3, \ldots, x_N, x_{N+1}$ defines a basis with $N$ many tophat functions. The $n$th basis function ($n = 0, 1, \ldots, N-1$) is defined as 
\begin{align}
{t}_{n} (x) &= 
\left\{
\begin{array}{r c l}
+T_{n} && x_n \leq x <  x_{n+1} ,
\\[4pt]
0 && \text{otherwise} ,
\end{array}
\right. 
&
T_{n} &= \sqrt{ \frac{d}{x_{n+1}^d - x_n^d} } .
\end{align}
The implementation of \texttt{Tophat} in Julia is discussed in Section~\ref{subsec:juproj}. 
In Python, the list of boundary points for a dimensionful parameter $u$ is included as an item \texttt{uiList} in the basis dictionary:
\begin{minted}[bgcolor=pink]{Python}
basisV = dict(uiList=[j*km_s for j in range(0,830,20)], type='tophat')
\end{minted}
The list \texttt{uiList} must be ordered, but it is not mandatory for the smallest value to be zero. The dimensionful quantity $u$ is converted to dimensionless $x$ via 
\begin{align}
x = \frac{u}{u_\text{max}}. 
\end{align}

\section{Python Example Code} \label{appx:PyExample}

Section~\ref{sec:python} provides multiple examples for each intermediate stage of the VSDM calculation, in more detail than is typically needed for a single analysis. In this section we provide a simpler summary, showing all the steps needed for a complete analysis. The result is simpler than it might appear in Section~\ref{sec:python}.

This example calculates $\langle g_\chi | n \ell m \rangle$ from an analytic function $g_\chi(\vec v)$ defined by the user, 
while importing the $\langle f_s^2 | n \ell m \rangle$ coefficients from a CSV file in the \texttt{vsdm} format. Such a file could have been generated by running \texttt{vsdm.Fnlm}; or, it could have been made outside of \texttt{vsdm}, e.g.~by the user's implementation of the discrete wavelet-harmonic transform of Section~\ref{sec:shortcuts}. 

\begin{minted}[bgcolor=pink]{Python}
from vsdm import *
import numpy as np
import quaternionic
# Step 1: project gX(v) onto wavelet-harmonic basis:
# 1a. define a gX(uSph) function:
def gX_v(vSph):
    v, theta, phi = vSph
    gX = ... # a function of v, theta, and phi 
    return gX # in units of (inverse velocity)**3
gX_v.phi_symmetric = False # True, if gX is azimuthally symmetric
# 1b. define velocity basis functions
basisV = dict(uMax=820*km_s, type='wavelet')
# 1c. get <gX|nlm> coefficients 
nlmlist = [(n,l,m) for n in range(2**9) for l in range(13) 
           for m in range(-l, l+1)]
integ_params = dict(method='gquad', rtol_f=1e-5, atol_f=1e-8)
gX = EvaluateFnlm(basisV, gX_v, integ_params, nlmlist=nlmlist, 
                  csvsave_name='gXnlm.csv', f_type='gX', 
                  use_gvar=False)
gX._makeFarray() # populate the f_lm_n arrays for later
# Step 2: import <fs2|nlm> coefficients from a CSV file
# 2a. define momentum basis functions (must match CSV file!)
basisQ = dict(uMax=10*qBohr, type='wavelet')
# 2b. initialize an empty Fnlm object and import CSV coefficients:
fs2 = Fnlm(basisQ, f_type='fs2', use_gvar=False)
fs2.center_Z2 = True # add symmetry properties of fs2 as needed
fs2.importFnlm_csv('fs2_6eV.csv')
fs2.DeltaE = 6.0*eV # optional: keep track of excitation energy here
fs2._makeFarray() # populate the f_lm_n arrays for later
# Step 3: make an McalI matrix for specific DM model 
modelDMSM = dict(mX=20*MeV, fdm=(-4,2), DeltaE=fs2.DeltaE, mSM=mElec)
mI = McalI(basisV, basisQ, modelDMSM, mI_shape=(13,512,fs2.nMax+1),
           do_mcalI=True, center_Z2=True) 
# Step 4: make the McalK partial rate matrix
mK = RateCalc(gX, fs2, mI, lmod=2) # center-symmetric: lmod=2
# Step 5: calculate Wigner G matrix for a list of rotations
# rotating the detector about an axis 'nHat' by angle 'beta'
nHat = quaternionic.array([0, 3.2, -1.2, 2.7])
nHat = nHat/nHat.abs # nHat: a 3d unit vector
qR = quaternionic.array([1, 0, 0, 0]) 
rotations = [np.cos(beta*np.pi/360)*qR + np.sin(beta*np.pi/360)*nHat 
             for beta in range(0, 360, 3)]
wG = WignerG(mK.ellMax, rotations=rotations, lmod=2)
# Step 6: calculate scattering rate for each rotation
mass_kg = 1.0 # 1 kg detector
time_yr = 1.0 / SECONDS_PER_YEAR # 1 second exposure time
N_events = mK.Nevents(wG, exp_kgyr=mass_kg*time_yr, mCell_g=125., 
                      sigma0_cm2=1e-37, rhoX_GeVcm3=0.4)
\end{minted}
The final line produces a list, \texttt{N\_events}, for the scattering rate (in events per second) for each orientation of the detector, in the same order as the \texttt{rotations} list. The hypothetical detector has a mass of 1.0\,kg, and the molar mass for the $f_S^2$ unit cell is $m_\text{cell}^{(\text{mol})} = 125\,\text{g}$. The Julia version of the \texttt{ExposureFactor} would have taken $N_T$, the number of unit cells, as an input; in this case, $N_T = N_A M_T / m_\text{cell}^{(\text{mol})} = 8.0 \, N_A$, where $N_A$ is the Avogadro constant. 

This example code works for either discrete or continuous final states. For discrete final states, \texttt{fs2\_6eV.csv} provides the function $f_{g \rightarrow s}^2(\vec q)$, and 
\begin{align}
f_S^2(\vec q, E) &= \texttt{fs2}(\vec q) \, \delta( E - \Delta E).
\end{align}
Here \texttt{mK.Nevents} is dimensionless, counting the total number of scattering events that produce this $6\,\text{eV}$ excited state. 

Alternatively, \texttt{fs2\_6eV.csv} could just as well provide the $\langle f_S^2(E) | n \ell m \rangle$ coefficients at fixed $E = 6\,\text{eV}$, in which case 
\begin{align}
\texttt{fs2}(\vec q)  &= f_S^2(\vec q, 6\,\text{eV})
\end{align}
has units of inverse energy. Likewise, \texttt{mK.Nevents} counts the number of events expected within the $6\,\text{eV}$ energy bin, and 
\begin{align}
\texttt{N\_events}(\mathcal R) &= T_\text{exp} \frac{dR(\mathcal R)}{dE} \bigg|_{E = 6\,\text{eV}}. 
\end{align}

\end{appendix}

\bibliography{vsdm_refs.bib}

\end{document}